\documentclass[a4paper]{report}
\pdfoutput=1
\usepackage[a4paper,left=2cm,right=2cm,top=2cm,bottom=2cm]{geometry}

\usepackage{multicol}
\usepackage{calc}
\usepackage{enumitem}
\usepackage{moreenum}
\usepackage{environ}
\usepackage[german,english]{babel}
\usepackage{amsmath}
\usepackage{amssymb}
\usepackage{amsthm}
\usepackage{mathtools}
\usepackage{stmaryrd}
\usepackage{amsthm}
\usepackage{wasysym}
\usepackage[utf8]{inputenc}
\usepackage{mathrsfs}
\usepackage{latexsym}
\usepackage{array}
\usepackage{color}
\usepackage{siunitx}
\usepackage{graphicx}
\usepackage{float}
\usepackage{pgfplotstable}
\usepackage{tikz}
\usepackage{nicefrac}
\usetikzlibrary{chains,arrows,matrix,cd,calc}
\usetikzlibrary{decorations.pathreplacing, decorations.pathmorphing, patterns}
\usepackage{epstopdf}
\usepackage{subfigure}
\usepackage{enumitem}
\usepackage[justification=centering]{caption}
\usepackage{import}
\usepackage{xparse}
\usepackage{polynom}
\allowdisplaybreaks

\DeclareMathOperator{\arcosh}{arcosh}

\DeclareMathOperator{\laplace}{\Delta}
\def\mathclap#1{\text{\hbox to 0pt{\hss$\mathsurround=0pt#1$\hss}}}

\newcommand{\ud}{\,\mathrm{d}}
\renewcommand{\d}{\,\mathrm{d}}
\newcommand\abs[1]{\left|#1\right|}

\NewDocumentCommand{\overarrow}{O{=} O{\uparrow} m}{%
    \overset{\makebox[0pt]{\begin{tabular}{@{}c@{}}#3\\[0pt]\ensuremath{#2}\end{tabular}}}{#1}
}
\NewDocumentCommand{\underarrow}{O{=} O{\downarrow} m}{%
    \underset{\makebox[0pt]{\begin{tabular}{@{}c@{}}\ensuremath{#2}\\[0pt]#3\end{tabular}}}{#1}
}
\theoremstyle{definition} \newtheorem{defn}{Definiton}
\theoremstyle{definition} \newtheorem{thm}[defn]{Theorem}
\theoremstyle{remark} \newtheorem{rem}[defn]{Remark}
\theoremstyle{remark} \newtheorem{cor}[defn]{Corollary}
\newcommand{\gdw}{\Leftrightarrow}

\newcommand{\f}{\Rightarrow}
\newcommand{\fa}[1]{\forall_{#1}\,}

\newcommand{\R}{\mathbb{R}}

\newcommand{\m}[1]{\ensuremath{\left\{#1\right\} }}

\newcommand{\sD}{\mathscr{D}}

\newcommand{\sF}{\mathscr{F}}

\newcommand{\sM}{\mathscr{M}}

\mathtoolsset{showonlyrefs}
\newcommand{\eq}[1]{\begin{align}#1\end{align}}
\newcommand{\ileq}[1]{$\begin{aligned}#1\end{aligned}$}
\newcommand{\fleq}[1]{\begin{flalign}#1\end{flalign}}
\newcommand{\piecewise}[1]{\begin{cases}#1\end{cases}}
\newcommand{\ubr}[2]{\underbrace{#1}_{#2}}
\newcommand{\obr}[2]{\overbrace{#1}^{#2}}
\newcommand{\ubs}[2]{\underset{#2}{#1}}

\newcommand{\landauO}[0]{\mathcal{O}}
\newcommand{\ctr}[0]{\mathrm{Tr}_\varphi}
\newcommand{\tchn}[1]{{\color{gray} #1 }}

\graphicspath{{.}}
\title{
}
\author{Timo Schmetzer}
\date{\today}
\usepackage{chngcntr}
\counterwithout{section}{chapter}
\begin{document}
\setlength{\parindent}{0pt}
\selectlanguage{English}
\begin{titlepage}
    \centering
    \huge Bachelorarbeit\\
    \Large zur Erlangung des akademischen Grades Bachelor of Science Physik\\
    \vspace{4cm}
    \textbf{Elektrostatische Wechselwirkung zwischen nicht identischen geladenen Teilchen an einer elektrolytischen Grenzfl\"ache}\\
    \vspace{0.5cm}
    \textbf{Electrostatic interaction between non-identical charged particles at an electrolyte interface}\\
    \vspace{4cm}
    Timo Schmetzer\\
    \vspace{4cm}
    Betreuer: Dr. Arghya Majee\\
    Pr\"ufer: Priv.-Doz. Dr. Markus Bier\\
    \vspace{2cm}
    Max-Planck-Institut f\"ur intelligente Systeme Stuttgart\\
    Universit\"at Stuttgart\\
    \vspace{1.25cm}
    15. September 2017\\
    \vspace{1.25cm}
    Korrigierte Version vom 15. Oktober 2017\\
    \vspace{.05cm}
    \normalsize In dieser Version wurden einige nach der Abgabe gefundene Fehler korrigiert.
\end{titlepage}
\begin{abstract}
    In this thesis we study the lateral electrostatic interaction between a pair of non-identical,
    moderately charged colloidal particles trapped at an electrolyte interface in the limit of short
    inter-particle separations. Using a simplified model system we solve the problem
    analytically within the framework of linearised Poisson-Boltzmann theory and classical
    density functional theory. In the first step, we calculate the electrostatic potential
    inside the system exactly as well as within the widely used superposition approximation.
    Then these results are used to calculate the surface and line interaction energy densities
    between the particles. Contrary to the case of identical particles, depending upon the
    parameters of the system, we obtain that both the surface and the line interaction can vary
    non-monotonically with varying separation between the particles and the superposition
    approximation fails to predict the correct qualitative behaviours in most cases. Additionally,
    the superposition approximation is unable to predict the energy contributions quantitatively
    even at large distances. We also provide expression for the constant (independent of the
    inter-particle separation) interaction parameters, i.e., the surface tension, the line tension and
    the interfacial tension. Our results are expected to be of use for modelling particle-interaction
    at fluid interfaces and, in particular, for emulsion stabilization using oppositely charged particles.
\end{abstract}
\begingroup
\selectlanguage{german}
\begin{abstract}
    In dieser Arbeit wird die laterale elektrostatische Wechselwirkung zwischen einem Paar
    nicht identischer, nicht zu stark geladener Kolloidteilchen, die sich an einer Grenzfl\"ache
    zwischen zwei elektrolytischen L\"osungen befinden im Grenzfall kleiner Teilchenabst\"ande
    diskutiert.
    Wir l\"osen das Problem in einem vereinfachten Modellsystem analytisch mithilfe von
    linearer Poisson-Boltzmann Theorie und klassischer Dichtefunktionaltheorie.
    Als erstes berechnen wir das elektrostatische Potential in dem System exakt
    und im Rahmen der h\"aufig verwendeten Superpositionsn\"aherung.
    Wir benutzen diese Ergebnisse, um die Oberfl\"achen- und Linienwechselwirkungsenergiedichten
    zwischen den Teilchen zu berechnen. Im Gegensatz zum Fall identischer Teilchen
    kann sowohl die Oberfl\"achenenergiedichte als auch die Linienenergiedichte
    eine nicht monotone Ver\"anderung bez\"uglich des Teilchenabstands aufweisen.
    Die Superpositionsn\"aherung kann das Verhalten der Energiebeitr\"age in
    den meisten F\"allen qualitativ nicht richtig wiedergeben.
    Die Superpositionsn\"aherung kann die Energiebeitr\"age quantitiv nicht einmal
    f\"ur gro{\ss}e Teilchenabst\"ande korrekt wiedergeben.
    Wir berechnen ebenfalls die Energiebeitr\"age,
    die nicht vom Teilchenabstand abh\"angen, wie die Oberfl\"achenspannung,
    die Linienspannung und die Grenzfl\"achenspannung.
    Die Ergebnisse sollten zur Modellierung der Teilcheninteraktion
    an Fl\"ussigkeitsgrenzschichten und der Emulsionsstabilisierung durch
    entgegengesetzt geladene Teilchen verwendet werden k\"onnen.
\end{abstract}
\endgroup

\tableofcontents
\chapter{Introduction}
\section{Charged colloids at an electrolyte interface}
Suspended colloidal particles can get trapped at a liquid-liquid interface
if the decrease in system energy due to the reduction of the interfacial area
is larger than the thermal energy.
Typically this adsorption energy is much larger than $k_BT$ and thus
particles are virtually irreversibly trapped at the interface \cite{lit}.
This effect was discovered by Ramsden in 1903 \cite{ramsden}.
The most popular application of this phenomenon is stabilization of emulsions
\cite{dickinson,tambe}. There is also a range of industrial applications,
as discussed in Ref. \cite{horozov}.
\\

The stability of such an effectively two dimensional system depends on the lateral
forces between the particles.
Forces acting on the particle generally include the van der Waals force,
which is the dominating force at small distances and attractive capillary interaction,
which is dominating at large distances.
These attractive interactions can, however, lead to accumulation and finally
coagulation of the particles. It is therefore desirable to have some additional
repulsive force for the stabilization of the particles, which is often
achieved by using charged colloids.
However, charges can also have an averse effect on the formation of 
such a system and can preclude its formation by an electrostatic
image force between a particle approaching the interface
and an image charge of the same sign, that is caused by the
dielectric jump at the interface \cite{imgch}.
\\

This electrostatic interaction between colloids has received much attention.
Pieranski has shown that for such charge stabilized colloids the
electrostatic force can be described as the interaction between
dipoles originating due to charge asymmetry around the particle \cite{pieranski}.
The overall electrostatic interaction comprises of a screened coulombic part
in polar media and a unscreened part in the non-polar media.
Hurd calculated both the power law and exponential contributions to such a
system within linear Poisson-Boltzmann theory for point-like particles \cite{hurd}.
\\

Later, this work has been extended in numerous directions by others
\cite{pieranski,park,aveyard,aveyard2},
but almost all studies on these systems have looked at the case of particles
being far away from each other, where linear superposition is a commonly used
approximation. However, Ref. \cite{lit} has shown that a superposition
approximation is unreliable even for large distances
and shows qualitative differences for short separations.
However, the limit of short particle separations is often encountered for
aggregating systems and systems with a high number density \cite{toor}.
\\

The results presented in Ref. \cite{lit} only consider equally charged particles based on the
prevalence of such systems in practice. However in recent years heteroaggregation
(systems featuring different sorts of particles that might vary in charge, size
and form etc.) have been proven to be successful in stabilizing emulsions
\cite{novelstab} and can be used as an alternative to other
more undesirable surfactants \cite{abend}.
From the experimental point of view, stabilization using oppositely charged
particles has also become a standard technique
\cite{novelstab,stabpick,stabpickpattern}.
Due to the reduced net charge carried by a system of differently charged particles they
are easily attached to an interface where stabilization with colloids
is otherwise precluded by the image charge effect \cite{stabpick,stabpickpattern}.
In such systems particles can come to distances even smaller than a nanometre
\cite{stabpick},
highlighting the case for the introduction of a theoretical description
of interaction between non-identical particles in the limit of small separations.
Therefore, in this thesis, we will generalize the model presented in Ref. \cite{lit}
to non-identical particles.
\\

In the case of identical particles the model in Ref. \cite{lit} predicts that
the superposition approximation underestimates interaction energies
asymptotically by a factor of two compared to the exact results.
The question is then if this factor is a result of the symmetry of the
system or of something else. This work seeks to answer this question as well
as to find out if there are any qualitative differences in the solution
of the non-identical case compared to the monotonic behaviour of the
solutions found in Ref. \cite{lit}.

\section{The system considered in this thesis}
In this thesis we focus on the electrostatic interaction between
particles that carry a surface charge at an electrolyte interface.
Surface charges can, for example, be generated by dissolving of charged molecules
from the particle to the electrolyte solution \cite{lit}. 
In general, different particles can carry different surface charges and a particle
can have different surface charges depending on which fluid it does contact.
\\

For simplicity we consider a system with a planar interface and a
contact angle of $\frac{\pi}{2}$ between the particle and the interface,
i.e. we neglect any possible curvature caused by the capillary interaction
and we also ignore the thickness of the interface, which is usually of the
order of the molecular
lengthscale, which is much smaller than the lengthscale of interest here.
\\

Experimental systems that fulfill these assumptions do exist \cite{lit}.

\begin{figure}[H]
    \centering
    \begin{tikzpicture}[
        scale = 1.3,
        wall/.style={
                postaction={draw,decorate,decoration={border,angle=-45,
                            amplitude=0.3cm,segment length=2mm}}},
        ]
        \draw[->] (0,-1.5) -- (0,2) node (xaxis) [above] {$x$};
        \draw[->] (-3,0) -- (3,0) node (zaxis) [right] {$z$};
        \draw[very thick] (0,0) -- (2,0);
        \draw[line width=1pt, pattern={north west lines}] (-1.5, 0) circle (1.5);
        \draw[line width=1pt, pattern={north west lines}] (3.5, 0) circle (1.5);
        \draw (-1.5, 1.75) node {$\sigma_1$};
        \draw (-1.5, -1.75) node {$\sigma_3$};
        \draw (3.5, 1.75) node {$\sigma_2$};
        \draw (3.5, -1.75) node {$\sigma_4$};
        \draw (1, 0.75) node {$\epsilon_1,\kappa_1$};
        \draw (1,-0.75) node {$\epsilon_2,\kappa_2$};
        \draw (0,-1.5) node [below] {$0$};
        \draw (2,-1.5) node [below] {$L$};
    \end{tikzpicture}
    \caption{
        Charged colloidal particles are trapped at a flat electrolyte interface.
        The spherical colloidal particles are separated by a distance of $L>0$ and
        can carry four different surface charge
        densities $\sigma_1$, $\sigma_2$, $\sigma_3$, and $\sigma_4$ depending
        on the particle and the contacted fluid.
        The electrolytic fluids are characterised by their dielectric constants
        $\varepsilon_1$, $\varepsilon_2$ and their inverse Debye lengths $\kappa_1$, $\kappa_2$.
    }
\end{figure}
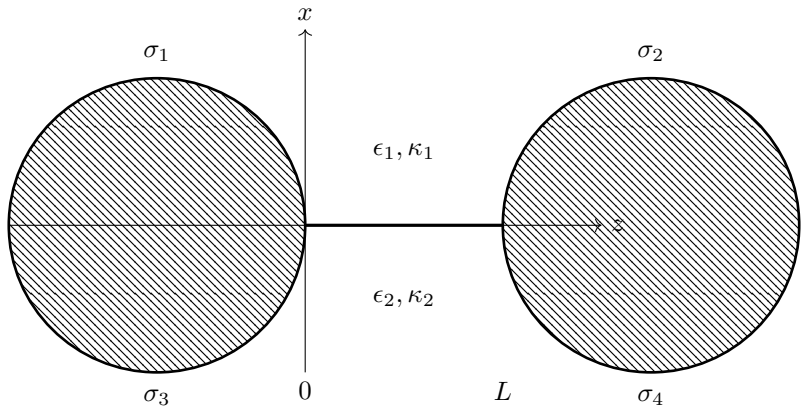

Solutions have been given for a single spherical particle \cite{sph}, but no exact solution
for two spherical particles is known. We focus on the limit of small distances
as this is important for aggregating particles and therefore for considerations about the
stability of such systems.
Because of this we simplify our System
and assume that for small separations between the colloidal particles the problem
can be approximated by using a model of two planar walls.

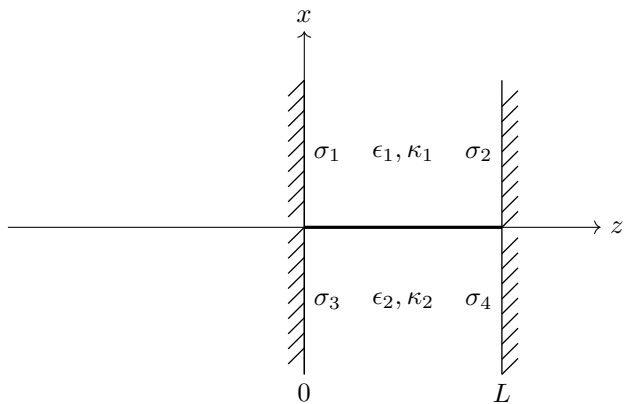
\begin{figure}[H]
    \centering
    \begin{tikzpicture}[
        scale = 1.3,
        wall/.style={
                postaction={draw,decorate,decoration={border,angle=-45,
                            amplitude=0.3cm,segment length=2mm}}},
        ]
        \draw[->] (0,-1.5) -- (0,2) node (xaxis) [above] {$x$};
        \draw[->] (-3,0) -- (3,0) node (zaxis) [right] {$z$};
        \draw[very thick] (0,0) -- (2,0);
        \draw[,  line width=.5pt, wall] (0,1.5)  -- node [right] {$\sigma_1$} (0,0)   ;
        \draw[, line width=.5pt, wall] (0,0)    -- node [right] {$\sigma_3$} (0,-1.5);
        \draw[, line width=.5pt, wall] (2,0)    -- node [left ] {$\sigma_2$} (2,1.5) ;
        \draw[,  line width=.5pt, wall] (2,-1.5) -- node [left ] {$\sigma_4$} (2,0)   ;
        \draw (1, 0.75) node {$\epsilon_1,\kappa_1$};
        \draw (1,-0.75) node {$\epsilon_2,\kappa_2$};
        \draw (0,-1.5) node [below] {$0$};
        \draw (2,-1.5) node [below] {$L$};
    \end{tikzpicture}
    \caption{
        Layout of the simplified model system. The system consists of
        two charged, planar walls at distance $L$ from each other,
        that can carry four different surface charges, depending on the wall and
        the contacted fluid. The electrolytes are separated by a planar interface
        and are characterised by their dielectric constants
        $\varepsilon_1$, $\varepsilon_2$ and their inverse Debye lengths $\kappa_1$, $\kappa_2$.
    }
\end{figure}

The solution strategy presented in this work is the same as in \cite{lit,supp}.
We use density functional theory so we can calculate the energies with scalar quantities.
The overall structure of the thesis is as follows:
\paragraph{Chapter \ref{ch:dft}}
        We use density functional theory to derive a density functional.
        From minimization of the density functional we can obtain the Poisson-Boltzmann equation.
        We linearise the density functional in order to be able to obtain an analytic solution.
        The minimization of this density functional leads to the Debye-H\"uckel equation and an
        expression of the grand canonical potential depending on the potential.
\paragraph{Chapter \ref{ch:pot}}
        We solve the Debye-H\"uckel equation for the boundary conditions of our system.
\paragraph{Chapter \ref{ch:en}}
        We use the potential from the last step and the expression obtained from density functional
        theory to calculate the energy of the system and separate the energy into several
        contributions that can be separately analyzed.
\paragraph{Chapter \ref{ch:conclusion}}
        We summarize our results.

\vspace{2cm}
The intermediate steps of some detailed calculations are given in grey.

\chapter{Density functional Theory}
\label{ch:dft}
\section{Classical density functional Theory}
In this chapter we will give a small introduction to classical density function theory (dft).
In this endeavor we mostly follow Ref. \cite{pdf}.
\begin{defn}
    If the fluid particles do not have any degrees of freedom
    other than position and momentum then we call the fluid a simple fluid.
    Otherwise we call it a complex fluid.
\end{defn}
\begin{cor}
    The microstate of a simple fluid with $N$ particles inside a volume $V\subset\R^d$ is given by
    $\varphi = (r_1,p_1,\dots,r_N,p_N)\in(V\times\R^d)^N$
\end{cor}
\begin{cor}
    The Hamiltonian of a simple fluid with particles of mass $m$ inside an external potential $V(r)$
    with pairwise interaction potential $U(r,r')$ is given by
    \eq{
        H(\varphi) = \frac{1}{2m}\sum_{i=1}^{N(\varphi)} p_i^2
        + \sum_{i=1}^{N(\varphi)} V(r_i)
        + \sum_{\substack{i,j=1\\i<j}}^{N(\varphi)} U(r_i, r_j).
    }
\end{cor}
\begin{defn}
    We define the (classical) trace as
    \eq{
        \ctr := 1 + \sum_{n=1}^\infty\frac{1}{h^{dN}N!}
        \int_{V^N}\d^{dN}r \int_{\R^{dN}}\d^{dN}p.
    }
\end{defn}
\begin{cor}
    The Boltzmann distribution for the grand canonical ensemble has the probability density
    \eq{
        p(\varphi) = \frac{e^{\beta(\mu N(\varphi)-H(\varphi))}}{Z}.
    }
    The grand canonical partition function of a fluid
    with chemical energy $\mu$ per particle and Hamiltonian $H$ is given by
    \eq{
        Z = \ctr\left(e^{\beta(\mu N(\varphi) -H(\varphi))}\right).
    }
\end{cor}
\begin{defn}
    If the single particle density observable is defined as
    \eq{
        \tilde{\rho}^{(1)}(r,\varphi) := \sum_{i=1}^{N(\varphi)} \delta(r-r_i)
    }
    then the single particle density can be defined as
    \eq{
        \rho^{(1)}(r):=\rho(r):= \left<\tilde{\rho}^{(1)}(r,\varphi)\right>
        = \ctr\left(p(\varphi)\tilde{\rho}^{(1)}(r,\varphi)\right).
    }
    Similarly, we define the two particle density observable as
    \eq{
        \tilde{\rho}^{(2)}(r,r',\varphi)
        := \sum_{\substack{i,j=1\\ i\neq j}}^{N(\varphi)} \delta(r-r_i)\delta(r'-r_j)
    }
    and the two particle density as
    \eq{
        \rho^{(2)}(r,r'):=\left<\tilde{\rho}^{(2)}(r,r',\varphi)\right>
        = \ctr\left(p(\varphi)\tilde{\rho}^{(1)}(r,r',\varphi)\right).
    }
    The pair distribution function is defined as
    \eq{
        g(r,r') := \frac{\rho^{(2)}(r,r')}{\rho^{(1)}(r)\rho^{(1)}(r')}.
    }
\end{defn}
\begin{rem}
    For fluids where we only have short range interactions it holds that
    \eq{
        g(r,r')\xrightarrow{\abs{r-r'}\to\infty}1.
    }
\end{rem}
\begin{thm}
    $p(\varphi)$ minimizes the Mermin functional
    \eq{
        \sM[\tilde{p}] := \ctr\left( \tilde{p}(\varphi)
        \left(\ln(\tilde{p}(\varphi)) - \beta\mu N(\varphi) + \beta H(\varphi)\right) \right)
    }
    and the minimum is given by $\beta\Omega=-\ln Z$.
\end{thm}
\begin{defn}
    Let $\rho:V\to\R^+$ a density function and $\tilde{p}(\varphi)$ a probability density
    \eq{
        \tilde{p}|\rho :\gdw \ctr\left(\tilde{p}(\varphi)\tilde{\rho}^{(1)}(r,\varphi)\right) = \rho(r).
    }
\end{defn}
\begin{thm}
    \eq{
        \beta\Omega_0 = \min_{\tilde{p}}\sM[\tilde{p}]
        = \min_{\rho}\min_{\substack{\tilde{p}\\\tilde{p}|\rho}}\sM[\tilde{p}]
    }
\end{thm}
\begin{defn}
    We define the Density functional
    \eq{
        \beta\Omega[\rho] := \min_{\substack{\tilde{p}\\\tilde{p}|\rho}}\sM[\tilde{p}]
    }
\end{defn}
\begin{thm}
    The single particle density in the equilibrium state $\rho_0$ minimizes the density functional
    $\beta\Omega[\rho]$ and $\beta\Omega[\rho_0]=\beta\Omega_0$.
\end{thm}
\begin{cor}
    It follows that
    \eq{
        \frac{\delta\beta\Omega[\rho]}{\delta\rho}\bigg|_{\rho=\rho_0} = 0
    }
    is a necessary condition for the equilibrium single particle density $\rho_0$.
\end{cor}
\begin{thm}
    The density functional of an ideal gas ($U(r_i,r_j) = 0$) is
    \eq{
        \beta\Omega^\mathrm{id}[\rho] =
        \int_V\d^dr \rho(r)(\ln(\rho(r)\Lambda^d) - 1 - \beta\mu + \beta V(r))
    }
    with the thermal de Broglie wavelength
    \eq{
        \Lambda = \sqrt{\frac{2\pi\hbar^2\beta}{m}}
    }
\end{thm}
\begin{rem}
    There is no general method for determining the form of the density functional for $U\neq 0$.
    Instead of computing the exact functional, approximation methods are frequently used.
\end{rem}
\begin{defn}
    We define the excess functional
    \eq{
        \beta F^\mathrm{ex}[\rho] = \beta\Omega[\rho] - \beta\Omega^\mathrm{id}[\rho]
    }
\end{defn}
\begin{rem}
    $\beta F^\mathrm{ex}[\rho]$ depends on $U(r,r')$ but does not depend on $\mu$ or $V(r)$.
\end{rem}
\begin{thm}
    \eq{
        \beta F^\mathrm{ex}[\rho, U] =
        \frac{\beta}{2} \int_V\d^dr\int_V\d^dr' \rho(r)\rho(r')U(r,r')\int_0^1\d\lambda g(r,r',[\rho,U^{(\lambda)}])
    }
    with $U^{(\lambda)} := \lambda U(r,r')$
\end{thm}
\begin{defn}
Random phase approximation (mean field type approximation): $g(r,r')=1$
\end{defn}
\begin{cor}
    For random phase approximation we can write the excess functional as
    \eq{
        \beta F^\mathrm{ex}(\rho, U) = \frac{1}{2}\int_V\d^dr\int_V\d^dr'\beta U(r,r') \rho(r)\rho(r')
    }
\end{cor}
\section{Derivation of the density functional}
In this section we derive a density functional for the following system.
\subsection{The system}
We describe the system in the usual three dimensional Cartesian coordinate system
with axes $x$,$y$,$z$.
Our system consists of two planar surfaces located at $z=0$ and $z=L$ and two
immiscible electrolyte solutions between the walls
forming a flat interface at $x=0$.
We call the medium at $x>0$ medium 1 and the medium at $x<0$ medium 2.
The two media are assumed to be homogeneous and structureless.
\\

\begin{figure}[H]
    \centering
    \begin{tikzpicture}[
        scale = 1.3,
        wall/.style={
                postaction={draw,decorate,decoration={border,angle=-45,
                            amplitude=0.3cm,segment length=2mm}}},
        ]
        \draw[->] (0,-1.5) -- (0,2) node (xaxis) [above] {$x$};
        \draw[->] (-3,0) -- (5,0) node (zaxis) [right] {$z$};
        \draw[very thick] (0,0) -- (4,0);
        \draw[,  line width=.5pt, wall] (0,1.5)  -- node [right] {$\sigma_1$} (0,0)   ;
        \draw[, line width=.5pt, wall] (0,0)    -- node [right] {$\sigma_3$} (0,-1.5);
        \draw[, line width=.5pt, wall] (4,0)    -- node [left ] {$\sigma_2$} (4,1.5) ;
        \draw[,  line width=.5pt, wall] (4,-1.5) -- node [left ] {$\sigma_4$} (4,0)   ;
        \draw (1.25, 1.5) node {$\varepsilon(r)=\varepsilon_1$};
        \draw (1.25,-0.5) node {$\varepsilon(r)=\varepsilon_2$};
        \draw (1.25, 1) node {$I(r)=I_1$};
        \draw (1.25,-1) node {$I(r)=I_2$};
        \draw (1.25, 0.5) node {$V_\pm(r)=0$};
        \draw (1.25,-1.5) node {$V_\pm(r)=f_\pm$};
        \draw (2.75,  1) node {$\zeta_+, \zeta_-$};
        \draw (2.75, -1) node {$\zeta_+, \zeta_-$};
        \draw (0,-1.5) node [below] {$0$};
        \draw (4,-1.5) node [below] {$L$};
    \end{tikzpicture}
    \caption{Layout of our system}
    \label{fig:dftsys}
\end{figure}
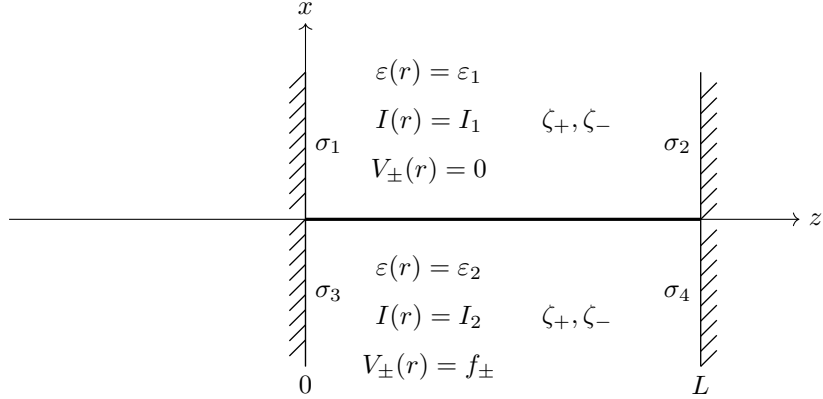

We assume that the walls carry a surface charge stemming from some chemical reaction with
the fluid. As such we can have four different surface charges depending at which wall we
look at and which fluid the wall is in contact with. We denote the surface charge at
$z=0,x>0$ as $\sigma_1$, $z=L,x>0$ as $\sigma_2$, $z=0,x<0$ as $\sigma_3$ and
$z=L,x<0$ as $\sigma_4$, as shown in Figure \ref{fig:dftsys}.

Each medium is assumed to be a linear dielectric medium and is characterized by
a constant electrical permittivity;
$\varepsilon_1$ in Medium 1 and $\varepsilon_2$ in Medium 2. Therefore the permittivity varies
steplike at the interface and
\eq{\varepsilon(r) = \piecewise{\varepsilon_1 & x > 0\\ \varepsilon_2 & x < 0}}

We assume that there are only two monovalent species of ions and that the ion species
contained in both fluids are the same and therefore have the same fugacities $\zeta_\pm$.

We assume that each fluid has a constant bulk ionic strength $I$, that is the concentration of ions without
any outside electrostatic influences. Therefore the bulk ionic strength also varies step like at the interface and
\eq{I(r) = \piecewise{I_1 & x > 0\\ I_2 & x < 0}}

Furthermore, we assume that there is a potential difference of $f_+$ ($f_-$) for the positive (negative) ions
between the two media. Since we are free to choose an offset for these potentials,
we simply set the potential $V_\pm$ to zero in medium 1. Therefore we have
\eq{V_\pm(r) = \piecewise{0 & x > 0\\ f_\pm & x < 0}}

For our derivation to work we have to additionally assume hat the resulting Debye screening length $1/\kappa(r)$
which is related to the screening thickness is larger than the size of the molecules contained in our system.

\subsection{The derivation}
The assumption that the Debye screening length is larger than the molecular scale allows us to neglect
layering effects like ion-ion correlation, screening of ions by dipolar solvent molecules
and Stern layers around objects and apply the (mean-field like) random phase approximation to derive
an approximate density functional for our system.
We mostly follow Ref. \cite{pdf2} with some changes to incorporate our wall charges.
\\

For two different particle sorts of ionic particles $i,j$ with different valencies $Z_i,Z_j$ the potential $U$ is
given by
\eq{
    \beta U(r,r') = \frac{Z_iZ_jl_B}{\abs{r-r'}}
}
with the Bjerrum length
\eq{
    l_B := \frac{\beta e^2}{4\pi\varepsilon}
}
which is the length where the electrostatic energy of two particles with a single elementary charge
is equal to the thermal energy $k_BT$.
The charge density due to the ions $\rho_{c,\mathrm{int}}$ is
\eq{
    \rho_{c,\mathrm{int}} = \sum_iZ_ie\rho_i.
}
Using this, we can write
\eq{
    \beta F^\mathrm{ex}(\rho)
    &= \frac{1}{2}\int_V\d^dr\int_V\d^dr'\sum_i\sum_j\frac{Z_iZ_jl_B}{\abs{r-r'}}\rho_i(r)\rho_j(r') &\\
    &= \frac{\beta}{2}\int_V\d^dr\sum_iZ_ie\rho_i(r)\frac{1}{4\pi\varepsilon}\int_V\d^dr'\frac{\sum_jZ_je\rho_j(r')}{\abs{r-r'}} &\\
    &= \frac{\beta}{2}\int_V\d^dr\rho_{c,\mathrm{int}}(r)\frac{1}{4\pi\varepsilon}\int_V\d^dr'\frac{\rho_{c,\mathrm{int}}(r')}{\abs{r-r'}} &\\
    &= \frac{\beta}{2}\int_V\d^dr\rho_{c,\mathrm{int}}(r)\phi_\mathrm{int}(\rho,r) &\\
}
where $\phi_\mathrm{int}$ is the potential due to the charge of the ions.\\
We can write the external potential as consiting of a electrostatic part
from charges outside the system volume or at its borders ($\rho_{c,\mathrm{ext}}$)
and other contributions
\eq{ V_i(r) = \bar{V}_i(r) + Z_ie\phi_\mathrm{ext}(r) }
where $\phi_\mathrm{ext}$ is the potential due to the external charge distribution $\rho_{c,\mathrm{ext}}$.\\
So the full approximated density functional is
\eq{
    \beta\Omega[\rho]
    &= \int_V\d^dr \sum_i\rho_i(r)(\ln(\rho_i(r)\Lambda_i^d) - 1 - \beta\mu_i + \beta V_i(r))
    + \frac{\beta}{2}\rho_{c,\mathrm{int}}(r)\phi_\mathrm{int}(\rho,r)&\\
    &= \int_V\d^dr \sum_i\rho_i(r)(\ln(\rho_i(r)\Lambda_i^d) - 1 - \beta\mu_i + \beta \bar{V}_i(r))
    + \beta\sum_iZ_ie\rho_i(r)\phi_\mathrm{ext}(r) + \frac{\beta}{2}\rho_{c,\mathrm{int}}(r)\phi_\mathrm{int}(\rho,r)&\\
    &= \int_V\d^dr \sum_i\rho_i(r)\left(\ln\left(\rho_i(r)\frac{\Lambda_i^d}{\exp(\beta\mu_i)}\right) - 1 + \beta \bar{V}_i(r)\right)
    + \beta\rho_{c,\mathrm{int}}\phi_\mathrm{ext}(r) + \frac{\beta}{2}\rho_{c,\mathrm{int}}(r)\phi_\mathrm{int}(\rho,r)&\\
    &= \int_V\d^dr \sum_i\rho_i(r)\left(\ln\left(\frac{\rho_i(r)}{\zeta_i}\right) - 1 + \beta \bar{V}_i(r)\right)
    + \beta\rho_{c.\mathrm{int}}\phi_\mathrm{ext}(r) + \frac{\beta}{2}\rho_{c,\mathrm{int}}(r)\phi_\mathrm{int}(\rho,r)&
}
with the fugacity
\eq{
    \zeta_i := \frac{\exp(\beta\mu_i)}{\Lambda_i^d}.
}
We rewrite the last part of the expression
\fleq{
    & \beta\int_V\d^dr \rho_{c.\mathrm{int}}\phi_\mathrm{ext}(r)
        + \frac{1}{2}\rho_{c,\mathrm{int}}(r)\phi_\mathrm{int}(\rho,r) &\\
    &\quad\text{since } \rho_{c,\mathrm{int}} = \nabla\cdot D_\mathrm{int}(\rho, r) &\\
    &= \beta\int_V\d^dr (\nabla\cdot D_\mathrm{int}(\rho,r))\phi_\mathrm{ext}(r)
        + \frac{1}{2}(\nabla\cdot D_\mathrm{int}(\rho,r))\phi_\mathrm{int}(\rho,r) &\\
    &\quad\text{since } (\nabla\cdot D_\mathrm{int}(\rho, r))\phi_\mathrm{int}(\rho,r) = 
    \nabla\cdot (\phi_\mathrm{int}(\rho, r)D_\mathrm{int}(\rho, r))
    - D_\mathrm{int}(\rho, r)\cdot \!\!\!\!\ubr{\nabla\phi_\mathrm{int}(\rho, r)}
    {\displaystyle =-\frac{D_\mathrm{int}(\rho,r)}{\varepsilon(r)}} \text{+ Gau\ss} &\\
    &= \beta\int_{\partial V} \phi_\mathrm{ext}(r) \ubr{n(r)\cdot D_\mathrm{int}(\rho,r)}{=0}\d^{d-1}r
    + \frac{1}{2}\beta\int_{\partial V} \phi_\mathrm{int}(r) \ubr{n(r)\cdot D_\mathrm{int}(\rho,r)}{=0}\d^{d-1}r &\\
    &+ \frac{1}{2}\beta\int_V\d^dr 2 D_\mathrm{int}(\rho,r))\frac{D_\mathrm{ext}(r)}{\varepsilon(r)}
    + \frac{1}{2}\beta\int_V\d^dr D_\mathrm{int}(\rho,r))\frac{D_\mathrm{int}(\rho,r)}{\varepsilon(r)} &\\
    &= \frac{1}{2}\beta\int_V\d^dr \frac{(D_\mathrm{int}(\rho,r)+D_\mathrm{ext}(r))^2}{\varepsilon(r)}
    - \frac{1}{2}\beta\int_V\d^dr \frac{D_\mathrm{ext}(r)^2}{\varepsilon(r)} &\\
    &= \frac{1}{2}\beta\int_V\d^dr \frac{D(\rho,r)^2}{\varepsilon(r)}
    - \frac{1}{2}\beta\int_V\d^dr \frac{D_\mathrm{ext}(r)^2}{\varepsilon(r)} &
}
Since the last term is constant we can ignore it for the energy calculation and we rename $\bar{V}$ to $V$.
Our functional therefore is
\eq{
    \beta\Omega[\rho]
    &= \int_V\d^dr \sum_i\rho_i(r)\left(\ln\left(\frac{\rho_i(r)}{\zeta_i}\right) - 1 + \beta V_i(r)\right)
    + \frac{\beta D(r,\rho)^2}{2\varepsilon(r)} &
}

\section{Derivation of the non-linear Poisson-Boltzmann equation}
For our system we assume that there are only two different singly and oppositely charged ion sorts, i.e.,
\eq{ i\in\m{+,-}, Z_+ = +1, Z_- = -1 .}
The density functional then has the form
\eq{
    \beta\Omega[\rho]
    &= \int_V\d^dr \sum_{i\in\pm}\rho_i(r)\left(\ln\left(\frac{\rho_i(r)}{\zeta_i}\right) - 1 + \beta V_i(r)\right)
    + \frac{\beta D(r,\rho)^2}{2\varepsilon(r)}
}
Therefore,
\eq{
    \delta_{\rho_i}\beta\Omega[\rho]
    &= \int_V\d^dr \left(\delta_{\rho_i}\rho_i(r)\right)\left(\ln\left(\frac{\rho_i(r)}{\zeta_i}\right) - 1 + \beta V_i(r)\right)
    + \rho_i(r) \frac{\delta_{\rho_i}\rho_i(r)}{\zeta_i}\frac{\zeta_i}{\rho_i(r)}
    + \frac{\beta D(r,\rho)}{\varepsilon(r)}\cdot\delta_{\rho_i}D(r,\rho) &\\
    &= \int_V\d^dr \left(\delta_{\rho_i}\rho_i(r)\right)\left(\ln\left(\frac{\rho_i(r)}{\zeta_i}\right) + \beta V_i(r)\right)
    + \frac{\beta D(r,\rho)}{\varepsilon(r)}\cdot\delta_{\rho_i}D(r,\rho). &
}
The last term can be expressed in the following way.
\eq{
       \int_V\d^dr \frac{\beta D(r,\rho)}{\varepsilon(r)}\cdot\delta_{\rho_i}D(r,\rho) 
    &= \int_V\d^dr \beta E(r,\rho)\cdot\delta_{\rho_i}D(r,\rho) &\\
    &= \int_V\d^dr \beta (-\nabla\Phi(r,\rho))\cdot\delta_{\rho_i}D(r,\rho) &\\
    &= -\beta\int_V\d^dr \nabla\Phi(r,\rho)\cdot\delta_{\rho_i}D(r,\rho) &\\
    &= -\beta\int_V\d^dr \nabla(\Phi(r,\rho)\cdot\delta_{\rho_i}D(r,\rho)) - \Phi(r,\rho)\nabla\cdot \delta_{\rho_i}D(r,\rho) &\\
    &= -\beta\int_{\partial V}\d^{d-1}r \Phi(r,\rho)n(r)\cdot\delta_{\rho_i}D(r,\rho)
    + \beta\int_V\d^dr \Phi(r,\rho)\nabla\cdot \delta_{\rho_i}D(r,\rho) &\\
    &= -\beta\int_{\partial V}\d^{d-1}r \Phi(r,\rho)\delta_{\rho_i}(n(r)\cdot D(r,\rho))
    + \beta\int_V\d^dr \Phi(r,\rho)\delta_{\rho_i}(\nabla\cdot D(r,\rho)) &\\
    &= -\beta\int_{\partial V}\d^{d-1}r \Phi(r,\rho)\delta_{\rho_i}(-\sigma(r))
    + \beta\int_V\d^dr \Phi(r,\rho)\delta_{\rho_i}\Big(\sum_{j\in\pm} eZ_j\rho_j(r)\Big) &\\
    &= -\beta\int_{\partial V}\d^{d-1}r \Phi(r,\rho)\cdot 0
    + \beta\int_V\d^dr \Phi(r,\rho) eZ_i\delta_{\rho_i}\rho_i(r) &\\
}
Using the Euler-Lagrange-Equation
\eq{
    \delta_{\rho_i} \beta\Omega[\rho]= 0
}
we get
\eq{
    \delta_{\rho_i}\beta\Omega[\rho]
    &= \int_V\d^dr \left(\delta_{\rho_i}\rho_i(r)\right)\left(\ln\left(\frac{\rho_i(r)}{\zeta_i}\right) + \beta V_i(r)
    + \beta e Z_i \Phi(r,\rho)\right) = 0,
}
which implies
\eq{
    \ln\left(\frac{\rho_i(r)}{\zeta_i}\right) + \beta V_i(r) + \beta e Z_i \Phi(r,\rho) = 0.
}
We denote the deviations of the ion number densities from the bulk ion density $I(r)$ by
\eq{
    \phi_\pm(r):=\rho_\pm(r) - I(r).
}
The introduction of this quantity is not strictly necessary for this section,
but it will be useful while deriving the linear theory in the next section. 
Using this we get
\eq{
    0 &= \ln\left(\frac{\phi_i(r) + I(r)}{\zeta_i}\right) + \beta V_i(r) + \beta e Z_i \Phi(r,\phi) &\\
    &= \ln\left(\frac{I(r)}{\zeta_i}\right) + \ln\left(1+\frac{\phi_i(r)}{I(r)}\right)
    + \beta V_i(r) + \beta e Z_i \Phi(r,\phi) \label{dft_1}&
}

\paragraph{Bulk of Phase 1}
In the bulk of medium 1,
$I(r) = I_1$, $\beta V_\pm(r) = 0$, $\phi_\pm(r) = 0$, $\Phi(r, \phi) = 0$.
Therefore, Eq. \eqref{dft_1} gives
\fleq{
    &\ln\left(\frac{I_1}{\zeta_i}\right) = 0 \quad\f\fa{i\in\pm} I_1 = \zeta_i & \label{bulk1}
}
\paragraph{Bulk of Phase 2}
In the bulk of medium 2, one has
$I(r) = I_2$, $\beta V_\pm(r) = \beta f_\pm$, $\phi_\pm(r) = 0$, $\Phi(r, \phi) = \Phi_D$, where
$\Phi_D$ is called the Donnan potential or Galvani potential difference \cite{felchem}.
Therefore, Eq. \eqref{dft_1} gives
\fleq{
     &\ln\left(\frac{I_2}{\zeta_i}\right) + 0 + \beta f_i + \beta e Z_i \Phi_D = 0 &\label{bulk2}\\
    \f &\ln\left(\frac{I_2}{I_1}\right) + \beta f_i + \beta e Z_i \Phi_D = 0 &\\
    \f &\ln\left(\frac{I_2}{I_1}\right) + \beta f_+ + \beta e \Phi_D = 0 \label{dft_2}&\\
       &\ln\left(\frac{I_2}{I_1}\right) + \beta f_- - \beta e \Phi_D = 0 \label{dft_3}&\\
    \eqref{dft_2}-\eqref{dft_3}\f &\beta(f_+ - f_-) + 2\beta e\Phi_D = 0 &\\
    \f &\Phi_D = -\frac{1}{2e}(f_+-f_-) &
}
Therefore, the Donnan potential is related to the 
difference in the solubilities of ions in the two liquids.
\fleq{
    \eqref{dft_2}+\eqref{dft_3} \f &2\ln\left(\frac{I_2}{I_1}\right) + \beta(f_++f_-) = 0 &\\
    \f & \frac{I_2}{I_1} = e^{-\frac{\beta(f_+ + f_-)}{2}} &
}
\paragraph{Back to non bulk}
Since $I(r)$, $V_i(r)$, $\zeta_i$ only depend on the bulk properties we can write in general that
\eq{
    \ln\left(\frac{I(r)}{\zeta_i}\right) + \beta V_i(r) + Z_i\beta e\varphi(r) = 0
    \label{Iphi}
}
with
\eq{
    \varphi(r) = \piecewise{0 &x> 0\\\Phi_D &x<0}\,.
}
So it follows that
\eq{
    \ln\left(1+\frac{\phi_i(r)}{I(r)}\right) + \beta e Z_i (\Phi(r,\phi) -\varphi(r)) = 0
}
\eq{
    \phi_i(r) = I(r)(\exp\left(-\beta e Z_i(\Phi(r,\phi) -\varphi(r))\right)-1)
}
Therefore,
\eq{
    \rho_i(r) = I(r)\exp\left(-\beta e Z_i(\Phi(r,\phi) -\varphi(r))\right)
}
Gauß
\eq{
    \nabla\cdot D(r) = \sum_i e Z_i \rho_i(r)
    = \sum_i e Z_i \phi_i(r) + e \ubr{\sum_i Z_i I(r)}{=0} = e\sum_iZ_i\phi_i(r)
}
\fleq{
    & D(r) = -\varepsilon(r)\nabla\Phi(r) &
}
\fleq{
    \f -\varepsilon(r)\laplace\Phi(r) &= e \sum_i Z_i \phi_i(r) &\\
    &= e I(r) (e^{-\beta e(\Phi(r,\phi) - \varphi(r))} - 1 - e^{+\beta e(\Phi(r,\phi) - \varphi(r))} + 1) &\\
    &= - e I(r) 2 \sinh(\beta e (\Phi(r, \phi) - \varphi(r)))
}
so we get the non-linear Poisson-Boltzmann equation
\eq{
    \laplace(\beta e\Phi(r)) =
    \ubr{\frac{2\beta e^2 I(r)}{\varepsilon(r)}}{:=\kappa^2(r)}\sinh(\beta e(\Phi(r) - \varphi(r)))
}

\section{Derivation of the linearisation and the Energies of the linearised System}
An analytical solution for the non-linear Poisson-Boltzmann equation is only known for a system with a single
wall \cite{coldisp}. In order to be able to solve the problem analytically we simplify the system by using
a linearisation.\\

Considering the deviations $\phi_\pm(r)$ of the ion number densities from the bulk ionic strength to be
small, we linearise the funcitonal
\eq{
    \beta\Omega[\rho]
    &= \int_V\d^dr \sum_{i\in\pm}\rho_i(r)\left(\ln\left(\frac{\rho_i(r)}{\zeta_i}\right) - 1 + \beta V_i(r)\right)
    + \frac{\beta D(r,\rho)^2}{2\varepsilon(r)} &\\
    &\color{gray}= \int_V\d^dr \sum_{i\in\pm}I(r)\left(\ln\left(\frac{\phi_i(r) + I(r)}{\zeta_i}\right) - 1 + \beta V_i(r)\right) &\\
    &\color{gray}+ \int_V\d^dr \sum_{i\in\pm}\phi_i(r)\left(\ln\left(\frac{\phi_i(r) + I(r)}{\zeta_i}\right) - 1 + \beta V_i(r)\right) &\\
    &\color{gray}+ \int_V\d^dr \frac{\beta D(r,\rho)^2}{2\varepsilon(r)} &\\
    &\color{gray}= \int_V\d^dr \sum_{i\in\pm}I(r)\left(\ln\left(\frac{I(r)}{\zeta_i}\right) +
       \ln\left(1+\frac{\phi_i(r)}{I(r)}\right) - 1 + \beta V_i(r)\right) &\\
    &\color{gray}+ \int_V\d^dr \sum_{i\in\pm}\phi_i(r)\left(\ln\left(\frac{I(r)}{\zeta_i}\right) +
       \ln\left(1+\frac{\phi_i(r)}{I(r)}\right) - 1 + \beta V_i(r)\right) &\\
    &\color{gray}+ \int_V\d^dr \frac{\beta D(r,\rho)^2}{2\varepsilon(r)} &\\
    &\color{gray}= \int_V\d^dr \sum_{i\in\pm}I(r)\left(\ln\left(\frac{I(r)}{\zeta_i}\right) +
       \frac{\phi_i(r)}{I(r)} - \frac{1}{2} \frac{\phi_i(r)^2}{I(r)^2} + \landauO(\phi_i^3) - 1 + \beta V_i(r)\right) &\\
    &\color{gray}+ \int_V\d^dr \sum_{i\in\pm}\phi_i(r)\left(\ln\left(\frac{I(r)}{\zeta_i}\right) +
       \frac{\phi_i(r)}{I(r)} - \frac{1}{2} \frac{\phi_i(r)^2}{I(r)^2} + \landauO(\phi_i^3) - 1 + \beta V_i(r)\right) &\\
    &\color{gray}+ \int_V\d^dr \frac{\beta D(r,\rho)^2}{2\varepsilon(r)} &\\
    &= \int_V\d^dr \sum_{i\in\pm}I(r)\left(\ln\left(\frac{I(r)}{\zeta_i}\right)
       - 1 + \beta V_i(r)\right) &\label{omegabulkcont}\\
    &+ \int_V\d^dr \sum_{i\in\pm}\phi_i(r)\left(\ln\left(\frac{I(r)}{\zeta_i}\right) +
       \frac{\phi_i(r)}{2I(r)} + \beta V_i(r)\right) &\\
    &+ \int_V\d^dr \frac{\beta D(r,\rho)^2}{2\varepsilon(r)} + \landauO(\phi_i^3)&\\
}
We define
\eq{
    \beta\mathcal{H}[\phi]
    &:= \int_V\d^dr \sum_{i\in\pm}\phi_i(r)\left(\ln\left(\frac{I(r)}{\zeta_i}\right) +
       \frac{\phi_i(r)}{2I(r)} + \beta V_i(r)\right) 
       + \frac{\beta D(r,\rho)^2}{2\varepsilon(r)} + \landauO(\phi_i^3)&
}
this term excludes the $\phi$ independent bulk-contribution \eqref{omegabulkcont} to the energy.
\\

The density then becomes
\eq{
    \phi_i(r) = - I(r)\beta e Z_i (\Phi((r,\phi) - \varphi(r))) + \landauO(\phi^2)
    \label{linphiPhi}
}
this can either be derived by using the linearised density and proceeding as in the non-linearised case
or by linearisation of the analogous equation in the non-linear case.
\\

In the same fashion the linearised Poisson-Boltzmann equation (also known as Debye-Hückel equation)
can either be derived by using the linearised density and proceeding as in the non-linearised case
or by linearisation of the analogous equation in the non-linear case.
The Debye-H\"uckel equation is given by
\eq{
    \laplace(\beta e\Phi(r)) = \kappa^2(r)(\beta e(\Phi(r) - \varphi(r)))
}
with
\eq{
    \kappa^2(r) = \frac{2\beta e^2 I(r)}{\varepsilon(r)}
    =\piecewise{\kappa_1^2 := \frac{2\beta e^2 I_1}{\varepsilon_1} & x > 0 \\
    \kappa_2^2 := \frac{2\beta e^2 I_2}{\varepsilon_2} & x < 0}\,.
}
The linearisation depends on
\eq{
    \abs{\frac{\phi_i(r)}{I(r)}}\ll 1 \qquad\text{or}\qquad \abs{\beta e(\Phi(r) - \varphi(r))} \ll 1
}
As \cite{lit2} shows this assumption can be quite problematic for some systems but
can be achieved by relatively small charges.
\\

We then proceed to rewrite the energy (without the bulk part) in terms of the potential
\fleq{
    \beta\mathcal{H}[\phi]
    &= \int_V\d^dr \sum_{i\in\pm}\phi_i(r)\left(\ln\left(\frac{I(r)}{\zeta_i}\right) +
       \frac{\phi_i(r)}{2I(r)} + \beta V_i(r)\right) 
       + \frac{\beta D(r,\rho)^2}{2\varepsilon(r)} + \landauO(\phi_i^3)&\\
    &\color{gray}\overset{\eqref{Iphi}}{=} \int_V\d^dr \sum_{i\in\pm}\phi_i(r)\left(
       \frac{\phi_i(r)}{2I(r)} -Z_i\beta e\varphi(r)\right) 
       + \frac{\beta D(r,\rho)^2}{2\varepsilon(r)} + \landauO(\phi_i^3)&\\
    &\color{gray}\overset{\eqref{linphiPhi}}{=} \int_V\d^dr \sum_{i\in\pm}\phi_i(r)\left(
       -Z_i\frac{\beta}{2}e(\Phi(\phi,r) - \varphi(r))-Z_i\beta e\varphi(r)\right) 
       + \frac{\beta D(r,\rho)^2}{2\varepsilon(r)} + \landauO(\phi_i^3)&\\
    &\color{gray}= \int_V\d^dr \left(\sum_{i\in\pm}Z_ie\phi_i(r)\right)\left(
       -\frac{\beta}{2}(\Phi(\phi,r) - \varphi(r))-\beta\varphi(r)\right) 
       + \frac{\beta D(r,\rho)^2}{2\varepsilon(r)} + \landauO(\phi_i^3)&\\
    &\color{gray}= \int_V\d^dr (\nabla\cdot D)\left(
       -\frac{\beta}{2}(\Phi(\phi,r) + \varphi(r))\right) 
       + \frac{\beta D(r,\rho)^2}{2\varepsilon(r)} + \landauO(\phi_i^3)&\\
    &\color{gray}= \int_V\d^dr (\nabla\cdot D)\left(
       -\frac{\beta}{2}(\Phi(\phi,r) + \varphi(r))\right) 
       - \frac{\beta}{2}D(r,\rho)\cdot\nabla\Phi(\phi, r) + \landauO(\phi_i^3)&\\
    &\color{gray}= -\frac{\beta}{2}\int_V\d^dr (\nabla\cdot D) (\Phi(\phi,r) + \varphi(r)) 
       + D(r,\rho)\cdot\nabla\Phi(\phi, r) + \landauO(\phi_i^3)&\\
    &\color{gray}\text{using}\quad \nabla\cdot(\Phi D) = \nabla\Phi\cdot D + \Phi(\nabla\cdot D) &\\
    &\color{gray}= -\frac{\beta}{2}\int_V\d^dr \nabla\cdot(\Phi(r) D(r))
    + \varphi(r)(\nabla\cdot D(r))+ \landauO(\phi_i^3)&\\
    &\color{gray}\text{using}\quad \nabla\cdot(\varphi D) = \nabla\varphi\cdot D + \varphi(\nabla\cdot D) &\\
    &\color{gray}= -\frac{\beta}{2}\int_V\d^dr \nabla\cdot(\Phi(r) D(r))
    + \nabla(\varphi(r)D(r)) - D(r)\cdot (\nabla \varphi(r)) + \landauO(\phi_i^3)&\\
    &\color{gray}\overset{\text{Gau\ss}}{=} -\frac{\beta}{2}\Bigg(
    \int_{\partial V}\d^{d-1}r n\cdot(\Phi(r) D(r) + \varphi(r)D(r)) 
    - \int_V\d^dr D(r)\cdot \ubr{(\nabla \varphi(r))}{=-\Phi_D\delta(x)e_x}\Bigg) + \landauO(\phi_i^3)&\\
    &\color{gray}= -\frac{\beta}{2}\Bigg(\int_{\partial V}\d^{d-1}r \ubr{n\cdot D(r)}{=-\sigma(r)} (\Phi(r) + \varphi(r)) 
    +\Phi_D \int_{x=0}\d^{d-1}r D(r)\cdot e_x\Bigg) + \landauO(\phi_i^3)&\\
    &= \frac{\beta}{2}\Bigg(\int_{\partial V}\d^{d-1}r \sigma(r) (\Phi(r) + \varphi(r)) 
    - \Phi_D \int_{x=0}\d^{d-1}r D(r)\cdot e_x\Bigg) + \landauO(\phi_i^3)&
    \label{dften}
}
We will use this expression to calculate the interaction energy parameters of our system in chapter \ref{ch:en}.

\chapter{Electrostatic Potentials}
\label{ch:pot}
\section{Problem}
In this chapter we solve the electrostatic problem for the system we have defined
in the previous chapter:
In a tree-dimensional Cartesian coordinate system we consider
two infinite walls at $z=0$ and $L$, and two electrolyte
solutions in between, the $x=0$ plane forming the interface between them.
The fluid at $x>0$ ($x<0$) is called medium 1 (medium 2) and is
characterized by its dielectric constant $\varepsilon_1$ ($\varepsilon_2$)
and its inverse Debye length $\kappa_1$ ($\kappa_2$).
The upper half ($x>0$) of the wall at $z=0$ ($L$) carries a surface-charge density
$\sigma_1$ ($\sigma_2$), the lower half ($x<0$) carries a surface-charge
$\sigma_3$ ($\sigma_4$).

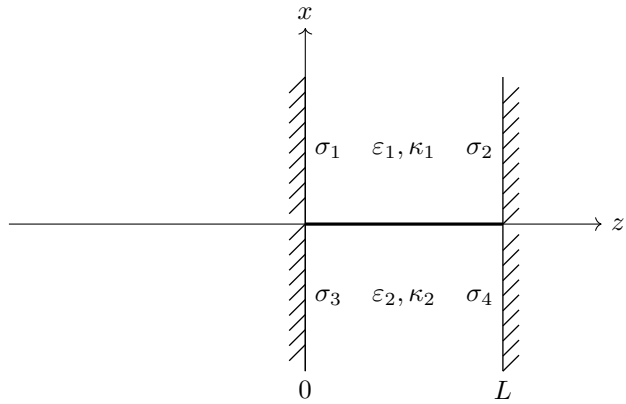
\begin{figure}[H]
    \centering
    \begin{tikzpicture}[
        scale = 1.3,
        wall/.style={
                postaction={draw,decorate,decoration={border,angle=-45,
                            amplitude=0.3cm,segment length=2mm}}},
        ]
        \draw[->] (0,-1.5) -- (0,2) node (xaxis) [above] {$x$};
        \draw[->] (-3,0) -- (3,0) node (zaxis) [right] {$z$};
        \draw[very thick] (0,0) -- (2,0);
        \draw[,  line width=.5pt, wall] (0,1.5)  -- node [right] {$\sigma_1$} (0,0)   ;
        \draw[, line width=.5pt, wall] (0,0)    -- node [right] {$\sigma_3$} (0,-1.5);
        \draw[, line width=.5pt, wall] (2,0)    -- node [left ] {$\sigma_2$} (2,1.5) ;
        \draw[,  line width=.5pt, wall] (2,-1.5) -- node [left ] {$\sigma_4$} (2,0)   ;
        \draw (1, 0.75) node {$\varepsilon_1,\kappa_1$};
        \draw (1,-0.75) node {$\varepsilon_2,\kappa_2$};
        \draw (0,-1.5) node [below] {$0$};
        \draw (2,-1.5) node [below] {$L$};
    \end{tikzpicture}
    \caption{Layout of the problem: Two planar charged walls with four different surface charges
    depending on the wall and the contacted fluid,
    and two different media characterized by their dielectric constants $\varepsilon_i$
    and inverse Debye lengths $\kappa_i$.}
    \label{fig:cp}
\end{figure}

\subsection{Differential equation}
In the previous chapter we have derived the Debye-H\"uckel equation, which we solve in
order to calculate the electrostatic potential for our problem:
\eq{
    \laplace\Phi(r) = \kappa^2(r)(\Phi(r) - \varphi(r))
    \label{dheq}
}
with
\eq{
    \kappa(r) = \piecewise{\kappa_1 & x > 0 \\ \kappa_2 & x < 0}
    \qquad\text{and}\qquad
    \varphi(r) = \piecewise{0 & x > 0 \\ \Phi_D & x < 0}
}
where $\Phi_D$ is the Donnan potential.

\subsection{Boundary conditions}
In order to obtain a unique solution of equation \eqref{dheq} we need to specify our boundary
conditions. We have Neumann boundary conditions at the walls and at the interface
\eq{n\cdot(D_2 - D_1) = \sigma}
which can be obtained using Gau{\ss} Law.
We consider a wall medium such that the electric field immediately vanishes inside the walls
like in an idealized metal-surface. Alternatively if we consider a non-metallic particle this
means we neglect the image charges of the ions in the solution that are inside the wall.
\\

Therefore we have the following boundary conditions:
\begin{enumerate}
    \item
        At the walls we have $D_1 = 0$, $D_2 = \varepsilon_i E = - \varepsilon_i \nabla \Phi$  
        and $n=\pm e_z$ therefore the electrostatic potential should satisfy the following
        condition at the walls:
    \eq{\varepsilon_i \partial_z  \Phi|_{z=0,L} = \mp\sigma_k}
    \item 
        At the interface we have $D_1=-\varepsilon_1\nabla\Phi$, $D_2=-\varepsilon_2\nabla\Phi$
        and $n=-e_x$ therefore the electortatic potential should satisfy
        \eq{\varepsilon_1 \partial_x \Phi|_{x=0} = \varepsilon_2 \partial_x \Phi|_{x=0}}
        i.e. the displacement field should be continous at the interface.
    \item
        Additionally $\Phi$ must be continuous at the interface
        \eq{\Phi(0^+,z) = \Phi(0^-,z)}
        because otherwise the displacement field would be undefined at the interface.
    \item
        Furthermore $\Phi$ should remain finite between the walls in the limit $x\to\pm\infty$
        \eq{\lim_{x\to\pm\infty} \abs{\Phi(x,z)} < \infty}
\end{enumerate}

\section{Ansatz}
In order to find the solution $\Phi(x,z)$ of our problem we follow
the strategy laid out in Ref. \cite{supp}.
We consider three subproblems:
\begin{enumerate}
    \item A fluid interface without any walls (See Fig. \ref{fig:sp1}).
        The potential $\phi$ we get from this problem fulfills the Debye-H\"uckel
        equation in each medium and satisfies the boundary conditions at the interface
        and for $x\to\pm\infty$.
        Due to the symmetry of the problem $\phi$ depends only on the $x$-coordinate,
        $\phi=\phi(x)$.
    \item Two walls with only one medium and two different surface charges. (See Fig. \ref{fig:sp2}).
        The potentials $\psi_1$ (with medium 1), $\psi_2$ (with medium 2) we get from this problem
        fulfill the Debye-H\"uckel
        equation in their respective medium and satisfy the boundary conditions at the walls.
        Due to the symmetry of the problem $\psi_i$ depends only on the $z$-coordinate,
        $\psi_i=\psi_i(z)$.\\
        
        If we would simply set
        $\Phi(x,z) = \phi(x) + \left\{\begin{smallmatrix}\psi_1(z) && x \geq 0 \\ \psi_2(z) && x < 0\end{smallmatrix}\right.$,
        then $\Phi$ would fulfill the Debye-H\"uckel equation since it is linear and both $\phi(x)$ and $\psi_i(z)$
        fulfill the equation.
        The condition for the limit to infinity holds since it is fulfilled by $\phi(x)$ and $\psi_i$ does not depend on $x$.
        Because $\phi$ does not depend on $z$ the boundary conditions at the walls that are
        by construction fulfilled by $\psi_i(z)$ are
        automatically preserved since $\partial_z\phi(x) = 0$.
        Because $\psi_i$ does not depend on $x$ the condition for the derivative of the potential at the
        interface that is by construction fulfilled by $\phi(x)$
        will be preserved since $\partial_x\psi_i(z)=0$,
        but the continuity at the interface would be violated.\\

        To overcome the last problem we introduce another subproblem:
    \item The calculation of a correction function $c_i$ that
        fulfills the homogeneous Debye-H\"uckel equation
        and restores continuity at the interface
        but leaves the other boundary conditions unchanged.
\end{enumerate}
For the final solution we set
\eq{\Phi(x,z) = \piecewise{\Phi_1(x,z) & x \ge 0 \\ \Phi_2(x,z) & x \le 0}}
with
\eq{\Phi_i(x,z) = \phi(x) + \psi_i(z) + c_i(x,z)}
since the Debye-H\"uckel equation  is linear 
and each function fulfills this homogeneous pde in its medium
and $\phi$ fulfills the inhomogeneous pde for medium 2,
the sum of the functions is also a solution of the Debye-H\"uckel equation:
\eq{\laplace\Phi_i(x,z) &= \laplace\phi(x) + \laplace\psi_i(z) + \laplace c_i(x,z)\\
    &= \kappa_i^2(\phi(x)+\delta_{i2}\Phi_D) + \kappa_i^2\psi_i(z) + \kappa_i^2 c_i(x,z) \\
    &= \kappa_i^2(\Phi_i(x,z)+\delta_{i2}\Phi_D)}
Furthermore
\eq{\partial_z\Phi_i(x,z)|_{z=0,L} &=\ubr{\partial_z\phi(x)|_{z=0,L}}{=0}
+ \partial_z\psi_i(z)|_{z=0,L} + \ubr{\partial_z c_i(x,z)|_{z=0,L}}{=0} \\
    &= \partial_z\psi_i(z)|_{z=0,L}
}
which implies that the boundary conditions at the walls are fulfilled, and
\eq{\partial_x\Phi_i(x,z)|_{z=0,L} &=\partial_x\phi(x)|_{z=0,L}
+ \ubr{\partial_x\psi_i(z)|_{z=0,L}}{=0} + \ubr{\partial_x c_i(x,z)|_{z=0,L}}{=0} \\
    &= \partial_x\phi(x)|_{z=0,L}
}
which implies that the boundary condition for the continuity of $D$ at the wall is satisfied as well.
Additionally, because of the way $c_i$ is constructed the continuity at the interface is satisfied.
\\

Thus if constructed this way $\Phi(x,z)$ will be a solution of our problem.

\section{Exact Solution}
Following the ansatz we will solve the three subproblems in order to derive the
exact solution of the potential, which shall be denoted by $\Phi^e$.
\subsection{Subproblem 1}
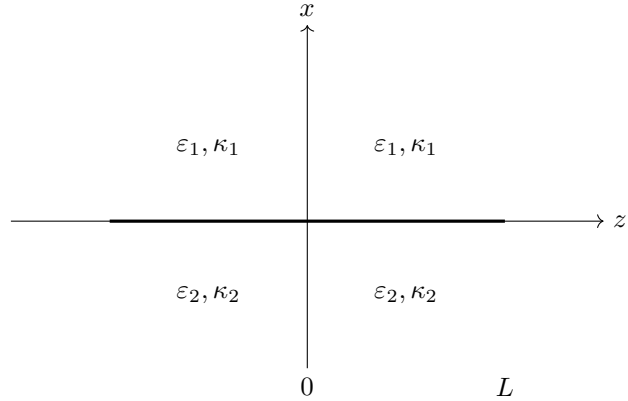
\begin{figure}[H]
    \centering
    \begin{tikzpicture}[
        scale = 1.3,
        wall/.style={
                postaction={draw,decorate,decoration={border,angle=-45,
                            amplitude=0.3cm,segment length=2mm}}},
        ]
        \draw[->] (0,-1.5) -- (0,2) node (xaxis) [above] {$x$};
        \draw[->] (-3,0) -- (3,0) node (zaxis) [right] {$z$};
        \draw[very thick] (-2,0) -- (2,0);
        \draw (1, 0.75) node {$\varepsilon_1,\kappa_1$};
        \draw (1,-0.75) node {$\varepsilon_2,\kappa_2$};
        \draw (-1, 0.75) node {$\varepsilon_1,\kappa_1$};
        \draw (-1,-0.75) node {$\varepsilon_2,\kappa_2$};
        \draw (0,-1.5) node [below] {$0$};
        \draw (2,-1.5) node [below] {$L$};
    \end{tikzpicture}
    \caption{
        Two dielectric fluids (medium 1 filling the space $x>0$
        and medium 2 filling the space $x<0$) separated by a flat interface
        at $x=0$ in abscence of any walls. The media are characterized by their
        dielectric constants $\varepsilon_i$, $i\in\m{1,2}$ and by their
        inverse Debye lengths $\kappa_i$, $i\in\m{1,2}$.
    }
    \label{fig:sp1}
\end{figure}
Here we consider a system with two fluid media separated by an interface
in the absence of any walls, as depicted in Fig. \ref{fig:sp1}.
Because of the symmetry of the problem the electrostatic potential $\phi$ only depends on
the $x$-coordinate:
\fleq{
   &\phi(x) = \piecewise{\phi_1(x) & x \ge 0 \\ \phi_2(x) & x \le 0} &
}
We obtain the potential by solving the Debye-H\"uckel equation
\fleq{
    &\laplace\phi_1 = \kappa_1^2 \phi_1 \label{sp1dh1}&\\
   &\laplace\phi_2 = \kappa_2^2 (\phi_2 - \Phi_D) \label{sp1dh2}&
}
in each medium. Here $\Phi_D$ is the Donnan potential.
The solution of Equations \eqref{sp1dh1} and \eqref{sp1dh2}
are given by
\fleq{
   &\phi_1(x) = Ae^{-\kappa_1x} + B e^{\kappa_1x} &\\
   &\phi_2(x) = \Phi_D + Ce^{-\kappa_2x} + \sD e^{\kappa_2x} &
}
where $A, B, C, \sD$ are constants. We can immediately reduce
the number of constants by using the following boundary conditions:
\fleq{
    &\lim_{x\to +\infty}\abs{\phi_1(x)}<\infty \f B=0 &\\
    &\lim_{x\to -\infty}\abs{\phi_2(x)}<\infty \f C=0 &
}
In order to determine the remaining constants $A$ and $\sD$, we use the continuity condition
for the electrostatic potential and the electric displacement vector at the interface.
\begin{enumerate}
    \item Continuity of $\phi$ at the interface
        \fleq{&\phi_1(0) = \phi_2(0)&\\
            &Ae^{-\kappa_1\cdot0} = \Phi_D + \sD e^{\kappa_2\cdot 0}&}
    \item Continuity of $D$ at the interface
        \fleq{&\varepsilon_1\partial_x\phi_1(x)|_{x=0} = \varepsilon_2\partial_x\phi_2(x)|_{x=0} &\\
            &\varepsilon_1A(-\kappa_1)e^{-\kappa_1\cdot0} = \varepsilon_2\sD\kappa_2 e^{\kappa_2\cdot 0}&}
\end{enumerate}
we can now solve this system of equations to obtain $A$ and $\sD$:
\fleq{
    &\f A = \Phi_D + \sD &\\
    &\f 0 = \varepsilon_1\kappa_1A+ \varepsilon_2\kappa_2\sD &\\
    &\f 0 = \varepsilon_1\kappa_1(\Phi_D + \sD)+ \varepsilon_2\kappa_2\sD &\\
    &\f \sD (\varepsilon_1\kappa_1 + \varepsilon_2\kappa_2) = - \varepsilon_1\kappa_1\Phi_D&\\
    &\f \sD = -\frac{\varepsilon_1\kappa_1\Phi_D}{\varepsilon_1\kappa_1+\varepsilon_2\kappa_2} &\\
    &\f A = \Phi_D + \sD = \Phi_D -\frac{\varepsilon_1\kappa_1\Phi_D}{\varepsilon_1\kappa_1+\varepsilon_2\kappa_2}
          = \frac{\varepsilon_2\kappa_2\Phi_D}{\varepsilon_1\kappa_1+\varepsilon_2\kappa_2}&
}
So, finally
\fleq{
    \f \phi_1(x) &= \frac{\varepsilon_2\kappa_2\Phi_D}{\varepsilon_1\kappa_1+\varepsilon_2\kappa_2}e^{-\kappa_1 x} &\label{ess3s}\\
    \f \phi_2(x) &= \Phi_D\left(1 - \frac{\varepsilon_1\kappa_1}{\varepsilon_1\kappa_1+\varepsilon_2\kappa_2}e^{\kappa_2 x}\right) &\\
}

\subsection{Subproblem 2}
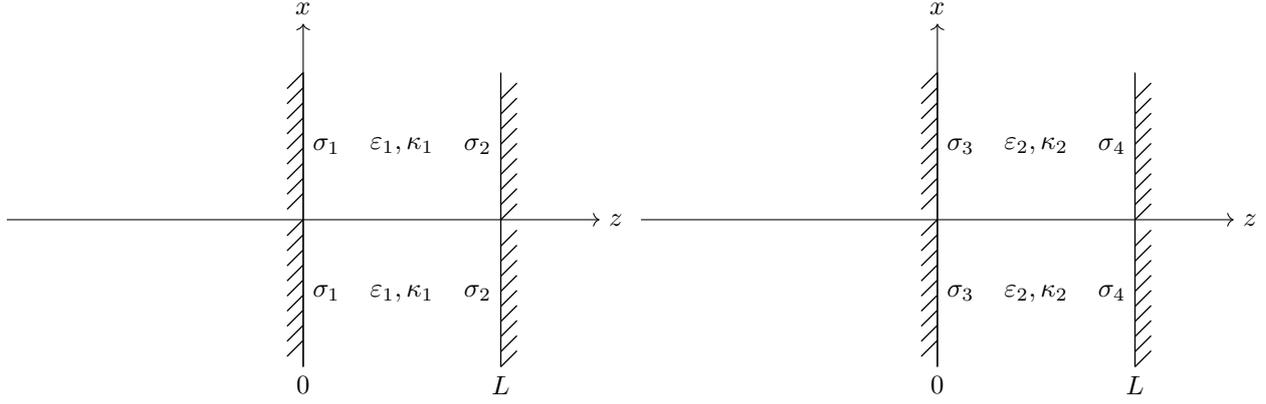
\begin{figure}[H]
    \centering
    \begin{tikzpicture}[
        scale = 1.3,
        wall/.style={
                postaction={draw,decorate,decoration={border,angle=-45,
                            amplitude=0.3cm,segment length=2mm}}},
        ]
        \draw[->] (0,-1.5) -- (0,2) node (xaxis) [above] {$x$};
        \draw[->] (-3,0) -- (3,0) node (zaxis) [right] {$z$};
        \draw[,  line width=.5pt, wall] (0,1.5)  -- node [right] {$\sigma_1$} (0,0)   ;
        \draw[,  line width=.5pt, wall] (0,0)    -- node [right] {$\sigma_1$} (0,-1.5);
        \draw[, line width=.5pt, wall] (2,0)    -- node [left ] {$\sigma_2$} (2,1.5) ;
        \draw[, line width=.5pt, wall] (2,-1.5) -- node [left ] {$\sigma_2$} (2,0)   ;
        \draw (1, 0.75) node {$\varepsilon_1,\kappa_1$};
        \draw (1,-0.75) node {$\varepsilon_1,\kappa_1$};
        \draw (0,-1.5) node [below] {$0$};
        \draw (2,-1.5) node [below] {$L$};
    \end{tikzpicture}
    \begin{tikzpicture}[
        scale = 1.3,
        wall/.style={
                postaction={draw,decorate,decoration={border,angle=-45,
                            amplitude=0.3cm,segment length=2mm}}},
        ]
        \draw[->] (0,-1.5) -- (0,2) node (xaxis) [above] {$x$};
        \draw[->] (-3,0) -- (3,0) node (zaxis) [right] {$z$};
        \draw[, line width=.5pt, wall] (0,1.5)  -- node [right] {$\sigma_3$} (0,0)   ;
        \draw[, line width=.5pt, wall] (0,0)    -- node [right] {$\sigma_3$} (0,-1.5);
        \draw[,  line width=.5pt, wall] (2,0)    -- node [left ] {$\sigma_4$} (2,1.5) ;
        \draw[,  line width=.5pt, wall] (2,-1.5) -- node [left ] {$\sigma_4$} (2,0)   ;
        \draw (1, 0.75) node {$\varepsilon_2,\kappa_2$};
        \draw (1,-0.75) node {$\varepsilon_2,\kappa_2$};
        \draw (0,-1.5) node [below] {$0$};
        \draw (2,-1.5) node [below] {$L$};
    \end{tikzpicture}
    \caption{
        The system left (right) is used to calculate $\psi_1$ ($\psi_2$).
        Two walls at $z=0$ and $z=L$ with a single medium filling the space between them.
        The wall at $z=0$ is homogeneously charged with charge density $\sigma_1$ ($\sigma_3$),
        the wall at $z=L$ is homogeneously charged with charge density $\sigma_2$ ($\sigma_4$),
        and the fluid is characterized by its dielectric constant $\varepsilon_1$ ($\varepsilon_2$)
        and its inverse Debye length $\kappa_1$ ($\kappa_2$).
    }
    \label{fig:sp2}
\end{figure}
Now we consider two walls in contact with a single fluid phase filling the space between them,
as depicted in Fig. \ref{fig:sp2}.
First we consider the case of medium 1. Because of the symmetry of the problem the solution only depends
on the $z$-coordinate. The electrostatic potential $\psi_1(z)$ can be obtained by solving the equation
\fleq{
    &\laplace\psi_1 =\kappa_1^2\psi_1 &
}
which has the general solution
\fleq{
    &\psi_1(z) = Ae^{-\kappa_1 z} + Be^{\kappa_1 z} &
}
with constants $A$ and $B$ that can be determined by using the boundary conditions at the two walls.
The boundary conditions for each wall are
\begin{enumerate}
    \item Boundary condition for $D$ at $z=0$
        \fleq{ 
            &n_1\cdot D|_{z=0} = -\varepsilon_1\partial_z\psi_1(z)|_{z=0} = \sigma_1 &\\
            & \varepsilon_1(-\kappa_1)Ae^{-\kappa_1 0} + \varepsilon_1\kappa_1Be^{\kappa_1 0} = -\sigma_1 &
        }
    \item Boundary condition for $D$ at $z=L$
        \fleq{ 
            &n_2\cdot D|_{z=L} = \varepsilon_1\partial_z\psi_1(z)|_{z=L} = \sigma_2 &\\
            & \varepsilon_1(-\kappa_1)Ae^{-\kappa_1 L} + \varepsilon_1\kappa_1Be^{\kappa_1 L} = \sigma_2 &
        }
\end{enumerate}
\fleq{
    & -\varepsilon_1\kappa_1A+\varepsilon_1\kappa_1 B = -\sigma_1 &\\
    & -\varepsilon_1\kappa_1A+\varepsilon_2\kappa_1 B e^{2\kappa_1L} = \sigma_2e^{\kappa_1L} &\\
    \f\quad & \varepsilon_1\kappa_1B(e^{2\kappa_1L} - 1) = \sigma_2e^{\kappa_1L} + \sigma_1 &\\
    \f\quad & B = \frac{1}{\varepsilon_1\kappa_1}\frac{\sigma_2 e^{\kappa_1L} + \sigma_1}{e^{2\kappa_1L} - 1}
    = \frac{1}{\varepsilon_1\kappa_1}\frac{\sigma_2+ \sigma_1 e^{-\kappa_1L}}{2\sinh(\kappa_1L)} &\\
    \f\quad & -\varepsilon_1\kappa_1 A + \frac{\sigma_2+ \sigma_1 e^{-\kappa_1L}}{2\sinh(\kappa_1L)} = -\sigma_1 &\\
    \f\quad & A = \frac{1}{\varepsilon_1\kappa_1}\frac{\sigma_2+\sigma_1(e^{-\kappa_1L}+2\sinh(\kappa_1L))}{2\sinh(\kappa_1L)}
    = \frac{1}{\varepsilon_1\kappa_1}\frac{\sigma_2+\sigma_1e^{\kappa_1L}}{2\sinh(\kappa_1L)} &
}
So, finally, we can write
\fleq{
    \psi_1(z) &= \frac{1}{\varepsilon_1\kappa_1\cdot 2\sinh(\kappa_1L)}
    \ubr{\left((\sigma_2+\sigma_1e^{\kappa_1L})e^{-\kappa_1 z} + (\sigma_2+\sigma_1e^{-\kappa_1L})e^{\kappa_1 z}\right)}
    {=2\sigma_2\cosh(\kappa_1z)+\sigma_1(e^{-\kappa_1(z-L)} + e^{\kappa_1(z-L)})} &\\
    & = \frac{1}{\varepsilon_1\kappa_1}\frac{\sigma_2\cosh(\kappa_1z) + \sigma_1\cosh(\kappa_1(z-L))}{\sinh(\kappa_1L)} &\\
}
The calculation for $\psi_2$ is exactly the same, with $\sigma_3$ in place of $\sigma_1$, $\sigma_4$ in place of $\sigma_2$
and $\varepsilon_2$, $\kappa_2$ in place of $\varepsilon_1$, $\kappa_1$, respectively.
Therefore, the potential in this case is given by
\fleq{
    \psi_2(z) &= \frac{1}{\varepsilon_2\kappa_2}\frac{\sigma_4\cosh(\kappa_2z) + \sigma_3\cosh(\kappa_2(z-L))}{\sinh(\kappa_2L)} &\\
}

\subsection{Subproblem 3}
In Ref. \cite{supp} the problem was symmetric, and thus $c_i(x,z)$ was symmetric in $z$ and thereby periodic in $z$,
so it could be written as a Fourier series in $z$. Because our problem is not symmetric in $z$ we
can not expect our $c_i(x,z)$ to be symmetric in $z$.\\

But to make the Fourier series approach work again, we can make the system symmetric by mirroring it on
the $x$-axis. The mirrored system is symmetric and thus we can expand $c_i(x,z)$ in a Fourier series
in $z$ within the interval $[-L,L]$.
And since our system is included in this larger system, $c_i(x,z)$ of our system is simply obtained as the restriction
of the $c_i(x,z)$ of the larger system to values of $z$ in $[0,L]$.
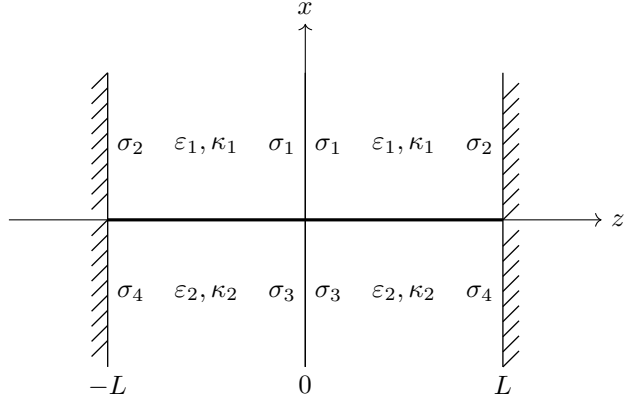
\begin{figure}[H]
    \centering
    \begin{tikzpicture}[
        scale = 1.3,
        wall/.style={
                postaction={draw,decorate,decoration={border,angle=-45,
                            amplitude=0.3cm,segment length=2mm}}},
        ]
        \draw[->] (0,-1.5) -- (0,2) node (xaxis) [above] {$x$};
        \draw[->] (-3,0) -- (3,0) node (zaxis) [right] {$z$};
        \draw[very thick] (-2,0) -- (2,0);
        \draw[,  line width=.5pt] (0,1.5)  -- node [right] {$\sigma_1$} node [left] {$\sigma_1$} (0,0)   ;
        \draw[, line width=.5pt] (0,0)    -- node [right] {$\sigma_3$} node [left] {$\sigma_3$} (0,-1.5);
        \draw[, line width=.5pt, wall] (2,0)    -- node [left ] {$\sigma_2$} (2,1.5) ;
        \draw[,  line width=.5pt, wall] (2,-1.5) -- node [left ] {$\sigma_4$} (2,0)   ;
        \draw[, line width=.5pt, wall] (-2,1.5)-- node [right] {$\sigma_2$} (-2,0)   ;
        \draw[,  line width=.5pt, wall] (-2,0)  -- node [right] {$\sigma_4$} (-2,-1.5);
        \draw (1, 0.75) node {$\varepsilon_1,\kappa_1$};
        \draw (1,-0.75) node {$\varepsilon_2,\kappa_2$};
        \draw (-1, 0.75) node {$\varepsilon_1,\kappa_1$};
        \draw (-1,-0.75) node {$\varepsilon_2,\kappa_2$};
        \draw (0,-1.5) node [below] {$0$};
        \draw (2,-1.5) node [below] {$L$};
        \draw (-2,-1.5) node [below] {$-L$};
    \end{tikzpicture}
    \caption{
        Symmetrically extended system.
        The system in Fig. \ref{fig:cp} was mirrored on the $x$-axis.
    }
    \label{fig:sp3}
\end{figure}
In our extended symmetric system we can extend $\psi_i(z)$ in the expected way:
\eq{\psi_i(z):=\piecewise{\psi_i(z) & z\ge 0 \\ \psi_i(-z) & z < 0}}
since $\phi$ is only a function of $x$ it does not need to be extended.\\

The conditions $c_i(x,z)$ has to fulfill in our extended system are
\begin{enumerate}
    \item\label{es3c1} $\laplace c_i(x,z) = \kappa_i^2(x,z)\cdot c_i(x,z)$ so the pde is fulfilled
    \item\label{es3c2} $\partial_z c_i(x,z) |_{z=0,\pm L} = 0$ so the boundary conditions for the surface charges stay valid
    \item\label{es3c3} $\displaystyle\lim_{x\to\pm\infty} \abs{c_i(x,z)}<\infty$ so $\Phi^e_i$ also stays finite in the limit $x\to\pm\infty$
    \item\label{es3c4} $c_1(0,z) + \psi_1(z) = c_2(0,z) + \psi_2(z)$ since $\phi_1(0)=\phi_2(0)$ is already satisfied
    \item\label{es3c5} $\varepsilon_1\partial_x c_1(x,z) |_{x=0} = \varepsilon_2\partial_x c_2(x,z) |_{x=0}$ so the boundary condition for the interface stays valid
\end{enumerate}
We write $c_i$ as a Fourier series in $z$:
\fleq{
    & c_i(x,z) = \frac{a_{0,i}(x)}{2}
    + \sum_{n=1}^\infty a_{n,i}(x)\cos\left(\frac{n\pi z}{L}\right)
    + \sum_{n=1}^\infty b_{n,i}(x)\sin\left(\frac{n\pi z}{L}\right) &
}
Due to the symmetry of $c_i(x,z)$ in $z$ it follows that
\fleq{
     & b_{n,i}(x) = \frac{1}{L}\int_{-L}^{L} c_i(x,z)\sin\left(\frac{n\pi z}{L}\right)\d z = 0 &
}
Therefore,
\fleq{
    & c_i(x,z) = \frac{a_{0,i}(x)}{2}
    + \sum_{n=1}^\infty a_{n,i}(x)\cos\left(\frac{n\pi z}{L}\right) \label{cifsc}
}

\paragraph{Treatment of Condition \ref{es3c2}}
The second condition listed above is automatically satisfied since
\fleq{
     &\partial_z c_i(x,z) = \sum_{n=1}^\infty -a_{n,i}(x) \left(\frac{n\pi}{L}\right)\ubr{\sin\left(\frac{n\pi z}{L}\right)}{=0 \text{ for } z = 0,\pm L} &\\
     &\f \partial_z c_i(x,z) |_{z=0,\pm L} = 0 &
}
\paragraph{Treatment of Condition \ref{es3c1}}
Plugging the expression for $c_i(x,z)$ from Eq. \eqref{cifsc} into the (homogeneous)
Debye-H\"uckel equation given in condition \ref{es3c1}, we obtain
\fleq{
    & \frac{a_{0,i}''(x)}{2}
    + \sum_{n=1}^\infty a_{n,i}''(x)\cos\left(\frac{n\pi z}{L}\right)
    + \sum_{n=1}^\infty -a_{n,i}(x)\left(\frac{n\pi}{L}\right)^2\cos\left(\frac{n\pi z}{L}\right)&\\
    &= \kappa_i^2 \frac{a_{0,i}(x)}{2} + \sum_{n=1}^\infty\kappa_i^2a_{n,i}(x)\cos\left(\frac{n\pi z}{L}\right) &
}
\fleq{
    \text{cos, sin, constant Basis} \f\quad &a_{0,i}''(x) = \kappa_i^2 a_{0,i}(x) &\\
    &a_{n,i}''(x) = \left(\left(\frac{n\pi}{L}\right)^2 + \kappa_i^2\right) a_{n,i}(x) &
}
Using condition \ref{es3c3}
the solutions of these equations can be written as
\fleq{
    \f\quad
    &a_{0,1}(x) = \sD e^{-\kappa_1x}  &\\
    &a_{0,2}(x) = C e^{\kappa_2x} &\\
    &a_{n,1}(x) = A_n e^{-\sqrt{\left(\frac{n\pi}{L}\right)^2 + \kappa_1^2}x} &\\
    &a_{n,2}(x) = B_n e^{\sqrt{\left(\frac{n\pi}{L}\right)^2 + \kappa_2^2}x}& 
}
For brevity we define
\fleq{&p_i := \sqrt{\left(\frac{n\pi}{L}\right)^2 + \kappa_i^2}&}

\paragraph{Treatment of Condition \ref{es3c5}}
\fleq{
    &\varepsilon_1\left(\frac{-\kappa_1 \sD}{2} + \sum_{n=1}^\infty -p_1A_n\cos\left(\frac{n\pi z}{L}\right) \right) &\\
    &=\varepsilon_2\left(\frac{ \kappa_2 C}{2} + \sum_{n=1}^\infty  p_2B_n\cos\left(\frac{n\pi z}{L}\right) \right) &
}
\fleq{
    \text{cos, sin, const Basis} \f\quad
    & -\varepsilon_1\kappa_1\sD = \varepsilon_2\kappa_2C \label{es3es1}&\\
    & -\varepsilon_1 p_1 A_n = \varepsilon_2p_2B_n \label{es3es2}
}
\paragraph{Treatment of Condition \ref{es3c4}}
\fleq{
    & c_1(0,z) = \frac{\sD}{2} + \sum_{n=1}^\infty A_n \cos\left(\frac{n\pi z}{L}\right) &\\
    & c_2(0,z) = \frac{C}{2} + \sum_{n=1}^\infty B_n \cos\left(\frac{n\pi z}{L}\right) &
}
We develop $\Psi_1(z)$ into a Fourier series
\fleq{
    & \Psi_1(x,z) = \frac{\alpha_{0,1}}{2}
    + \sum_{n=1}^\infty \alpha_{n,1}\cos\left(\frac{n\pi z}{L}\right)
    + \sum_{n=1}^\infty \beta_{n,1} \sin\left(\frac{n\pi z}{L}\right) &
}
Due to the Symmetry of the extended $\Psi_1(z)$ if follows that $\beta_{n,1}=0$ for all $n$.
\tchn{
\fleq{
    \alpha_{0,1} &= \frac{1}{L}\int_{-L}^{L}\Psi_1(z)\d z &\\
    &\overset{\text{symmetry}}{=} \frac{2}{L}\int_0^L \Psi_1(z)\d z&\\
    & = \ubr{\frac{2}{L}\frac{1}{\varepsilon_1\kappa_1\sinh(\kappa_1L)}}{:=\gamma_1}
    \Bigg(\sigma_2\ubs{\int_0^L\cosh(\kappa_1z)\ud z}{\text{subst. }z' = \kappa_1z} +
    \sigma_1\ubs{\int_0^L\cosh(\kappa_1(z-L))\ud z}{\text{subst. }z' = \kappa_1(z-L)}
    \Bigg) &\\
    &= \gamma_1\left(\frac{\sigma_2}{\kappa_1}\int_0^{\kappa_1L}\cosh(z')\d z'
    + \frac{\sigma_1}{\kappa_1}\int_{-\kappa_1L}^0\cosh(z')\d z'\right)&\\
    &= \frac{\gamma_1}{\kappa_1}\left(\sigma_2\sinh(\kappa_1L) + \sigma_1\sinh(\kappa_1L)\right)&\\
    &=\frac{2}{L\varepsilon_1\kappa_1^2}(\sigma_1+\sigma_2)&
}
Theorem 2.671.4 of \cite{grad} states that
\fleq{
    \int\cosh(ax+b)\cos(cx+d) \d x &=
    \frac{a}{a^2 + c^2}\sinh(ax + b)\cos(cx + d) &\\ &+ 
    \frac{c}{a^2 + c^2}\cosh(ax + b)\sin(cx + d) &
}
\fleq{
    \f & \int_0^L\cosh(az + b)\cos\left(\frac{n\pi z}{L}\right)\d z =
    \frac{a}{a^2 + \left(\frac{n\pi}{L}\right)^2}\left(\sinh(aL+b)\cdot (-1)^n + - \sinh(b)\right)&
}
\fleq{
    \alpha_{n,1} &= \frac{1}{L}\int_{-L}^{L}\Psi_1(z)\cos\left(\frac{n\pi z}{L}\right)\d z &\\
    &\overset{\text{symmetry}}{=} \frac{2}{L}\int_0^L \Psi_1(z)\cos\left(\frac{n\pi z}{L}\right)\d z&\\
    &= \gamma_1\left(\sigma_2\int_0^L\cosh(\kappa_1z)\cos\left(\frac{n\pi z}{L}\right)\d z
    + \sigma_1\int_0^L \cosh(\kappa_1(z-L))\cos\left(\frac{n\pi z}{L}\right)\d z \right) &\\
    &= \gamma_1 \sigma_2 \frac{\kappa_1}{\kappa_1^2 + \left(\frac{n\pi}{L}\right)^2} (-1)^n \sinh(\kappa_1L)
    +  \gamma_1 \sigma_1 \frac{\kappa_1}{\kappa_1^2 + \left(\frac{n\pi}{L}\right)^2} \sinh(\kappa_1L)&\\
    &= \frac{2}{L\varepsilon_1} \frac{\sigma_1 + (-1)^n\sigma_2}{p_1^2} &
}
}
Condition \ref{es3c4} then becomes
\fleq{
    &\frac{\sD}{2}
    + \sum_{n=1}^\infty A_n \cos\left(\frac{n\pi z}{L}\right)
    + \frac{1}{L\varepsilon_1\kappa_1^2}(\sigma_1+\sigma_2)
    + \sum_{n=1}^\infty \frac{2}{L\varepsilon_1} \frac{\sigma_1 + (-1)^n\sigma_2}{p_1^2}\cos\left(\frac{n\pi z}{L}\right) &\\
    =&\frac{C}{2}
    + \sum_{n=1}^\infty B_n \cos\left(\frac{n\pi z}{L}\right)
    + \frac{1}{L\varepsilon_2\kappa_2^2}(\sigma_3+\sigma_4)
    + \sum_{n=1}^\infty \frac{2}{L\varepsilon_2} \frac{\sigma_3 + (-1)^n\sigma_4}{p_2^2}\cos\left(\frac{n\pi z}{L}\right)&
}
\fleq{
    \text{cos, sin, constant Basis} \f\quad &
    \frac{\sD}{2}
    + \frac{1}{L\varepsilon_1\kappa_1^2}(\sigma_1+\sigma_2)
    =\frac{C}{2}
    + \frac{1}{L\varepsilon_2\kappa_2^2}(\sigma_3+\sigma_4)
    \label{es3es3}&\\ &
      A_n 
    + \frac{2}{L\varepsilon_1} \frac{\sigma_1 + (-1)^n\sigma_2}{p_1^2}
    = B_n 
    + \frac{2}{L\varepsilon_2} \frac{\sigma_3 + (-1)^n\sigma_4}{p_2^2}
    \label{es3es4}&
}
\paragraph{Conditions}
To calculate $C$, $\sD$ we use Eq. \eqref{es3es1} and \eqref{es3es3}:
\fleq{
    -\varepsilon_1\kappa_1\sD &= \varepsilon_2\kappa_2C \\
    \frac{\sD}{2} + \frac{1}{L\varepsilon_1\kappa_1^2}(\sigma_1+\sigma_2)
    &=\frac{C}{2} + \frac{1}{L\varepsilon_2\kappa_2^2}(\sigma_3+\sigma_4)
}
\tchn{
\fleq{
    \sD &= -\frac{\varepsilon_2\kappa_2}{\varepsilon_1\kappa_1}C &\\
    C &= \sD + \frac{2}{L\varepsilon_1\kappa_1^2}(\sigma_1+\sigma_2) -\frac{2}{L\varepsilon_2\kappa_2^2}(\sigma_3+\sigma_4) &\\
    C &= \frac{1}{1+\frac{\varepsilon_2\kappa_2}{\varepsilon_1\kappa_1}}
    \left(\frac{2}{L\varepsilon_1\kappa_1^2}(\sigma_1+\sigma_2) -\frac{2}{L\varepsilon_2\kappa_2^2}(\sigma_3+\sigma_4)\right) &\\
    C &= \frac{\varepsilon_1\kappa_1}{\varepsilon_1\kappa_1+\varepsilon_2\kappa_2}
    \left(\frac{2}{L\varepsilon_1\kappa_1^2}(\sigma_1+\sigma_2) -\frac{2}{L\varepsilon_2\kappa_2^2}(\sigma_3+\sigma_4)\right) &
}
}
\fleq{
    C &= \frac{2}{L} \frac{1}{\varepsilon_1\kappa_1+\varepsilon_2\kappa_2}
    \left(\frac{1}{\kappa_1}(\sigma_1+\sigma_2) -\frac{\varepsilon_1\kappa_1}{\varepsilon_2\kappa_2^2}(\sigma_3+\sigma_4)\right) &\\
    \sD &= \frac{2}{L} \frac{1}{\varepsilon_1\kappa_1+\varepsilon_2\kappa_2}
    \left(\frac{1}{\kappa_2}(\sigma_3+\sigma_4) -\frac{\varepsilon_2\kappa_2}{\varepsilon_1\kappa_1^2}(\sigma_1+\sigma_2)\right) &
}
To calculate $A_n$, $B_n$ we use Eq. \eqref{es3es2} and \eqref{es3es4}:
\fleq {
    -\varepsilon_1 p_1 A_n &= \varepsilon_2p_2B_n \\
    A_n + \frac{2}{L\varepsilon_1} \frac{\sigma_1 + (-1)^n\sigma_2}{p_1^2}
    &= B_n + \frac{2}{L\varepsilon_2} \frac{\sigma_3 + (-1)^n\sigma_4}{p_2^2}
}
\tchn{
\fleq{
    &A_n = -\frac{\varepsilon_2p_2}{\varepsilon_1p_1} &\\
    &B_n - A_n = \frac{2}{L\varepsilon_1} \frac{\sigma_1 + (-1)^n\sigma_2}{p_1^2}
    - \frac{2}{L\varepsilon_2} \frac{\sigma_3 + (-1)^n\sigma_4}{p_2^2} &\\
    &B_n = \frac{1}{1+\frac{\varepsilon_2p_2}{\varepsilon_1p_1}}\left(\frac{2}{L\varepsilon_1} \frac{\sigma_1 + (-1)^n\sigma_2}{p_1^2}
    - \frac{2}{L\varepsilon_2} \frac{\sigma_3 + (-1)^n\sigma_4}{p_2^2}\right) &
}
}
\fleq{
    &B_n = \frac{2}{L}\frac{\varepsilon_1p_1}{\varepsilon_1p_1+\varepsilon_2p_2}\left(\frac{\sigma_1 + (-1)^n\sigma_2}{\varepsilon_1} \frac{1}{p_1^2}
    - \frac{\sigma_3 + (-1)^n\sigma_4}{\varepsilon_2} \frac{1}{p_2^2}\right) &\\
    &A_n = \frac{2}{L}\frac{\varepsilon_2p_2}{\varepsilon_1p_1+\varepsilon_2p_2}\left(\frac{\sigma_3 + (-1)^n\sigma_4}{\varepsilon_2} \frac{1}{p_2^2}
    - \frac{\sigma_1 + (-1)^n\sigma_2}{\varepsilon_1} \frac{1}{p_1^2}\right) &
}
\paragraph{Finally}
\fleq{
    c_1(x,z) &= \frac{1}{L} \frac{1}{\kappa_1\varepsilon_1+\kappa_2\varepsilon_2}\left(\frac{\sigma_3 + \sigma_4}{\kappa_2} - \frac{1}{\kappa_1} \frac{\kappa_2\varepsilon_2}{\kappa_1\varepsilon_1}(\sigma_1+\sigma_2)\right) &\\
    &+ \sum_{n=1}^\infty \frac{2}{L} \frac{\varepsilon_2p_2}{\varepsilon_1p_1 + \varepsilon_2p_2}\left(\frac{\sigma_3 + (-1)^n \sigma_4}{\varepsilon_2}\frac{1}{p_2^2} - \frac{\sigma_1 + (-1)^n \sigma_2}{\varepsilon_1}\frac{1}{p_1^2}\right)e^{-p_1x}\cos\left(\frac{n\pi z}{L}\right)
}
\fleq{
    c_2(x,z) &= \frac{1}{L} \frac{1}{\kappa_1\varepsilon_1+\kappa_2\varepsilon_2}\left(\frac{\sigma_1 + \sigma_2}{\kappa_1} - \frac{1}{\kappa_2} \frac{\kappa_1\varepsilon_1}{\kappa_2\varepsilon_2}(\sigma_3+\sigma_4)\right) &\\
    &+ \sum_{n=1}^\infty \frac{2}{L} \frac{\varepsilon_1p_1}{\varepsilon_1p_1 + \varepsilon_2p_2}\left(\frac{\sigma_1 + (-1)^n \sigma_2}{\varepsilon_1}\frac{1}{p_1^2} - \frac{\sigma_3 + (-1)^n \sigma_4}{\varepsilon_2}\frac{1}{p_2^2}\right)e^{p_2x}\cos\left(\frac{n\pi z}{L}\right)
}

\subsection{Solution}
Adding the solutions of the three subproblems, one finally obtains the expressions for the potentials in the two media:
\fleq{
    \Phi_1^e(x,z) &= \frac{\sigma_1\cosh(\kappa_1(L-z)) + \sigma_2\cosh(\kappa_1z)}{\varepsilon_1\kappa_1\sinh(\kappa_1L)} &\\
    &+ \frac{\kappa_2\varepsilon_2\Phi_D}{\kappa_1\varepsilon_1+\kappa_2\varepsilon_2}e^{-\kappa_1x} &\\
    &+ \frac{1}{L} \frac{1}{\kappa_1\varepsilon_1+\kappa_2\varepsilon_2}\left(\frac{\sigma_3 + \sigma_4}{\kappa_2} - \frac{1}{\kappa_1} \frac{\kappa_2\varepsilon_2}{\kappa_1\varepsilon_1}(\sigma_1+\sigma_2)\right)e^{-\kappa_1x} &\\
    &+ \sum_{n=1}^\infty \frac{2}{L} \frac{\varepsilon_2p_2}{\varepsilon_1p_1 + \varepsilon_2p_2}\left(\frac{\sigma_3 + (-1)^n \sigma_4}{\varepsilon_2}\frac{1}{p_2^2} - \frac{\sigma_1 + (-1)^n \sigma_2}{\varepsilon_1}\frac{1}{p_1^2}\right)e^{-p_1x}\cos\left(\frac{n\pi z}{L}\right)
}
\fleq{
    \Phi_2^e(x,z) &= \frac{\sigma_3\cosh(\kappa_2(L-z)) + \sigma_4\cosh(\kappa_2z)}{\varepsilon_2\kappa_2\sinh(\kappa_2L)} &\\
    &+ \Phi_D  - \frac{\kappa_1\varepsilon_1\Phi_D}{\kappa_1\varepsilon_1+\kappa_2\varepsilon_2}e^{\kappa_2x} &\\
    &+ \frac{1}{L} \frac{1}{\kappa_1\varepsilon_1+\kappa_2\varepsilon_2}\left(\frac{\sigma_1 + \sigma_2}{\kappa_1} - \frac{1}{\kappa_2} \frac{\kappa_1\varepsilon_1}{\kappa_2\varepsilon_2}(\sigma_3+\sigma_4)\right)e^{\kappa_2x} &\\
    &+ \sum_{n=1}^\infty \frac{2}{L} \frac{\varepsilon_1p_1}{\varepsilon_1p_1 + \varepsilon_2p_2}\left(\frac{\sigma_1 + (-1)^n \sigma_2}{\varepsilon_1}\frac{1}{p_1^2} - \frac{\sigma_3 + (-1)^n \sigma_4}{\varepsilon_2}\frac{1}{p_2^2}\right)e^{p_2x}\cos\left(\frac{n\pi z}{L}\right)
}
\subsection{Consistency with literature}
We test our expression for consistency with the result of Ref. \cite{supp}.
Since in Ref. \cite{supp} the walls are located at $z=\pm L$, we need to do a transformation
in order to compare the results.
For this we have to set
\fleq{
    &L = 2L' &\\
    &z = z' + L' &\\
    &\sigma_1'=\sigma_1=\sigma_2 &\\
    &\sigma_2'=\sigma_3=\sigma_4 &
}
We transform each line of our expression for $\Phi^e(x,z)$ separately and compare with Ref. \cite{supp}.
\paragraph{Line 1}
\fleq{
    &\kappa_1(L-z) = \kappa_1(2L'-(z'+L')) = \kappa_1(L'-z') &\\
    &\kappa_1z = \kappa_1(z'+L') &\\
    &\frac{\sigma_1\cosh(\kappa_1(L-z)) + \sigma_2\cosh(\kappa_1z)}{\varepsilon_1\kappa_1\sinh(\kappa_1L)} &\\
    &= \frac{\sigma_1'}{\kappa_1\varepsilon_1}\frac{\cosh(\kappa_1(L'-z'))+\cosh(\kappa_1(z' + L'))}{\sinh(2\kappa_1L')} &\\
    &= \frac{\sigma_1'}{\kappa_1\varepsilon_1}\frac{1}{2\sinh(\kappa_1L')\cosh(\kappa_1L')}
        \cdot\frac{1}{2}\left(e^{\kappa_1L'-\kappa_1z'} + e^{-\kappa_1L'+\kappa_1z'} + e^{\kappa_1L'+\kappa_1z'} + e^{-\kappa_1L'-\kappa_1z'}\right) &\\
    &= \frac{\sigma_1'}{\kappa_1\varepsilon_1}\frac{1}{2\sinh(\kappa_1L')\cosh(\kappa_1L')}
        \cdot\frac{1}{2}\left(e^{\kappa_1L'}2\cosh(\kappa_1z') + e^{-\kappa_1L'}2\cosh(\kappa_1z') \right) &\\
    &= \frac{\sigma_1'}{\kappa_1\varepsilon_1}\frac{\cosh(\kappa_1z')}{\sinh(\kappa_1L')} \quad\checkmark&
}
\paragraph{Line 2}
Nothing to be done.
\paragraph{Line 3}
\fleq{
    &  \frac{1}{L} \frac{1}{\kappa_1\varepsilon_1+\kappa_2\varepsilon_2}\left(\frac{\sigma_3 + \sigma_4}{\kappa_2} - \frac{1}{\kappa_1} \frac{\kappa_2\varepsilon_2}{\kappa_1\varepsilon_1}(\sigma_1+\sigma_2)\right) e^{-\kappa_1x}&\\
    &= \frac{1}{L'} \frac{\frac{1}{\kappa_2\varepsilon_2}}{1+\frac{\kappa_1\varepsilon_1}{\kappa_2\varepsilon_2}} \left(\frac{\sigma_2'}{\kappa_2} - \frac{1}{\kappa_1} \frac{\kappa_2\varepsilon_2}{\kappa_1\varepsilon_1}\sigma_1'\right)e^{-\kappa_1x} &\\
    &= \frac{1}{L'} \frac{\frac{\sigma_2'}{\varepsilon_2\kappa_2} - \frac{\sigma_1'}{\varepsilon_1\kappa_1^2}}{1+\frac{\kappa_1\varepsilon_1}{\kappa_2\varepsilon_2}}e^{-\kappa_1x} \quad\checkmark &
}
\paragraph{Line 4}
\fleq{
    &\sigma_1 + (-1)^n \sigma_2 = \sigma_1' + (-1)^n \sigma_1' = \piecewise{2\sigma_1' & n \text{ even} \\ 0 & n \text{ odd}} &\\
    &\sigma_3 + (-1)^n \sigma_4 = \sigma_2' + (-1)^n \sigma_2' = \piecewise{2\sigma_2' & n \text{ even} \\ 0 & n \text{ odd}} &
}
so we can define $n' := \frac{n}{2}$ and write
\fleq{
    &p_i = \sqrt{\left(\frac{n\pi}{L}\right)^2 + \kappa_i^2}
    = \sqrt{\left(\frac{n\pi}{2L'}\right)^2 + \kappa_i^2}
    = \sqrt{\left(\frac{n'\pi}{L'}\right)^2 + \kappa_i^2} &\\
    &\cos\left(\frac{n\pi z}{L}\right)
    =\cos\left(\frac{n\pi (z' + L')}{2L'}\right)
    =\cos\left(\frac{n'\pi (z' + L')}{L'}\right)
    =\cos\left(\frac{n'\pi z'}{L'} + n'\pi\right)
    =(-1)^{n'}\cos\left(\frac{n'\pi z'}{L'}\right) &
}
and
\fleq{
    &\sum_{n=1}^\infty \frac{2}{L} \frac{\varepsilon_2p_2}{\varepsilon_1p_1 + \varepsilon_2p_2}\left(\frac{\sigma_3 + (-1)^n \sigma_4}{\varepsilon_2}\frac{1}{p_2^2} - \frac{\sigma_1 + (-1)^n \sigma_2}{\varepsilon_1}\frac{1}{p_1^2}\right)e^{-p_1x}\cos\left(\frac{n\pi z}{L}\right) &\\
    &=\sum_{n'=1}^\infty \frac{2}{2L'} \frac{\varepsilon_2p_2}{\varepsilon_1p_1 + \varepsilon_2p_2}\left(\frac{2\sigma_2'}{\varepsilon_2}\frac{1}{p_2^2} - \frac{2\sigma_1'}{\varepsilon_1}\frac{1}{p_1^2}\right)e^{-p_1x}(-1)^{n'}\cos\left(\frac{n'\pi z}{L'}\right) &\\
    &=\frac{2}{L'}\sum_{n'=1}^\infty (-1)^{n'} \frac{1}{1+\frac{\varepsilon_1p_1}{\varepsilon_2p_2}}\left(\frac{\sigma_2'}{\varepsilon_2}\frac{1}{p_2^2} - \frac{\sigma_1'}{\varepsilon_1}\frac{1}{p_1^2}\right)e^{-p_1x}\cos\left(\frac{n'\pi z}{L'}\right) \quad\checkmark &
}
Therefore we can obtain the result for the case of
identical particles given in Ref. \cite{supp} from our general expressions.

\section{Superposition Approximation}
\begin{figure}[H]
    \centering
    \begin{tikzpicture}[
        baseline=-0.1cm,
        scale = 1.3,
        wall/.style={
                postaction={draw,decorate,decoration={border,angle=-45,
                            amplitude=0.3cm,segment length=2mm}}},
        ]
        \draw[->] (0,-1.5) -- (0,2) node (xaxis) [above] {$x$};
        \draw[->] (-3,0) -- (3,0) node (zaxis) [right] {$z$};
        \draw[very thick] (0,0) -- (3,0);
        \draw[,  line width=.5pt, wall] (0,1.5)  -- node [right] {$\sigma_1$} (0,0)   ;
        \draw[, line width=.5pt, wall] (0,0)    -- node [right] {$\sigma_3$} (0,-1.5);
        \draw (1, 0.75) node {$\varepsilon_1,\kappa_1$};
        \draw (1,-0.75) node {$\varepsilon_2,\kappa_2$};
        \draw (0,-1.5) node [below] {$0$};
        \draw (2,-1.5) node [below] {$L$};
    \end{tikzpicture}
    +
    \begin{tikzpicture}[
        baseline=-0.1cm,
        scale = 1.3,
        wall/.style={
                postaction={draw,decorate,decoration={border,angle=-45,
                            amplitude=0.3cm,segment length=2mm}}},
        ]
        \draw[->] (0,-1.5) -- (0,2) node (xaxis) [above] {$x$};
        \draw[->] (-3,0) -- (3,0) node (zaxis) [right] {$z$};
        \draw[very thick] (-3,0) -- (2,0);
        \draw[, line width=.5pt, wall] (2,0)    -- node [left ] {$\sigma_2$} (2,1.5) ;
        \draw[,  line width=.5pt, wall] (2,-1.5) -- node [left ] {$\sigma_4$} (2,0)   ;
        \draw (1, 0.75) node {$\varepsilon_1,\kappa_1$};
        \draw (1,-0.75) node {$\varepsilon_2,\kappa_2$};
        \draw (0,-1.5) node [below] {$0$};
        \draw (2,-1.5) node [below] {$L$};
    \end{tikzpicture}
    \caption{
        We consider two different systems. One system (left) with a wall at $z=0$
        where the half space $z>0$ is filled with two media forming an interface at $x=0$.
        The other system (right) with a wall at $z=L$
        where the half space $z<L$ is filled with two media forming an interface at $x=0$.
        In both cases the medium residing in the space $x>0$ ($x<0$) is called medium 1 (2) and is
        characterized by its dielectric constant $\varepsilon_1$ ($\varepsilon_2$).
        In the first case (left) the wall carries a surface charge density of $\sigma_1$ for $x>0$ and $\sigma_3$ for $x<0$.
        In the second case (right) the wall carries a surface charge density of $\sigma_2$ for $x>0$ and $\sigma_4$ for $x<0$.
    }
    \label{fig:ssp}
\end{figure}
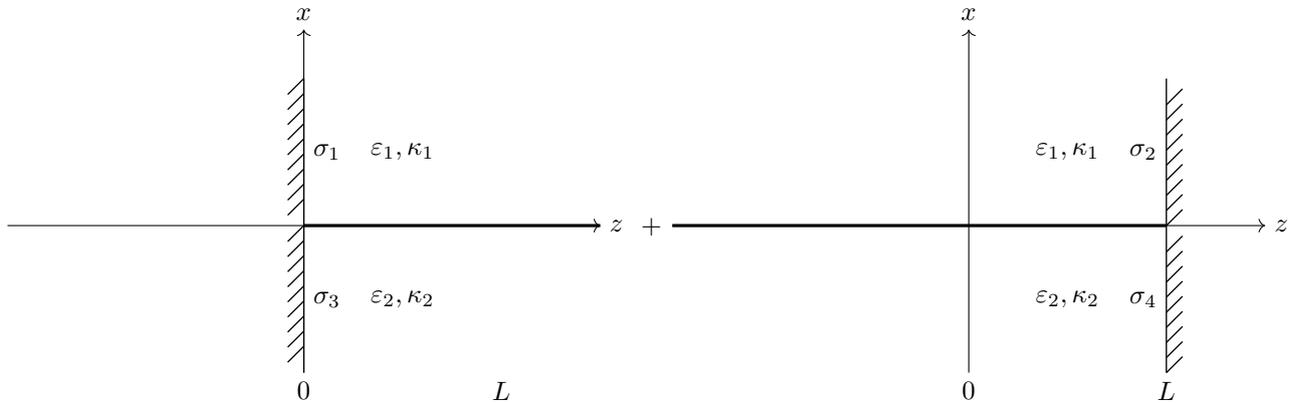

In the superposition approximation the potential is approximated
using the sum of the potentials of two systems (see Fig. \ref{fig:ssp}) with one wall each.
We will call the superposition potential $\Phi^s$.

Because we simply add the solutions, $\Phi^s$ will fulfill
boundary conditions at the interface and at infinity.

But the Debye-H\"uckel equation will only be exactly fulfilled for $x>0$.
If the solution of the problem depicted in the left side of Fig. \ref{fig:ssp} is
denoted by $\Phi^{\sigma_1,\sigma_3}(x,z)$ and the solution of the problem depicted in the
right side of Fig. \ref{fig:ssp} is denoted by $\Phi^{\sigma_2,\sigma_4}(x,-(z-L))$,
then
\fleq{
    \laplace\Phi^s(x,z) &= \laplace\Phi^{\sigma_1,\sigma_3}(x,z) + \laplace\Phi^{\sigma_2,\sigma_4}(x,-(z-L))&\\
    &= \kappa^2(x,z)(\Phi^{\sigma_1,\sigma_3}(x,z) - \varphi(x,z))
        + \kappa^2(x,-(z-L))(\Phi^{\sigma_2,\sigma_4}(x,-(z-L)) - \varphi(x,-(z-L)))&\\
    &= \kappa^2(x,z)(\Phi^s(x,z) - 2\varphi(x,z)) .&
}
It doesn't behave correctly in the ``bulk limit'' $\sigma_1=\sigma_2=\sigma_3=\sigma_4=0$ and $L\to\infty$
($\Phi^{s,\text{bulk}}(r)=2\varphi(r)$ instead of $\varphi(r)$) for the same reason.
The boundary condition
at the walls will also be violated because each solution adds some value to the normal derivative at the
wall in the other solution. Since $\Phi^s(x,z) \to 0$ for $x\to\pm\infty$, 
this error will become smaller for larger $L$. That is why it may be a better approximation for
large values of $L$ but certainly not for small $L$.
\\

Please note that we only add the potentials of the systems. The calculation of the energies
will not use such an addition and will be entirely based on the expression of $\Phi^s(x,z)$
derived here.
\\

\begin{figure}[H]
    \centering
    \begin{tikzpicture}[
        baseline=-0.1cm,
        scale = 1.3,
        wall/.style={
                postaction={draw,decorate,decoration={border,angle=-45,
                            amplitude=0.3cm,segment length=2mm}}},
        ]
        \draw[->] (0,-1.5) -- (0,2) node (xaxis) [above] {$x$};
        \draw[->] (-3,0) -- (3,0) node (zaxis) [right] {$z$};
        \draw[very thick] (0,0) -- (3,0);
        \draw[,  line width=.5pt, wall] (0,1.5)  -- node [right] {$\sigma_k$} (0,0)   ;
        \draw[, line width=.5pt, wall] (0,0)    -- node [right] {$\sigma_l$} (0,-1.5);
        \draw (1, 0.75) node {$\varepsilon_1,\kappa_1$};
        \draw (1,-0.75) node {$\varepsilon_2,\kappa_2$};
        \draw (0,-1.5) node [below] {$0$};
    \end{tikzpicture}
    \caption{
        The system has a wall at $z=0$, and the half space $z>0$
        is filled with two media forming an interface at $x=0$.
        The medium residing in the space $x>0$ ($x<0$) is called medium 1 (2) and is
        characterized by its dielectric constant $\varepsilon_1$ ($\varepsilon_2$).
        The wall carries a surface charge density of $\sigma_k$ for $x>0$ and $\sigma_l$ for $x<0$.
    }
    \label{fig:ssp2}
\end{figure}
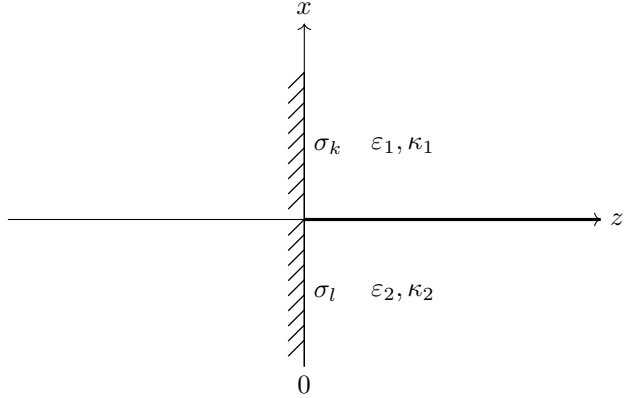

We calculate the potential for the first system depicted in \ref{fig:ssp2}.
We will call this solution $\Phi^{\sigma_k,\sigma_l}$.
The solution for the complete System can then be determined by coordinate transformation
and exchange of $\sigma$'s (see Eq. \eqref{ssprfinal}).

We again employ the same ansatz as used for the exact solution, with the systems changed
appropriatly. So our subproblems will be
\begin{enumerate}
    \item A fluid interface without any walls. (See Fig. \ref{fig:sp1}).
        The potential $\phi$ we get from this problem fulfills the Poisson-Boltzmann
        equation in each medium and satisfies the boundary conditions at the interface
        and for $x\to\pm\infty$.
        Due to the symmetry of the problem $\phi$ depends only on the $x$-coordinate,
        $\phi=\phi(x)$. This problem is exactly the same as the first subproblem from
        the exact solution.
    \item A wall with only one medium and a single surface charge. (See Fig. \ref{fig:sssp2}).
        The potentials $\psi_1$ (with medium 1), $\psi_2$ (with medium 2) we get from this problem
        fulfill the Poisson-Boltzmann
        equation in their respective medium and satisfy the boundary conditions at the wall.
        Due to the symmetry of the problem $\psi_i$ depends only on the $z$-coordinate,
        $\psi_i=\psi_i(z)$.
    \item The calculation of a correction function $c_i$ that
        fulfills the homogenous Poisson-Boltzmann equation
        and restores continuity at the interface
        but leaves the other boundary conditions unchanged.
\end{enumerate}
For the solution of the system we set
\eq{\Phi^{\sigma_k,\sigma_l}(x,z) = \piecewise{\Phi^{\sigma_k,\sigma_l}_1(x,z) & x \ge 0 \\ \Phi^{\sigma_k,\sigma_l}_2(x,z) & x \le 0}}
with
\eq{\Phi^{\sigma_k,\sigma_l}_i(x,z) = \phi(x) + \psi_i(z) + c_i(x,z)}
and the final solution will be given by
\eq{\Phi^s(x,z) = \Phi^{\sigma_1,\sigma_3}(x,z) + \Phi^{\sigma_2,\sigma_4}(x,-(z-L))\label{ssprfinal}}

\subsection{Subproblem 1}
Subproblem 1 is the same as subroblem 1 of the exact solution. Its solutions are given by Eq. \eqref{ess3s}.
\fleq{
    \phi_1(x) &= \frac{\varepsilon_2\kappa_2\Phi_D}{\varepsilon_1\kappa_1+\varepsilon_2\kappa_2}e^{-\kappa_1 x} &\\
    \phi_1(x) &= \Phi_D\left(1 - \frac{\varepsilon_1\kappa_1}{\varepsilon_1\kappa_1+\varepsilon_2\kappa_2}e^{\kappa_2 x}\right)
}

\subsection{Subproblem 2}

\begin{figure}[H]
    \centering
    \begin{tikzpicture}[
        scale = 1.3,
        wall/.style={
                postaction={draw,decorate,decoration={border,angle=-45,
                            amplitude=0.3cm,segment length=2mm}}},
        ]
        \draw[->] (0,-1.5) -- (0,2) node (xaxis) [above] {$x$};
        \draw[->] (-3,0) -- (3,0) node (zaxis) [right] {$z$};
        \draw[,  line width=.5pt, wall] (0,1.5)  -- node [right] {$\sigma_k$} (0,0)   ;
        \draw[,  line width=.5pt, wall] (0,0)    -- node [right] {$\sigma_k$} (0,-1.5);
        \draw (1, 0.75) node {$\varepsilon_1,\kappa_1$};
        \draw (1,-0.75) node {$\varepsilon_1,\kappa_1$};
        \draw (0,-1.5) node [below] {$0$};
    \end{tikzpicture}
    \begin{tikzpicture}[
        scale = 1.3,
        wall/.style={
                postaction={draw,decorate,decoration={border,angle=-45,
                            amplitude=0.3cm,segment length=2mm}}},
        ]
        \draw[->] (0,-1.5) -- (0,2) node (xaxis) [above] {$x$};
        \draw[->] (-3,0) -- (3,0) node (zaxis) [right] {$z$};
        \draw[, line width=.5pt, wall] (0,1.5)  -- node [right] {$\sigma_l$} (0,0)   ;
        \draw[, line width=.5pt, wall] (0,0)    -- node [right] {$\sigma_l$} (0,-1.5);
        \draw (1, 0.75) node {$\varepsilon_2,\kappa_2$};
        \draw (1,-0.75) node {$\varepsilon_2,\kappa_2$};
        \draw (0,-1.5) node [below] {$0$};
    \end{tikzpicture}
    \caption{
        The system left (right) is used to calculate $\psi_1$ ($\psi_2$).
        The system has a wall at $z=0$ and a single medium filling the half space $z>0$.
        The wall at is homogenously charged with charge density $\sigma_k$ ($\sigma_l$),
        and the fluid is characterized by its dielectric constant $\varepsilon_1$ ($\varepsilon_2$)
        and its inverse Debye length $\kappa_1$ ($\kappa_2$).
    }
    \label{fig:sssp2}
\end{figure}
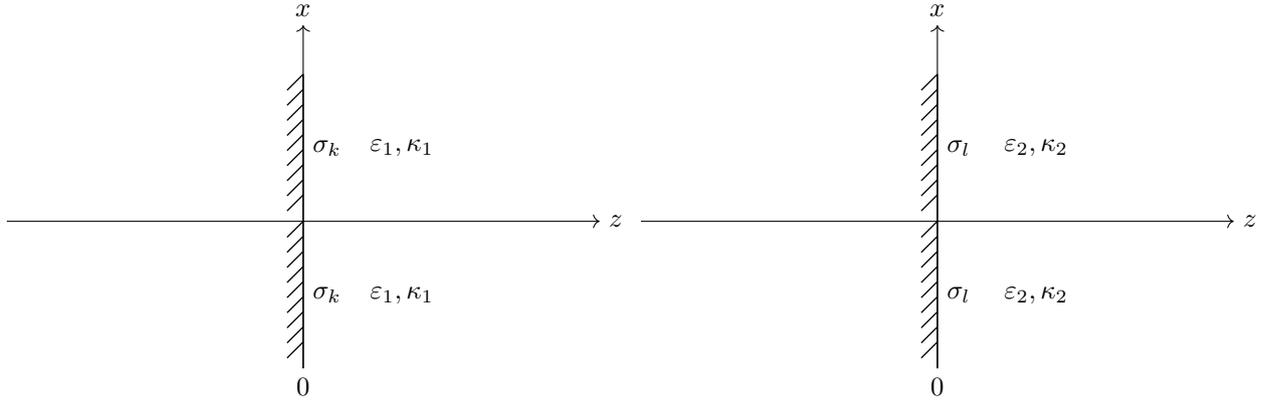
We first solve the problem for $\psi_1$, as depicted and described in Fig. \ref{fig:sssp2}.
The electrostatic potential is obtained by solving the following Debye-H\"uckel equation
\fleq{
    &\laplace\psi_1 =\kappa_1^2\psi_1 &
}
The general solution is
\fleq{
    &\psi_1(z) = Ee^{-\kappa_1 z} + Fe^{\kappa_1 z} &
}
with constants $E$ and $F$.
The potential $\psi$ should be finite for $x\to\infty$. Therefore, $F=0$.
With the boundary condition at the wall
\fleq{
    &\varepsilon_1\partial_z\psi_1(z)|_{z=0} = -\sigma_k &
}
we obtain
\fleq{
    &\varepsilon_1E(-\kappa_1) = \sigma_k \f E = \frac{\sigma_k}{\varepsilon_1\kappa_1} &
}
So, finally
\fleq{
    &\psi_1(z) = \frac{\sigma_k}{\varepsilon_1\kappa_1}e^{-\kappa_1 z} &
}
The calculation for $\psi_2$ is exactly the same, with $\sigma_l$ in place of $\sigma_k$
and $\varepsilon_2$, $\kappa_2$ in place of $\varepsilon_1$, $\kappa_1$, respectively.
Therefore, the potential in this case is given by
\fleq{
    & \psi_2(z) = \frac{\sigma_l}{\varepsilon_2\kappa_2}e^{-\kappa_2 z} &
}
\subsection{Subproblem 3}
The conditions $c_i(x,z)$ has to fulfill in our extended system are as follows:
\begin{enumerate}
    \item\label{ss3c1} $\laplace c_i(x,z) = \kappa_i^2(x,z)$ so the pde is fulfilled
    \item\label{ss3c2} $\partial_z c_i(x,z) |_{z=0} = 0$ so the boundary conditions for the surface charges stay valid
    \item\label{ss3c3} $\lim_{x\to\pm\infty}\abs{c_i(x,z)}<\infty$ so the same will hold for $\Phi^e_i$.
    \item\label{ss3c4} $c_1(0,z) + \psi_1(z) = c_2(0,z) + \psi_2(z)$ since $\phi_1(0)=\phi_2(0)$ is already satisfied
    \item\label{ss3c5} $\varepsilon_1\partial_x c_1(x,z) |_{x=0} = \varepsilon_2\partial_x c_2(x,z) |_{x=0}$
        so the boundary condition for the interface stays valid
\end{enumerate}
Since we now have a semi-infinite system we use a Fourier transform in $z$
instead of the Fourier series used in the exact solution.
In the following calculation we will transform all the conditions for $c_i(x,z)$
to conditions for the Fourier transformation $\hat{c}_i(x,q)=\sF[c_i(x,z)]$
and then derive an explicit solution for $\hat{c}_i(x,q)$.
\paragraph{Treatment of Condition \ref{ss3c1}}
\fleq{
    & (\partial_x^2 + \partial_z^2 - \kappa_i^2)c_i(x,z) = 0 \qquad|\sF &\\
    & (\partial_x^2 - q^2 - \kappa_i^2)\hat{c}_i(x,q) = 0 \qquad \hat{c}_i(x,q) = \int_{-\infty}^\infty c_i(x,z)e^{-iqz}\d z &
}
\fleq{
    \overset{\ref{ss3c3}}{\f} & \hat{c}_1(x,q) = M_1(q) e^{-p_1x} \qquad\text{with } p_i := \sqrt{q^2+\kappa_i^2} &\\
                              & \hat{c}_2(x,q) = M_2(q) e^{ p_2x} &
}
\paragraph{Treatment of Condition \ref{ss3c5}}
\fleq{
    & \varepsilon_1 \partial_x c_1(x,z)|_{x=0} = \varepsilon_2\partial_x c_2(x,z) |_{x=0} \qquad|\sF &\\
    & \varepsilon_1 \partial_x \hat{c}_1(x,q)|_{x=0} = \varepsilon_2\partial_x \hat{c}_2(x,q) |_{x=0}&\\
    & \f \varepsilon_1 (-p_1) M_1(q) = \varepsilon_2 p_2 M_2(q) \label{ssc5r}&
}
\paragraph{Treatment of Condition \ref{ss3c4}}
\fleq{
    & c_1(0,z) + \psi_1(z) = c_2(0,z) + \psi_2(z) \qquad|\sF&\\
    & \hat{c}_1(0,q) + \hat{\psi}_1(q) = \hat{c}_2(0,q) + \hat{\psi}_2(q) \label{ssc4r}&
}
To proceed further, we need to first calculate the Fourier transform $\hat{\psi_i}(q)$ of $\psi_i(z)$.
\tchn{
\fleq{
    \hat{\psi}_1(q) &= \int_{-\infty}^\infty \psi_1(z) e^{-iqz} \d z &\\
    &= \frac{\sigma_k}{\varepsilon_1\kappa_1}\left(\int_0^\infty e^{-\kappa_1z} e^{-iqz} \d z + \int_{-\infty}^0 e^{\kappa_1z} e^{-iqz} \d z \right)&\\
    &= \frac{\sigma_k}{\varepsilon_1\kappa_1}\left(\int_0^\infty e^{-(\kappa_1 + iq)z} \d z + \int_{-\infty}^0 e^{-(-\kappa_1 + iq)z} \d z \right)&\\
    &= \frac{\sigma_k}{\varepsilon_1\kappa_1}\left(\left[\frac{1}{-(\kappa_1+iq)}e^{-(\kappa_1 + iq)z}\right]_0^\infty + \left[\frac{1}{-(-\kappa_1+iq)}e^{-(-\kappa_1 + iq)z}\right]_{-\infty}^0 \right)&\\
    &= \frac{\sigma_k}{\varepsilon_1\kappa_1}\left(\frac{1}{\kappa_1+iq} + \frac{1}{-(-\kappa_1+iq)}\right) &\\
    &= \frac{\sigma_k}{\varepsilon_1\kappa_1}\frac{2\kappa_1}{\kappa_1^2 + q^2} &
}
}
Therefore, condition \eqref{ssc4r} gives:
\fleq{
    & M_1(q) + \frac{\sigma_k}{\varepsilon_1\kappa_1}\frac{2\kappa_1}{\kappa_1^2 + q^2} = 
    M_2(q) + \frac{\sigma_l}{\varepsilon_2\kappa_2}\frac{2\kappa_2}{\kappa_2^2 + q^2} \label{ssc42r}&
}
$M_1(q)$ and $M_2(q)$ can be obtained by solving the Eqs. \eqref{ssc5r} and \eqref{ssc42r}.
\\

\tchn{
Shorthand
\fleq{
    & aM_1(q) = bM_2(q) \qquad M_1(q) + c = M_2(q) + d&\\
    & \f \frac{b}{a} M_2(q) + c = M_2(q) + d &\\
    & \f \left(\frac{b}{a} - 1\right) M_2(q) = d - c &\\
    & \f M_2(q) = \frac{a}{b - a} (d - c) &\\
    & \f M_1(q) = \frac{b}{a} M_2(q) = \frac{b}{b - a} (d - c) &
}
\fleq{
    \f M_1(q) &= \frac{\varepsilon_2p_2}{\varepsilon_1p_1 + \varepsilon_2p_2}
    \left(-\frac{\sigma_k}{\varepsilon_1\kappa_1}\frac{2\kappa_1}{\kappa_1^2 + q^2}
    + \frac{\sigma_l}{\varepsilon_2\kappa_2}\frac{2\kappa_2}{\kappa_2^2 + q^2} \right)
    = 2\frac{\varepsilon_2p_2}{\varepsilon_1p_1 + \varepsilon_2p_2}
    \left(-\frac{\sigma_k}{\varepsilon_1}\frac{1}{p_1^2}
    + \frac{\sigma_l}{\varepsilon_2}\frac{1}{p_2^2} \right)
    &\\
    M_2(q) &= -\frac{\varepsilon_1p_1}{\varepsilon_1p_1 + \varepsilon_2p_2}
    \left(-\frac{\sigma_k}{\varepsilon_1\kappa_1}\frac{2\kappa_1}{\kappa_1^2 + q^2}
    + \frac{\sigma_l}{\varepsilon_2\kappa_2}\frac{2\kappa_2}{\kappa_2^2 + q^2} \right)
    = -2\frac{\varepsilon_1p_1}{\varepsilon_1p_1 + \varepsilon_2p_2}
    \left(-\frac{\sigma_k}{\varepsilon_1}\frac{1}{p_1^2}
    + \frac{\sigma_l}{\varepsilon_2}\frac{1}{p_2^2} \right) &
}
}
So we have derived an expression for the Fourier transform of $c_i(x,z)$.
Applying the back transformation formula, we get the following expressions for $c_i(x,z)$:
\fleq{
    & c_1(x,z) = \frac{1}{\pi}\int_{-\infty}^\infty 
    \frac{\varepsilon_2p_2}{\varepsilon_1p_1 + \varepsilon_2p_2}
    \left(-\frac{\sigma_k}{\varepsilon_1}\frac{1}{p_1^2}
    + \frac{\sigma_l}{\varepsilon_2}\frac{1}{p_2^2} \right)
    e^{-p_1x} e^{iqz} \d q &\\
    & c_2(x,z) = -\frac{1}{\pi}\int_{-\infty}^\infty 
    \frac{\varepsilon_1p_1}{\varepsilon_1p_1 + \varepsilon_2p_2}
    \left(-\frac{\sigma_k}{\varepsilon_1}\frac{1}{p_1^2}
    + \frac{\sigma_l}{\varepsilon_2}\frac{1}{p_2^2} \right)
    e^{p_2x} e^{iqz} \d q &
}
$e^{iqz} = \cos(qz) + i\sin(qz)$ and because the integrand is even $\f\int\dots\sin = 0$, so we can write
\fleq{
    & c_1(x,z) = \frac{1}{\pi}\int_{-\infty}^\infty 
    \frac{\varepsilon_2p_2}{\varepsilon_1p_1 + \varepsilon_2p_2}
    \left(-\frac{\sigma_k}{\varepsilon_1}\frac{1}{p_1^2}
    + \frac{\sigma_l}{\varepsilon_2}\frac{1}{p_2^2} \right)
    e^{-p_1x} \cos(qz) \d q &\\
    & c_2(x,z) = -\frac{1}{\pi}\int_{-\infty}^\infty 
    \frac{\varepsilon_1p_1}{\varepsilon_1p_1 + \varepsilon_2p_2}
    \left(-\frac{\sigma_k}{\varepsilon_1}\frac{1}{p_1^2}
    + \frac{\sigma_l}{\varepsilon_2}\frac{1}{p_2^2} \right)
    e^{p_2x} \cos(qz) \d q &
}
\paragraph{Finally,}
using these expressions for $c_1(x,z)$ and $c_2(x,z)$, one can write
\fleq{
    \Phi_1^{\sigma_k,\sigma_l}(x,z) &= \frac{\sigma_k}{\varepsilon_1\kappa_1}e^{-\kappa_1z} &\\
    &+ \frac{\kappa_2\varepsilon_2\Phi_D}{\kappa_1\varepsilon_1+\kappa_2\varepsilon_2}e^{-\kappa_1x} &\\
    &+ \frac{1}{\pi}\int_{-\infty}^{\infty} \frac{\varepsilon_2p_2}{\varepsilon_1p_1+\varepsilon_2p_2}\left(-\frac{\sigma_k}{\varepsilon_1}\frac{1}{p_1^2} + \frac{\sigma_l}{\varepsilon_2}\frac{1}{p_2^2}\right)e^{-p_1x}\cos(qz)\d q &
}
\fleq{
    \Phi_2^{\sigma_k,\sigma_l}(x,z) &= \frac{\sigma_l}{\varepsilon_2\kappa_2}e^{-\kappa_2z} &\\
    &+ \Phi_D  - \frac{\kappa_1\varepsilon_1\Phi_D}{\kappa_1\varepsilon_1+\kappa_2\varepsilon_2}e^{\kappa_2x} &\\
    &+ \frac{1}{\pi}\int_{-\infty}^{\infty} -\frac{\varepsilon_1p_1}{\varepsilon_1p_1+\varepsilon_2p_2}\left(-\frac{\sigma_k}{\varepsilon_1}\frac{1}{p_1^2} + \frac{\sigma_l}{\varepsilon_2}\frac{1}{p_2^2}\right)e^{p_2x}\cos(qz)\d q &
}
Using Eq. \eqref{ssprfinal} we finally obtain the superposition potentials in the two media:
\fleq{
    \Phi_1^s(x,z) &= \frac{\sigma_1}{\varepsilon_1\kappa_1}e^{-\kappa_1z} + \frac{\sigma_2}{\varepsilon_1\kappa_1}e^{\kappa_1(z-L)}&\\
    &+ \frac{2\kappa_2\varepsilon_2\Phi_D}{\kappa_1\varepsilon_1+\kappa_2\varepsilon_2}e^{-\kappa_1x} &\\
    &+ \frac{1}{\pi}\int_{-\infty}^{\infty} \frac{\varepsilon_2p_2}{\varepsilon_1p_1+\varepsilon_2p_2}\left(-\frac{\sigma_1}{\varepsilon_1}\frac{1}{p_1^2} + \frac{\sigma_3}{\varepsilon_2}\frac{1}{p_2^2}\right)e^{-p_1x}\cos(qz)\d q &\\
    &+ \frac{1}{\pi}\int_{-\infty}^{\infty} \frac{\varepsilon_2p_2}{\varepsilon_1p_1+\varepsilon_2p_2}\left(-\frac{\sigma_2}{\varepsilon_1}\frac{1}{p_1^2} + \frac{\sigma_4}{\varepsilon_2}\frac{1}{p_2^2}\right)e^{-p_1x}\cos(-q(z-L))\d q &
    \label{phis1}
}
\fleq{
    \Phi_2^s(x,z) &= \frac{\sigma_3}{\varepsilon_2\kappa_2}e^{-\kappa_2z} + \frac{\sigma_4}{\varepsilon_2\kappa_2}e^{\kappa_2(z-L)}&\\
    &+ 2\Phi_D  - 2\frac{\kappa_1\varepsilon_1\Phi_D}{\kappa_1\varepsilon_1+\kappa_2\varepsilon_2}e^{\kappa_2x} &\\
    &+ \frac{1}{\pi}\int_{-\infty}^{\infty} -\frac{\varepsilon_1p_1}{\varepsilon_1p_1+\varepsilon_2p_2}\left(-\frac{\sigma_1}{\varepsilon_1}\frac{1}{p_1^2} + \frac{\sigma_3}{\varepsilon_2}\frac{1}{p_2^2}\right)e^{p_2x}\cos(qz)\d q &\\
    &+ \frac{1}{\pi}\int_{-\infty}^{\infty} -\frac{\varepsilon_1p_1}{\varepsilon_1p_1+\varepsilon_2p_2}\left(-\frac{\sigma_2}{\varepsilon_1}\frac{1}{p_1^2} + \frac{\sigma_4}{\varepsilon_2}\frac{1}{p_2^2}\right)e^{p_2x}\cos(-q(z-L))\d q &
    \label{phis2}
}

\subsection{Consistency with literature}
We test our expression for consistency with the result of Ref. \cite{supp}.
Since in Ref. \cite{supp} the walls are located at $z=\pm L$, we need to do a transformation
in order to compare the results.
For this we have to set
\fleq{
    &L = 2L' &\\
    &z = z' + L' &\\
    &\sigma_1'=\sigma_1=\sigma_2 &\\
    &\sigma_2'=\sigma_3=\sigma_4 &
}
We transform each line of our expression for $\Phi^s(x,z)$ separately and compare with Ref. \cite{supp}.
\paragraph{Line 1}
\fleq{
    &  \frac{\sigma_1}{\varepsilon_1\kappa_1}e^{-\kappa_1z} + \frac{\sigma_2}{\varepsilon_1\kappa_1}e^{\kappa_1(z-L)} &\\
    &= \frac{\sigma_1'}{\varepsilon_1\kappa_1}\left(e^{-\kappa_1(z'+L')} + e^{\kappa_1(z'+L'-2L')}\right) &\\
    &= \frac{\sigma_1'}{\varepsilon_1\kappa_1}\left(e^{-\kappa_1z' - \kappa_1L'} + e^{\kappa_1z'-\kappa_1L'}\right) &\\
    &= \frac{\sigma_1'}{\varepsilon_1\kappa_1}e^{- \kappa_1L'}2\cosh(\kappa_1z') \quad\checkmark &
}
\paragraph{Line 2}
Nothing to be done.
\paragraph{Line 3}
\fleq{
    &  \cos(-q(z-L)) &\\
    &= \cos(-q(z'+L'-2L')) &\\
    &= \cos(-qz' + qL') &\\
    &= \cos(-qz')\cos(qL')-\sin(-qz')\sin(qL')&\\
    &= \cos(qz')\cos(qL')+\sin(qz')\sin(qL')&
}
\fleq{
    &  \cos(qz) &\\
    &= \cos(q(z'+L')) &\\
    &= \cos(qz' + qL') &\\
    &= \cos(qz')\cos(qL')-\sin(qz')\sin(qL')&\\
    &= \cos(qz')\cos(qL')-\sin(qz')\sin(qL')&
}
\fleq{
    & \frac{1}{\pi}\int_{-\infty}^{\infty} \frac{\varepsilon_2p_2}{\varepsilon_1p_1+\varepsilon_2p_2}\left(-\frac{\sigma_1}{\varepsilon_1}\frac{1}{p_1^2} + \frac{\sigma_3}{\varepsilon_2}\frac{1}{p_2^2}\right)e^{-p_1x}\cos(qz)\d q &\\
    &+ \frac{1}{\pi}\int_{-\infty}^{\infty} \frac{\varepsilon_2p_2}{\varepsilon_1p_1+\varepsilon_2p_2}\left(-\frac{\sigma_2}{\varepsilon_1}\frac{1}{p_1^2} + \frac{\sigma_4}{\varepsilon_2}\frac{1}{p_2^2}\right)e^{-p_1x}\cos(-q(z-L))\d q &\\
    &= \frac{1}{\pi}\int_{-\infty}^{\infty} \frac{\varepsilon_2p_2}{\varepsilon_1p_1+\varepsilon_2p_2}\left(-\frac{\sigma_1'}{\varepsilon_1}\frac{1}{p_1^2} + \frac{\sigma_2'}{\varepsilon_2}\frac{1}{p_2^2}\right)e^{-p_1x}\left(\cos(qz')\cos(qL')-\sin(qz')\sin(qL')\right)\d q &\\
    &+ \frac{1}{\pi}\int_{-\infty}^{\infty} \frac{\varepsilon_2p_2}{\varepsilon_1p_1+\varepsilon_2p_2}\left(-\frac{\sigma_1'}{\varepsilon_1}\frac{1}{p_1^2} + \frac{\sigma_2'}{\varepsilon_2}\frac{1}{p_2^2}\right)e^{-p_1x}\left(\cos(qz')\cos(qL')+\sin(qz')\sin(qL')\right)\d q &\\
    &= \frac{2}{\pi}\int_{-\infty}^{\infty} \frac{\varepsilon_2p_2}{\varepsilon_1p_1+\varepsilon_2p_2}\left(-\frac{\sigma_1'}{\varepsilon_1}\frac{1}{p_1^2} + \frac{\sigma_2'}{\varepsilon_2}\frac{1}{p_2^2}\right)e^{-p_1x}\cos(qz')\cos(qL')\d q \quad\checkmark&
}
Therefore we can obtain the result for the case of
identical particles given in Ref. \cite{supp} from our general expressions.

\section{Plots}

\newcommand{\phieSi}{\SI{0.02}{e/nm^2}}
\newcommand{\phieSii}{\SI{0.03}{e/nm^2}}
\newcommand{\phieSiii}{\SI{0.0004}{e/nm^2}}
\newcommand{\phieSiv}{\SI{0.0002}{e/nm^2}}
\newcommand{\phieEi}{\SI{80}{\varepsilon_0}}
\newcommand{\phieEii}{\SI{2}{\varepsilon_0}}
\newcommand{\phieKi}{\SI{0.1}{nm^{-1}}}
\newcommand{\phieKii}{\SI{0.03}{nm^{-1}}}
\newcommand{\phiePhiD}{\SI{1}{k_BT/e}}

For all the following plots of this chapter we use
$\varepsilon_1=\phieEi$, $\varepsilon_2=\phieEii$, $\kappa_1=\phieKi$, $\kappa_2=\phieKii$, and $\Phi_D=\phiePhiD$.
\subsection{Comparison}
Here, we compare the exact and superpositon potentials $\Phi^e$ and $\Phi^s$ for different separations between the walls.
As expected, and as it can be seen below, 
the superposition aprroximation increasingly deviates from the exact solutions for small values of $L$.

    \begin{figure}[H]
        \centering \small\input{cmp1_1.tex}\input{cmp1_1x.tex}\normalsize\caption{Comparison of $\Phi^e$ and $\Phi^s$ for $L=\SI{10}{nm}$ and $\sigma_1=\phieSi$, $\sigma_2=-\phieSii$, $\sigma_3=-\phieSiii$, $\sigma_4=\phieSiv$} \label{plot:phi_cmp10_1}
    \end{figure}

    \begin{figure}[H]
        \centering \small\input{cmp1_3.tex}\input{cmp1_3x.tex}\normalsize\caption{Comparison of $\Phi^e$ and $\Phi^s$ for $L=\SI{100}{nm}$ and $\sigma_1=\phieSi$, $\sigma_2=-\phieSii$, $\sigma_3=-\phieSiii$, $\sigma_4=\phieSiv$} \label{plot:phi_cmp100_1}
    \end{figure}

    \begin{figure}[H]
        \centering \small\input{cmp2_1.tex}\input{cmp2_1x.tex}\normalsize\caption{Comparison of $\Phi^e$ and $\Phi^s$ for $L=\SI{10}{nm}$ and $\sigma_1=\phieSi$, $\sigma_2=\phieSii$, $\sigma_3=\phieSiii$, $\sigma_4=\phieSiv$} \label{plot:phi_cmp10_2}
    \end{figure}

    \begin{figure}[H]
        \centering \small\input{cmp2_3.tex}\input{cmp2_3x.tex}\normalsize\caption{Comparison of $\Phi^e$ and $\Phi^s$ for $L=\SI{100}{nm}$ and $\sigma_1=\phieSi$, $\sigma_2=\phieSii$, $\sigma_3=\phieSiii$, $\sigma_4=\phieSiv$} \label{plot:phi_cmp100_2}
    \end{figure}

\subsection{Exact Solution}
Some cross sections of the exact solution Potential

    \begin{figure}[H]
        \centering \small\input{phie_pppp_1.tex}\input{phie_pppp_2.tex}\normalsize\caption{$\Phi^e$ for $L=\SI{100}{nm}$ and $\sigma_1=\phieSi$, $\sigma_2=\phieSii$, $\sigma_3=\phieSiii$, $\sigma_4=\phieSiv$} \label{plot:phie_pppp}
    \end{figure}

    \begin{figure}[H]
        \centering \small\input{phie_pppm_1.tex}\input{phie_pppm_2.tex}\normalsize\caption{$\Phi^e$ for $L=\SI{100}{nm}$ and $\sigma_1=\phieSi$, $\sigma_2=\phieSii$, $\sigma_3=\phieSiii$, $\sigma_4=-\phieSiv$} \label{plot:phie_pppm}
    \end{figure}

    \begin{figure}[H]
        \centering \small\input{phie_ppmp_1.tex}\input{phie_ppmp_2.tex}\normalsize\caption{$\Phi^e$ for $L=\SI{100}{nm}$ and $\sigma_1=\phieSi$, $\sigma_2=\phieSii$, $\sigma_3=-\phieSiii$, $\sigma_4=\phieSiv$} \label{plot:phie_ppmp}
    \end{figure}

    \begin{figure}[H]
        \centering \small\input{phie_ppmm_1.tex}\input{phie_ppmm_2.tex}\normalsize\caption{$\Phi^e$ for $L=\SI{100}{nm}$ and $\sigma_1=\phieSi$, $\sigma_2=\phieSii$, $\sigma_3=-\phieSiii$, $\sigma_4=-\phieSiv$} \label{plot:phie_ppmm}
    \end{figure}

    \begin{figure}[H]
        \centering \small\input{phie_pmpp_1.tex}\input{phie_pmpp_2.tex}\normalsize\caption{$\Phi^e$ for $L=\SI{100}{nm}$ and $\sigma_1=\phieSi$, $\sigma_2=-\phieSii$, $\sigma_3=\phieSiii$, $\sigma_4=\phieSiv$} \label{plot:phie_pmpp}
    \end{figure}

    \begin{figure}[H]
        \centering \small\input{phie_pmpm_1.tex}\input{phie_pmpm_2.tex}\normalsize\caption{$\Phi^e$ for $L=\SI{100}{nm}$ and $\sigma_1=\phieSi$, $\sigma_2=-\phieSii$, $\sigma_3=\phieSiii$, $\sigma_4=-\phieSiv$} \label{plot:phie_pmpm}
    \end{figure}

    \begin{figure}[H]
        \centering \small\input{phie_pmmp_1.tex}\input{phie_pmmp_2.tex}\normalsize\caption{$\Phi^e$ for $L=\SI{100}{nm}$ and $\sigma_1=\phieSi$, $\sigma_2=-\phieSii$, $\sigma_3=-\phieSiii$, $\sigma_4=\phieSiv$} \label{plot:phie_pmmp}
    \end{figure}

    \begin{figure}[H]
        \centering \small\input{phie_pmmm_1.tex}\input{phie_pmmm_2.tex}\normalsize\caption{$\Phi^e$ for $L=\SI{100}{nm}$ and $\sigma_1=\phieSi$, $\sigma_2=-\phieSii$, $\sigma_3=-\phieSiii$, $\sigma_4=-\phieSiv$} \label{plot:phie_pmmm}
    \end{figure}

The cases with inverted signs do not have the same solutions but the plots are qualitatily not much different, so we omit them.

\chapter{Interaction Energies}
\label{ch:en}
\section{Calculation of Energies}
Using the electrostatic potentials derived in the last chapter we can now proceed
to calculate the interaction energies of our system.
\\

To better understand the behavior of the system we
split the total energy into different contributions:
\eq{
    \Omega(L) &= \Omega_{b,1}V_1 + \Omega_{b,2}V_2 + \gamma_{1,2}A_{1,2}
        + (\gamma_1 + \gamma_2)\frac{A_1}{2}
        + \omega_{\gamma, 1}(L)A_1\\
        &+ (\gamma_3 + \gamma_4)\frac{A_2}{2}
        + \omega_{\gamma, 2}(L)A_2
        + (\tau_1 + \tau_2)\frac{l}{2}
        + \omega_\tau(L)l
        \label{omegasplit}
}
Here $V_i$ denotes the volume containing medium $i$, $A_{1,2}$ denotes the area of
the liquid-liquid interface, $A_i$ denotes the total area of the walls in contact
with medium $i$ and $l$ denotes the total length of the two three-phase contact
lines (in contact with medium 1, medium 2 and the wall).
\begin{itemize}
    \item
        $\Omega_{b,i}$ is the bulk energy density in medium $i$.
        \fleq{
            \beta\Omega_{b,1} &\stackrel{\eqref{omegabulkcont}}{=}
            \frac{1}{V_1}\int_{V_1} \sum_{i\in\m{\pm}} \ubr{I(r)}{=I_1}
            \bigg(\ubr{\ln\left(\frac{I(r)}{\zeta_i}\right)}
                {\overset{\eqref{bulk1}}{=}0} - 1 + \ubr{\beta V_i(r)}{=0}\bigg)
            = \frac{1}{V_1}\int_{V_1} \sum_{i\in\m{\pm}} -I_1
            = -2 I_1 &
        }
        \fleq{
            \beta\Omega_{b,2} &\stackrel{\eqref{omegabulkcont}}{=}
            \frac{1}{V_2}\int_{V_2} \sum_{i\in\m{\pm}} \ubr{I(r)}{=I_2}
            \bigg(\ln\left(\frac{I(r)}{\zeta_i}\right) - 1 + \ubr{\beta V_i(r)}{=\beta f_i}\bigg)&\\
            &= \frac{1}{V_2}\int_{V_2} \sum_{i\in\m{\pm}} I_2
            \bigg(\ubr{\ln\left(\frac{I_2}{\zeta_i}\right) + \beta f_i}
                {\overset{\eqref{bulk2}}{=}-\beta eZ_i\Phi_D} - 1\bigg)
            = \frac{1}{V_2}\int_{V_2} -2I_2
            = -2 I_2 &
        }
        Therefore the bulk energy density is simply given by the negative osmotic pressure.
    \item 
        $\gamma_i$ is the surface tension between the part of the wall with surface charge $\sigma_i$
        and the neighbouring liquid. 
        It is the energy per surface area of the part of the wall with a charge
        of $\sigma_i$ in a single medium
        (1 if $i\in\m{1,2}$, 2 if $i\in\m{3,4}$) without interface
        (system (b)/(c)/(e)/(f) in Fig. \ref{fig:enallsys}) minus the bulk energy.
    \item
        $\omega_{\gamma,i}(L)$ is the surface interaction energy per total surface area of the walls
        in contact with medium $i$ at distance $L$. 
        $\omega_{\gamma, i}(L)A_i$ is the energy required to bring two walls together to distance $L$,
        i.e. the energy in a system with only two walls charged with
        $\sigma_1, \sigma_2$ in medium 1 ($i=1$) or $\sigma_3, \sigma_4$ in medium 2 ($i=2$) at distance $L$
        (system (d)/(g) in Fig. \ref{fig:enallsys}) minus $(\gamma_1 + \gamma_2)\frac{A_1}{2} + \Omega_{b,1}V_1$
        or $(\gamma_3 + \gamma_4)\frac{A_2}{2} + \Omega_{b,2}V_2$.
    \item
        $\gamma_{1,2}$ is the interfacial tension,
        i.e. the energy per interface area in a system without any walls (system (a) in Fig. \ref{fig:enallsys})
        minus the sum of bulk contributions $\Omega_{b,1}V_1+\Omega_{b,2}V_2$.
    \item
        $\tau_i$ is the line tension acting at the left ($i=1$) or right ($i=2$) three-phase contact line.
        $\tau_i$ can be calculated by calculating the energy of the systems depicted in
        (h) and (i) of Fig. \ref{fig:enallsys} and subtracting the bulk, interface and surface contributions
        calculated previously.
    \item
        $\omega_{\tau}(L)$ is the interaction energy per total length of the two three contact lines at distance $L$.
        $\omega_{\tau}(L)$ is the energy per length required for bringing left and right wall to distance $L$
        that was not yet accounted for in the previous energies.
\end{itemize}
\newpage
\begin{figure}[H]
    \centering
    \begin{tabular}{ccc|ccc|ccc}
        &(a)&&&&&&&\\
        &
        \begin{tikzpicture}[
            scale = 1.3,
            wall/.style={
                    postaction={draw,decorate,decoration={border,angle=-45,
                                amplitude=0.3cm,segment length=2mm}}},
            ]
            \draw[very thick] (0,0) -- (2,0);
            \draw (1, 0.25) node {$\varepsilon_1,\kappa_1$};
            \draw (1,-0.25) node {$\varepsilon_2,\kappa_2$};
            \draw (0,0) node [below] {$0$};
            \draw (2,0) node [below] {$L$};
            \draw (2,-0.7) node {$$};
            \draw (2,0.7) node {$$};
        \end{tikzpicture}
        &&&
        &&&
        &
        \\
        &\ileq{\Omega = \gamma_{1,2}A_{1,2}&+\Omega_{b,1}V_1\\&+\Omega_{b,2}V_2}
        &&&
        &&&
        &
        \\&&&&&&&&\\\hline&&&&&&&&\\&(b)&&&(c)&&&(d)&\\
        &
        \begin{tikzpicture}[
            scale = 1.3,
            wall/.style={
                    postaction={draw,decorate,decoration={border,angle=-45,
                                amplitude=0.3cm,segment length=2mm}}},
            ]
            \draw[line width=.5pt, wall] (0,1.5)  -- node [right] {$\sigma_1$} (0,0)   ;
            \draw (1, 0.75) node {$\varepsilon_1,\kappa_1$};
            \draw (0,0) node [below] {$0$};
            \draw (2,0) node [below] {$L$};
        \end{tikzpicture}
        &&&
        \begin{tikzpicture}[
            scale = 1.3,
            wall/.style={
                    postaction={draw,decorate,decoration={border,angle=-45,
                                amplitude=0.3cm,segment length=2mm}}},
            ]
            \draw[line width=.5pt, wall] (2,0)    -- node [left ] {$\sigma_2$} (2,1.5) ;
            \draw (1, 0.75) node {$\varepsilon_1,\kappa_1$};
            \draw (0,0) node [below] {$0$};
            \draw (2,0) node [below] {$L$};
        \end{tikzpicture}
        &&&
        \begin{tikzpicture}[
            scale = 1.3,
            wall/.style={
                    postaction={draw,decorate,decoration={border,angle=-45,
                                amplitude=0.3cm,segment length=2mm}}},
            ]
            \draw[line width=.5pt, wall] (0,1.5)  -- node [right] {$\sigma_1$} (0,0)   ;
            \draw[line width=.5pt, wall] (2,0)    -- node [left ] {$\sigma_2$} (2,1.5) ;
            \draw (1, 0.75) node {$\varepsilon_1,\kappa_1$};
            \draw (0,0) node [below] {$0$};
            \draw (2,0) node [below] {$L$};
        \end{tikzpicture}
        &
        \\
        &\ileq{\Omega = \gamma_1\frac{A_1}{2} +\Omega_{b,1}V_1}
        &&&\ileq{\Omega = \gamma_2\frac{A_1}{2} +\Omega_{b,1}V_1}
        &&&\ileq{\Omega &= (\gamma_1+\gamma_2)\frac{A_1}{2} +\omega_{\gamma,1}A_1\\&+\Omega_{b,1}V_1}
        &
        \\&&&&&&&&\\\hline&&&&&&&&\\&(e)&&&(f)&&&(g)&\\
        &
        \begin{tikzpicture}[
            scale = 1.3,
            wall/.style={
                    postaction={draw,decorate,decoration={border,angle=-45,
                                amplitude=0.3cm,segment length=2mm}}},
            ]
            \draw[line width=.5pt, wall] (0,1.5)  -- node [right] {$\sigma_3$} (0,0)   ;
            \draw (1, 0.75) node {$\varepsilon_2,\kappa_2$};
            \draw (0,0) node [below] {$0$};
            \draw (2,0) node [below] {$L$};
        \end{tikzpicture}
        &&&
        \begin{tikzpicture}[
            scale = 1.3,
            wall/.style={
                    postaction={draw,decorate,decoration={border,angle=-45,
                                amplitude=0.3cm,segment length=2mm}}},
            ]
            \draw[line width=.5pt, wall] (2,0)    -- node [left ] {$\sigma_4$} (2,1.5) ;
            \draw (1, 0.75) node {$\varepsilon_2,\kappa_2$};
            \draw (0,0) node [below] {$0$};
            \draw (2,0) node [below] {$L$};
        \end{tikzpicture}
        &&&
        \begin{tikzpicture}[
            scale = 1.3,
            wall/.style={
                    postaction={draw,decorate,decoration={border,angle=-45,
                                amplitude=0.3cm,segment length=2mm}}},
            ]
            \draw[line width=.5pt, wall] (0,1.5)  -- node [right] {$\sigma_3$} (0,0)   ;
            \draw[line width=.5pt, wall] (2,0)    -- node [left ] {$\sigma_4$} (2,1.5) ;
            \draw (1, 0.75) node {$\varepsilon_2,\kappa_2$};
            \draw (0,0) node [below] {$0$};
            \draw (2,0) node [below] {$L$};
        \end{tikzpicture}
        &
        \\
        &\ileq{\Omega = \gamma_3\frac{A_2}{2} +\Omega_{b,2}V_2}
        &&&\ileq{\Omega = \gamma_4\frac{A_2}{2} +\Omega_{b,2}V_2}
        &&&\ileq{\Omega &= (\gamma_3+\gamma_4)\frac{A_2}{2} +\omega_{\gamma,2}A_2\\&+\Omega_{b,2}V_2}
        &
        \\&&&&&&&&\\\hline&&&&&&&&\\&(h)&&&(i)&&&(j)&\\
        &
        \begin{tikzpicture}[
            scale = 1.3,
            wall/.style={
                    postaction={draw,decorate,decoration={border,angle=-45,
                                amplitude=0.3cm,segment length=2mm}}},
            ]
            \draw[very thick] (0,0) -- (2,0);
            \draw[line width=.5pt, wall] (0,1.5)  -- node [right] {$\sigma_1$} (0,0)   ;
            \draw[line width=.5pt, wall] (0,0)    -- node [right] {$\sigma_3$} (0,-1.5);
            \draw (1, 0.75) node {$\varepsilon_1,\kappa_1$};
            \draw (1,-0.75) node {$\varepsilon_2,\kappa_2$};
            \draw (0,-1.5) node [below] {$0$};
            \draw (2,-1.5) node [below] {$L$};
        \end{tikzpicture}
        &&&
        \begin{tikzpicture}[
            scale = 1.3,
            wall/.style={
                    postaction={draw,decorate,decoration={border,angle=-45,
                                amplitude=0.3cm,segment length=2mm}}},
            ]
            \draw[very thick] (0,0) -- (2,0);
            \draw[line width=.5pt, wall] (2,0)    -- node [left ] {$\sigma_2$} (2,1.5) ;
            \draw[line width=.5pt, wall] (2,-1.5) -- node [left ] {$\sigma_4$} (2,0)   ;
            \draw (1, 0.75) node {$\varepsilon_1,\kappa_1$};
            \draw (1,-0.75) node {$\varepsilon_2,\kappa_2$};
            \draw (0,-1.5) node [below] {$0$};
            \draw (2,-1.5) node [below] {$L$};
        \end{tikzpicture}
        &&&
        \begin{tikzpicture}[
            scale = 1.3,
            wall/.style={
                    postaction={draw,decorate,decoration={border,angle=-45,
                                amplitude=0.3cm,segment length=2mm}}},
            ]
            \draw[very thick] (0,0) -- (2,0);
            \draw[line width=.5pt, wall] (0,1.5)  -- node [right] {$\sigma_1$} (0,0)   ;
            \draw[line width=.5pt, wall] (0,0)    -- node [right] {$\sigma_3$} (0,-1.5);
            \draw[line width=.5pt, wall] (2,0)    -- node [left ] {$\sigma_2$} (2,1.5) ;
            \draw[line width=.5pt, wall] (2,-1.5) -- node [left ] {$\sigma_4$} (2,0)   ;
            \draw (1, 0.75) node {$\varepsilon_1,\kappa_1$};
            \draw (1,-0.75) node {$\varepsilon_2,\kappa_2$};
            \draw (0,-1.5) node [below] {$0$};
            \draw (2,-1.5) node [below] {$L$};
        \end{tikzpicture}
        &
        \\
        &\ileq{\Omega &= \gamma_1\frac{A_1}{2} + \gamma_3\frac{A_2}{2}\\&+\gamma_{1,2}A_{1,2}
        + \tau_1\frac{l}{2}\\&+\Omega_{b,1}V_1+\Omega_{b,2}V_2}
        &&&\ileq{\Omega &= \gamma_2\frac{A_1}{2} + \gamma_4\frac{A_2}{2}\\&+\gamma_{1,2}A_{1,2}
        + \tau_2\frac{l}{2}\\&+\Omega_{b,1}V_1+\Omega_{b,2}V_2}
        &&&$\Omega = \eqref{omegasplit}$
    \end{tabular}
    \caption{Different sytems and their energies}
    \label{fig:enallsys}
\end{figure}
\begin{figure}[H]
    \centering
    \begin{tikzpicture}[
        scale = 1.3,
        wall/.style={
                postaction={draw,decorate,decoration={border,angle=-45,
                            amplitude=0.3cm,segment length=2mm}}},
        ]
        \draw[->] (0,-1.5) -- (0,2) node (xaxis) [above] {$x$};
        \draw[->] (-3,0) -- (3,0) node (zaxis) [right] {$z$};
        \draw[very thick] (0,0) -- (2,0);
        \draw (1, 0.75) node {$\varepsilon_1,\kappa_1$};
        \draw (1,-0.75) node {$\varepsilon_2,\kappa_2$};
        \draw (0,-1.5) node [below] {$0$};
        \draw (2,-1.5) node [below] {$L$};
        \draw (0,1.3) rectangle (2,-1.3);
        \draw (-0.1,-1.3) node [left] {$-L_x$};
        \draw (-0.1,1.3) node [left] {$L_x$};
        \draw [decorate,decoration={brace,amplitude=7pt}]
        (2,-1.3)+(0.4,0.4) -- (2,-1.3) node [black,midway,xshift=0.4cm,yshift=-0.25cm] {$L_y$};
        \draw (0,1.3) -- +(0.4,0.4);
        \draw (0,-1.3) -- +(0.4,0.4);
        \draw[very thick] (0,0) -- +(0.4,0.4);
        \draw (2,1.3) -- +(0.4,0.4);
        \draw (2,-1.3) -- +(0.4,0.4);
        \draw[very thick] (2,0) -- +(0.4,0.4);
        \draw[very thick] (0,0)+(0.4,0.4) -- (2.4,0.4);
        \draw (0,-1.3)+(0.4,0.4) rectangle +(2.4,3);
        \draw (0.2, 0.8) node {$\sigma_1$};
        \draw (2.2, 0.8) node {$\sigma_2$};
        \draw (0.2, -0.5) node {$\sigma_3$};
        \draw (2.2, -0.5) node {$\sigma_4$};
    \end{tikzpicture}
    \caption{
        Schematic of the system volume considered in the following calculations.
        The volume contains an equal volume of both media (each $L_xL_yL$), the
        interface at $x=0$ with a surface of $L_yL$, four charged wall parts with
        a surface of $L_xL_y$ each and two three-phase contact lines, each of length $L_y$.
    }
    \label{fig:encp}
\end{figure}
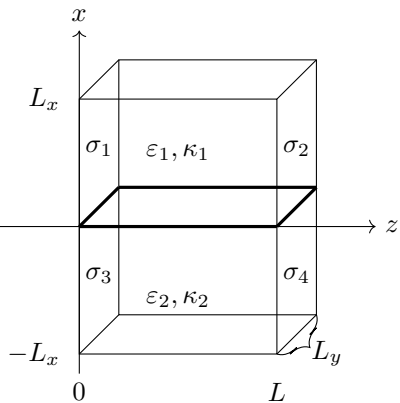
We calculate the energy contained in the volume
$[-L_x,L_x]\times [0,L_y]\times [0,L]$, as illustrated in Figure \ref{fig:encp}.
After subtracting the bulk contributions from the total energy $\Omega$, we obtain
$\mathcal{H}$ which is in this case given by
\fleq{
    \mathcal{H}
    &= (\gamma_1 + \gamma_2 + \gamma_3 + \gamma_4) L_x L_y
        + (\omega_{\gamma,1}(L) + \omega_{\gamma,2}(L)) 2L_xL_y
        + \gamma_{1,2} L_yL
        + (\tau_1 + \tau_2)L_y
        + \omega_\tau(L) 2L_y &
        \label{spliten}
}
Since we can calculate an analytical solution for the total energy,
we can avoid having to calculate the energies
for all the systems outlined in Fig. \ref{fig:enallsys} by making use of the following method.
\\

We check if a term in our solution is proportional to $L_xL_y$.
If it is proportional then we check if it has an $L$-dependence.
If it has an $L$-dependence we separate the parts containing contributions
from $\sigma_1$ and $\sigma_2$, half them and add them to $\tilde{\omega}_{\gamma,1}$
and add half of the parts involving $\sigma_3$ and $\sigma_4$ to $\tilde{\omega}_{\gamma,2}$.
We split up the non-$L$-dependent part into parts containing only one of the $\sigma$'s
and then add the term involving $\sigma_i$ to $\tilde{\tau}_i$.\\

We identify the other contributions in the same fashion, i.e. through the proportionality
with $L_yL$ or $L_y$ and then splitting in $L$-dependent and non-$L$-dependent
parts and identifying the appropriate terms in \eqref{spliten}.
From that we obtain the following quantities
\fleq{
    \mathcal{H}
    &= (\tilde{\gamma_1} + \tilde{\gamma_2} + \tilde{\gamma_3} + \tilde{\gamma_4}) L_x L_y
        + (\tilde{\omega}_{\gamma,1}(L) + \tilde{\omega}_{\gamma,2}(L)) 2L_xL_y
        + \tilde{\gamma}_{1,2} L_yL
        + (\tilde{\tau_1} + \tilde{\tau_2})L_y
        + \tilde{\omega}_\tau(L) 2L_y &
}
but since $L$-dependent and non-$L$-dependent parts have the same kind of proportionality to our lengths
we need to separate them by using the relation that the $L$-dependent part needs to go to zero for $L$
to infinity. This can be done in the following way:
\eq{
    x(L) = \tilde{x}(L) - \lim_{L\to\infty}\tilde{x}(L)
    \qquad x = \tilde{x} + 2\lim_{L\to\infty}\tilde{x}(L) \label{limsubtract}
}
where $x(L)$ is any of $\omega_{\gamma,1}(L)$ or $\omega_{\gamma,1}(L)$ or $\omega_\tau(L)$ and $x$ is
$\gamma_1 + \gamma_2$ or $\gamma_3 + \gamma_4$ or $\tau_1+\tau_2$. The last expressions can be split by
looking at the dependence on the respective charges in the expression.
\\

Before presenting the detailed calculation we first simplify \eqref{dften}.
We can use the symmetry of the Potential with regards to $y$ for the integration in \eqref{dften}.
\fleq{
    \mathcal{H}[\phi]
    &= \frac{1}{2}\Bigg(\int_{\partial V}\d^2r \sigma(r) (\Phi(r) + \varphi(r)) 
    - \Phi_D \int_{x=0}\d^2r D(r)\cdot e_x\Bigg) + \landauO(\phi_i^3)&\\
    &= \frac{L_y}{2}\Bigg(
      \int_0^{L_x} \sigma_1(\Phi_1(x, 0) + 0) \d x
    + \int_0^{L_x} \sigma_2(\Phi_1(x, L) + 0) \d x &\\
    & + \int_{-L_x}^{0} \sigma_3(\Phi_2(x, 0) + \Phi_D) \d x
    + \int_{-L_x}^{0} \sigma_4(\Phi_2(x, L) + \Phi_D) \d x
    + \Phi_D\int_0^L\varepsilon_1\partial_x\Phi_1(0,z) \d z
    \Bigg) &\\
    &= \frac{L_y}{2} \int_0^{L_x}\sigma_1\Phi_1(x,0) + \sigma_2\Phi_1(x,L) \d x &\\
    &+ \frac{L_y}{2} \int_{-L_x}^0\sigma_3\Phi_3(x,0) + \sigma_4\Phi_2(x,L) \d x &\\
    &+ \frac{L_xL_y\Phi_D}{2}(\sigma_3+\sigma_4) &\\
    &+ \frac{\varepsilon_1\Phi_DL_y}{2}\int_0^L (\partial_x\Phi_1)(0,z) \d z \label{Hform}&
}

\section{Exact Solution}
We introduce abbreviations $a_i$, $b_i$, $c_i(L)$, $d_{n,i}(L)$
for terms in $\Phi^e(x,z)$ that are not depending on $x$ or $z$:
\fleq{
    \Phi_1(x,z) &= a_1(L)(\sigma_1\cosh(\kappa_1(L-z)) + \sigma_2\cosh(\kappa_1z)) &\\
    &+ b_1 e^{-\kappa_1 x} &\\
    &+ c_1(L) e^{-\kappa_1 x} &\\
    &+ \sum_{n=1}^\infty d_{1,n}(L) e^{-p_1x}\cos\left(\frac{n\pi z}{L}\right)
}
\fleq{
    \partial_x\Phi_1(x,z) &= -\kappa_1 b_1 e^{-\kappa_1 x} -\kappa_1 c_1(L) e^{-\kappa_1 x} &\\
    &+ \sum_{n=1}^\infty d_{1,n}(L) (-p_1) e^{-p_1x}\cos\left(\frac{n\pi z}{L}\right)
}
\fleq{
    \Phi_2(x,z) &= a_2(L)(\sigma_3\cosh(\kappa_2(L-z)) + \sigma_4\cosh(\kappa_2z)) &\\
    &+ \Phi_D + b_2 e^{\kappa_2 x} &\\
    &+ c_2(L) e^{\kappa_2 x} &\\
    &+ \sum_{n=1}^\infty d_{2,n}(L) e^{p_2x}\cos\left(\frac{n\pi z}{L}\right)
}
We then proceed to insert these expressions into Eq. \eqref{Hform} in order to calculate the energy.
\fleq{
    \f \mathcal{H} &= \frac{L_y}{2}\sigma_1\Bigg(a_1(L)(\sigma_1\cosh(\kappa_1L)+\sigma_2)L_x
        + b_1\frac{1}{-\kappa_1}(e^{-\kappa_1 L_x} - 1) &\\
        &\phantom{+}\qquad+ c_1(L)\frac{1}{-\kappa_1}(e^{-\kappa_1L_x} - 1)
        + \sum_{n=1}^\infty d_{1,n}(L) \frac{1}{-p_1}(e^{-p_1L_x}-1)\Bigg) &\\
    &+ \frac{L_y}{2}\sigma_2\Bigg(a_1(L)(\sigma_1+\sigma_2\cosh(\kappa_1L))L_x
        + b_1\frac{1}{-\kappa_1}(e^{-\kappa_1 L_x} - 1) &\\
        &\phantom{+}\qquad+ c_1(L)\frac{1}{-\kappa_1}(e^{-\kappa_1L_x} - 1)
        + \sum_{n=1}^\infty (-1)^n d_{1,n}(L) \frac{1}{-p_1}(e^{-p_1L_x}-1)\Bigg) &\\
    &+ \frac{L_y}{2}\sigma_3\Bigg(a_2(L)(\sigma_3\cosh(\kappa_2L)+\sigma_4)L_x
        + b_2\frac{1}{\kappa_2}(1-e^{-\kappa_2 L_x}) + \Phi_DL_x &\\
        &\phantom{+}\qquad+ c_2(L)\frac{1}{\kappa_2}(1-e^{-\kappa_2L_x})
        + \sum_{n=1}^\infty d_{2,n}(L) \frac{1}{p_2}(1-e^{-p_2L_x})\Bigg) &\\
    &+ \frac{L_y}{2}\sigma_4\Bigg(a_2(L)(\sigma_3 + \sigma4\cosh(\kappa_2L))L_x
        + b_2\frac{1}{\kappa_2}(1-e^{-\kappa_2 L_x}) + \Phi_DL_x &\\
        &\phantom{+}\qquad+ c_2(L)\frac{1}{\kappa_2}(1-e^{-\kappa_2L_x})
        + \sum_{n=1}^\infty (-1)^n d_{2,n}(L) \frac{1}{p_2}(1-e^{-p_2L_x})\Bigg) &\\
    &+ \frac{L_xL_y\Phi_D}{2}(\sigma_3+\sigma_4) &\\
    &+ \frac{\varepsilon_1\Phi_D L_y}{2}\Bigg(-\kappa_1b_1L-\kappa_1c_1(L)L
        + \sum_{n=1}^\infty d_{1,n}(L) (-p_1)\frac{L}{n\pi}
        \ubr{\left(\sin\left(\frac{n\pi L}{L}\right) - \sin(0)\right)}{=0}\Bigg)
}
We are interested in the case of large $L_x$ so we approx. $e^{-aL_x}\approx 0$
for all factors $a\in\R$ not depending on $L_x$.
\fleq{
    \mathcal{H} &\approx
    \frac{\sigma_1}{2}a_1(L)(\sigma_1\cosh(\kappa_1L)+\sigma_2)L_xL_y
        + \frac{\sigma_1}{2}\frac{b_1}{\kappa_1}L_y + \frac{\sigma_1}{2}\frac{c_1(L)}{\kappa_1}L_y
        + \frac{\sigma_1}{2}\left(\sum_{n=1}^\infty\frac{d_{1,n}(L)}{p_1}\right)L_y &\\
    &+ \frac{\sigma_2}{2}a_1(L)(\sigma_1 + \sigma_2\cosh(\kappa_1L))L_xL_y
        + \frac{\sigma_2}{2}\frac{b_1}{\kappa_1}L_y + \frac{\sigma_2}{2}\frac{c_1(L)}{\kappa_1}L_y
        + \frac{\sigma_2}{2}\left(\sum_{n=1}^\infty(-1)^n\frac{d_{1,n}(L)}{p_1}\right)L_y &\\
    &+ \frac{\sigma_3}{2}a_2(L)(\sigma_3\cosh(\kappa_2L)+\sigma_4)L_xL_y
        + \frac{\sigma_3}{2}\frac{b_2}{\kappa_2}L_y + \frac{\sigma_3}{2}\frac{c_1(L)}{\kappa_2}L_y
        + \frac{\sigma_3}{2}\Phi_DL_xL_y + \frac{\sigma_3}{2}\left(\sum_{n=1}^\infty\frac{d_{2,n}(L)}{p_2}\right)L_y &\\
    &+ \frac{\sigma_4}{2}a_2(L)(\sigma_3+\sigma_4\cosh(\kappa_2L))L_xL_y
        + \frac{\sigma_4}{2}\frac{b_2}{\kappa_2}L_y + \frac{\sigma_4}{2}\frac{c_1(L)}{\kappa_2}L_y
        + \frac{\sigma_4}{2}\Phi_DL_xL_y + \frac{\sigma_4}{2}\left(\sum_{n=1}^\infty(-1)^n\frac{d_{2,n}(L)}{p_2}\right)L_y &\\
    &+ \frac{\Phi_D}{2}(\sigma_3+\sigma_4)L_xL_y&\\
    &+ \frac{\varepsilon_1\Phi_D }{2}(-\kappa_1b_1)L_yL + \frac{\varepsilon_1\Phi_D }{2}(-\kappa_1)c_1(L)L_yL &\\
    &= \Big[ \frac{a_1(L)}{2}((\sigma_1^2+\sigma_2^2)\cosh(\kappa_1L)+2\sigma_1\sigma_2) &\\
    &+ \frac{a_2(L)}{2}(\sigma_3^2+\sigma_4^2)\cosh(\kappa_2L)+2\sigma_3\sigma_4) &\\
    &+ \Phi_D(\sigma_3+\sigma_4) \Big]L_xL_y &\\
    &+ \Bigg[ \frac{b_1}{2\kappa_1}(\sigma_1+\sigma_2) + \frac{c_1(L)}{2\kappa_1}(\sigma_1+\sigma_2)
        + \frac{1}{2}\left(\sum_{n=1}^\infty\frac{(\sigma_1+(-1)^n\sigma_2)d_{1,n}(L)}{p_1}\right) &\\
    &+ \phantom{\Bigg[}\frac{b_2}{2\kappa_2}(\sigma_3+\sigma_4) + \frac{c_2(L)}{2\kappa_2}(\sigma_3+\sigma_4)
        + \frac{1}{2}\left(\sum_{n=1}^\infty\frac{(\sigma_3+(-1)^n\sigma_4)d_{2,n}(L)}{p_2}\right) \Bigg]L_y &\\
    &+ \left[-\frac{\varepsilon_1\Phi_D}{2}\kappa_1b_1 - \frac{\varepsilon_1\Phi_D}{2}\kappa_1c_1(L)\right]L_yL &
}
We then proceed to identify the different terms by their proportionality, $\sigma$ and $L$-dependence, as outlined above.
\fleq{
    & \tilde{\gamma_1} = 0 &\\
    & \tilde{\gamma_2} = 0 &\\
    & \tilde{\gamma_3} = \Phi_D\sigma_3 &\\
    & \tilde{\gamma_4} = \Phi_D\sigma_4 &\\
    & \tilde{\omega}_{\gamma,1}(L) = \frac{1}{4\varepsilon_1\kappa_1\sinh(\kappa_1L)}
        ((\sigma_1^2+\sigma_2^2)\cosh(\kappa_1L)+2\sigma_1\sigma_2) &\\
    & \tilde{\omega}_{\gamma,2}(L) = \frac{1}{4\varepsilon_2\kappa_2\sinh(\kappa_2L)}
        ((\sigma_3^2+\sigma_4^2)\cosh(\kappa_1L)+2\sigma_3\sigma_4) &\\
    & \tilde{\tau_1} =
          \frac{1}{2\kappa_1}\frac{\kappa_2\varepsilon_2\Phi_D}{\kappa_1\varepsilon_1+\kappa_2\varepsilon_2}\sigma_1
        - \frac{1}{2\kappa_2}\frac{\kappa_1\varepsilon_1\Phi_D}{\kappa_1\varepsilon_1+\kappa_2\varepsilon_2}\sigma_3
        - \frac{\Phi_D}{2} \frac{1}{\kappa_1\varepsilon_1+\kappa_2\varepsilon_2}
        \left(\frac{\varepsilon_1\kappa_1}{\kappa_2}\sigma_3 - \frac{\varepsilon_2\kappa_2}{\kappa_1}\sigma_1\right) &\\
        &= \frac{\Phi_D}{\kappa_1\varepsilon_1+\kappa_2\varepsilon_2}
        \left(\frac{\varepsilon_2\kappa_2}{\kappa_1}\sigma_1 - \frac{\varepsilon_1\kappa_1}{\kappa_2}\sigma_3\right) &\\
    & \tilde{\tau_2} =
          \frac{1}{2\kappa_1}\frac{\kappa_2\varepsilon_2\Phi_D}{\kappa_1\varepsilon_1+\kappa_2\varepsilon_2}\sigma_2
        - \frac{1}{2\kappa_2}\frac{\kappa_1\varepsilon_1\Phi_D}{\kappa_1\varepsilon_1+\kappa_2\varepsilon_2}\sigma_4
        - \frac{\Phi_D}{2} \frac{1}{\kappa_1\varepsilon_1+\kappa_2\varepsilon_2}
        \left(\frac{\varepsilon_1\kappa_1}{\kappa_2}\sigma_4 - \frac{\varepsilon_2\kappa_2}{\kappa_1}\sigma_2\right) &\\
        &= \frac{\Phi_D}{\kappa_1\varepsilon_1+\kappa_2\varepsilon_2}
        \left(\frac{\varepsilon_2\kappa_2}{\kappa_1}\sigma_2 - \frac{\varepsilon_1\kappa_1}{\kappa_2}\sigma_4\right) &\\
    & \tilde{\omega}_\tau = \frac{c_1(L)}{4\kappa_1}(\sigma_1 + \sigma_2) + \frac{c_2(L)}{4\kappa_2}(\sigma_3 + \sigma_4)
        +\frac{1}{4}\sum_{n=1}^\infty (\sigma_1+(-1)^n\sigma_2)\frac{d_{1,n}(L)}{p_1} + (\sigma_3+(-1)^n\sigma_4)\frac{d_{2,n}(L)}{p_2} &\\
    &\color{gray}= \frac{1}{4\kappa_1L} \frac{1}{\kappa_1\varepsilon_1+\kappa_2\varepsilon_2}
    \left(\frac{\sigma_3 + \sigma_4}{\kappa_2} - \frac{1}{\kappa_1} \frac{\kappa_2\varepsilon_2}{\kappa_1\varepsilon_1}(\sigma_1+\sigma_2)\right)
    (\sigma_1+\sigma_2) &\\
    &\color{gray}+ \frac{1}{4\kappa_2L} \frac{1}{\kappa_1\varepsilon_1+\kappa_2\varepsilon_2}
    \left(\frac{\sigma_1 + \sigma_2}{\kappa_1} - \frac{1}{\kappa_2} \frac{\kappa_1\varepsilon_1}{\kappa_2\varepsilon_2}(\sigma_3+\sigma_4)\right)
    (\sigma_3+\sigma_4) &\\
    &\color{gray}+ \frac{1}{2L}\sum_{n=1}^\infty \frac{1}{\varepsilon_1p_1+\varepsilon_2p_2}\left(
    \frac{2(\sigma_1+(-1)^n\sigma_2)(\sigma_3+(-1)^n\sigma_4)}{p_1p_2}
    - \frac{1}{p_1}\frac{\varepsilon_2p_2}{\varepsilon_1p_1^2}(\sigma_1+(-1)^n\sigma_2)^2
    - \frac{1}{p_2}\frac{\varepsilon_1p_1}{\varepsilon_2p_2^2}(\sigma_3+(-1)^n\sigma_4)^2\right) &\\
    &= \frac{1}{L} \frac{1}{\kappa_1\varepsilon_1+\kappa_2\varepsilon_2} \left(\frac{(\sigma_1+\sigma_2)(\sigma_3 +\sigma_4)}{2\kappa_1\kappa_2}
    - \frac{1}{4\kappa_1^2}\frac{\kappa_2\varepsilon_2}{\kappa_1\varepsilon_1}(\sigma_1+\sigma_2)^2
    - \frac{1}{4\kappa_2^2}\frac{\kappa_1\varepsilon_1}{\kappa_2\varepsilon_2}(\sigma_3+\sigma_4)^2\right) &\\
    &+ \frac{1}{2L}\sum_{n=1}^\infty \frac{1}{\varepsilon_1p_1+\varepsilon_2p_2}\left(
    \frac{2(\sigma_1+(-1)^n\sigma_2)(\sigma_3+(-1)^n\sigma_4)}{p_1p_2}
    - \frac{1}{p_1}\frac{\varepsilon_2p_2}{\varepsilon_1p_1^2}(\sigma_1+(-1)^n\sigma_2)^2
    - \frac{1}{p_2}\frac{\varepsilon_1p_1}{\varepsilon_2p_2^2}(\sigma_3+(-1)^n\sigma_4)^2\right) &\\
    & \gamma_{1,2} = -\frac{\varepsilon_1\kappa_1\Phi_Db_1}{2}
        = -\frac{\varepsilon_1\varepsilon_2\kappa_1\kappa_2\Phi_D^2}{2(\kappa_1\varepsilon_1+\kappa_2\varepsilon_2)}&
}
As explained earlier, we now calculate the limits of the $L$-dependent terms so we can obtain our final quantities
\fleq{
    \lim_{L\to\infty}\tilde{\omega}_{\gamma,1} &= \lim_{L\to\infty}\frac{1}{4\varepsilon_1\kappa_1}
        \left((\sigma_1^2+\sigma_2^2)\frac{\cosh(\kappa_1L)}{\sinh(\kappa_1L)}+\frac{2\sigma_1\sigma_2}{\sinh(\kappa_1L)}\right) &\\
    &= \frac{\sigma_1^2+\sigma_2^2}{4\kappa_1\varepsilon_1}
}
Similarly,
\fleq{
    & \lim_{L\to\infty}\tilde{\omega}_{\gamma,2} = \frac{\sigma_3^2+\sigma_3^2}{4\kappa_2\varepsilon_2} &
}
with this limit and the expression calculated above we can use Eq. \eqref{limsubtract} to obtain
\fleq{
    &\gamma_1 = \frac{\sigma_1^2}{2\kappa_1\varepsilon_1} &\\
    &\gamma_2 = \frac{\sigma_2^2}{2\kappa_1\varepsilon_1} &\\
    &\gamma_3 = \frac{\sigma_3^2}{2\kappa_2\varepsilon_2}+\sigma_3\Phi_D &\\
    &\gamma_4 = \frac{\sigma_4^4}{2\kappa_2\varepsilon_2}+\sigma_4\Phi_D &\\
    &\omega_{\gamma, 1}(L) = \frac{1}{4\kappa_1\varepsilon_1}\left((\sigma_1^2 + \sigma_2^2)(\coth(\kappa_1L) - 1)
        + \frac{2\sigma_1\sigma_2}{\sinh(\kappa_1L)}\right) &\\
    &\omega_{\gamma, 2}(L) = \frac{1}{4\kappa_2\varepsilon_2}\left((\sigma_3^2 + \sigma_4^2)(\coth(\kappa_2L) - 1)
        + \frac{2\sigma_3\sigma_4}{\sinh(\kappa_2L)}\right) &
}
for the following, we define
\fleq{
    p_{i,n} := \sqrt{\left(\frac{n\pi}{L}\right)^2 + \kappa_i^2}
    \quad\text{and}\quad
    p_i(x) := \sqrt{x^2\pi^2 + \kappa_i^2}
}
With this definition $\tilde{\omega}_\tau$ can be written in a more compact form, which is useful for calculating
the following limit:
\fleq{
    \lim_{L\to\infty} \tilde{\omega}_\tau
    &= \lim_{L\to\infty}\frac{1}{L} \frac{1}{\kappa_1\varepsilon_1+\kappa_2\varepsilon_2} \left(\frac{(\sigma_1+\sigma_2)(\sigma_3 +\sigma_4)}{2\kappa_1\kappa_2}
    - \frac{1}{4\kappa_1^2}\frac{\kappa_2\varepsilon_2}{\kappa_1\varepsilon_1}(\sigma_1+\sigma_2)^2
    - \frac{1}{4\kappa_2^2}\frac{\kappa_1\varepsilon_1}{\kappa_2\varepsilon_2}(\sigma_3+\sigma_4)^2\right) &\\
    &+ \lim_{L\to\infty}\frac{1}{2L}\sum_{n=1}^\infty \frac{1}{\varepsilon_1p_{1,n}+\varepsilon_2p_{2,n}}\Big(
    \frac{2(\sigma_1+(-1)^n\sigma_2)(\sigma_3+(-1)^n\sigma_4)}{p_{1,n}p_{2,n}} &\\&\qquad
    - \frac{1}{p_{1,n}}\frac{\varepsilon_2p_{2,n}}{\varepsilon_1p_{1,n}^2}(\sigma_1+(-1)^n\sigma_2)^2
    - \frac{1}{p_{2,n}}\frac{\varepsilon_1p_{1,n}}{\varepsilon_2p_{2,n}^2}(\sigma_3+(-1)^n\sigma_4)^2\Big) &\\
    &\color{gray}= \lim_{L\to\infty}\frac{1}{2L}\sum_{n=1}^\infty \frac{1}{\varepsilon_1p_{1,n}+\varepsilon_2p_{2,n}}\Big(
    \frac{2(\sigma_1+(-1)^n\sigma_2)(\sigma_3+(-1)^n\sigma_4)}{p_{1,n}p_{2,n}} &\\&\color{gray}\qquad
    - \frac{1}{p_{1,n}}\frac{\varepsilon_2p_{2,n}}{\varepsilon_1p_{1,n}^2}(\sigma_1+(-1)^n\sigma_2)^2
    - \frac{1}{p_{2,n}}\frac{\varepsilon_1p_{1,n}}{\varepsilon_2p_{2,n}^2}(\sigma_3+(-1)^n\sigma_4)^2\Big) &\\
    &\color{gray}= \lim_{L\to\infty}\frac{1}{2L}\sum_{k=1}^\infty \frac{1}{\varepsilon_1p_{1,2k}+\varepsilon_2p_{2,2k}}\Big(
    \frac{2(\sigma_1+\sigma_2)(\sigma_3+\sigma_4)}{p_{1,2k}p_{2,2k}} &\\&\color{gray}\qquad
    - \frac{1}{p_{1,2k}}\frac{\varepsilon_2p_{2,2k}}{\varepsilon_1p_{1,2k}^2}(\sigma_1+\sigma_2)^2
    - \frac{1}{p_{2,2k}}\frac{\varepsilon_1p_{1,2k}}{\varepsilon_2p_{2,2k}^2}(\sigma_3+\sigma_4)^2\Big) &\\
    &\color{gray}+ \lim_{L\to\infty}\frac{1}{2L}\sum_{l=1}^\infty \frac{1}{\varepsilon_1p_{1,2l-1}+\varepsilon_2p_{2,2l-1}}\Big(
    \frac{2(\sigma_1-\sigma_2)(\sigma_3-\sigma_4)}{p_{1,2l-1}p_{2,2l-1}} &\\&\color{gray}\qquad
    - \frac{1}{p_{1,2l-1}}\frac{\varepsilon_2p_{2,2l-1}}{\varepsilon_1p_{1,2l-1}^2}(\sigma_1-\sigma_2)^2
    - \frac{1}{p_{2,2l-1}}\frac{\varepsilon_1p_{1,2l-1}}{\varepsilon_2p_{2,2l-1}^2}(\sigma_3-\sigma_4)^2\Big) &\\
    &\color{gray}= \frac{1}{2L}\frac{L}{2}\int_0^\infty \frac{1}{\varepsilon_1p_1(x)+\varepsilon_2p_2(x)}\Big(
    \frac{2(\sigma_1+\sigma_2)(\sigma_3+\sigma_4)}{p_1(x)p_2(x)} &\\&\color{gray}\qquad
    - \frac{1}{p_1(x)}\frac{\varepsilon_2p_2(x)}{\varepsilon_1p_1(x)^2}(\sigma_1+\sigma_2)^2
    - \frac{1}{p_2(x)}\frac{\varepsilon_1p_1(x)}{\varepsilon_2p_2(x)^2}(\sigma_3+\sigma_4)^2\Big)\d x &\\
    &\color{gray}+ \frac{1}{2L}\frac{L}{2}\int_0^\infty \frac{1}{\varepsilon_1p_1(x)+\varepsilon_2p_2(x)}\Big(
    \frac{2(\sigma_1-\sigma_2)(\sigma_3-\sigma_4)}{p_1(x)p_2(x)} &\\&\color{gray}\qquad
    - \frac{1}{p_1(x)}\frac{\varepsilon_2p_2(x)}{\varepsilon_1p_1(x)^2}(\sigma_1-\sigma_2)^2
    - \frac{1}{p_2(x)}\frac{\varepsilon_1p_1(x)}{\varepsilon_2p_2(x)^2}(\sigma_3-\sigma_4)^2\Big)\d x &\\
    &\color{gray}= \frac{1}{4}\int_0^\infty\frac{1}{\varepsilon_1p_1(x) + \varepsilon_2p_2(x)}\bigg(
    \frac{2((\sigma_1+\sigma_2)(\sigma_3+\sigma_4)+(\sigma_1 - \sigma_2)(\sigma_3 - \sigma_4))}{p_1(x)p_2(x)} &\\
    &\color{gray}- \frac{\varepsilon_2p_2(x)}{\varepsilon_1p_1(x)^3}\left((\sigma_1+\sigma_2)^2 + (\sigma_1-\sigma_2)^2\right)
    - \frac{\varepsilon_1p_1(x)}{\varepsilon_2p_2(x)^3}\left((\sigma_3+\sigma_4)^2 + (\sigma_3-\sigma_4)^2\right)\bigg)\d x &\\
    &\color{gray}= \frac{1}{2}\int_0^\infty\frac{1}{\varepsilon_1p_1(x) + \varepsilon_2p_2(x)}\bigg(
    \frac{2(\sigma_1\sigma_3 + \sigma_2\sigma_4)}{p_1(x)p_2(x)} &\\
    &\color{gray}- \frac{\varepsilon_2p_2(x)}{\varepsilon_1p_1(x)^3}\left(\sigma_1^2+\sigma_2^2\right)
    - \frac{\varepsilon_1p_1(x)}{\varepsilon_2p_2(x)^3}\left(\sigma_3^2 + \sigma_4^2\right)\bigg)\d x &\\
    &= \frac{1}{2}\int_0^\infty\frac{1}{\varepsilon_1p_1(x) + \varepsilon_2p_2(x)}\bigg(
    \frac{2\sigma_1\sigma_3}{p_1(x)p_2(x)}
    - \frac{\varepsilon_2p_2(x)}{\varepsilon_1p_1(x)^3}\sigma_1^2
    - \frac{\varepsilon_1p_1(x)}{\varepsilon_2p_2(x)^3}\sigma_3^2\bigg)\d x &\\
    &+ \frac{1}{2}\int_0^\infty\frac{1}{\varepsilon_1p_1(x) + \varepsilon_2p_2(x)}\bigg(
    \frac{2\sigma_2\sigma_4}{p_1(x)p_2(x)}
    - \frac{\varepsilon_2p_2(x)}{\varepsilon_1p_1(x)^3}\sigma_2^2
    - \frac{\varepsilon_1p_1(x)}{\varepsilon_2p_2(x)^3}\sigma_4^2\bigg)\d x &
}
With this limit and the expression calculated above we can use Eq. \eqref{limsubtract} to obtain
\fleq{
    \tau_1
        &= \frac{\Phi_D}{\kappa_1\varepsilon_1+\kappa_2\varepsilon_2}
        \left(\frac{\varepsilon_2\kappa_2}{\kappa_1}\sigma_1 - \frac{\varepsilon_1\kappa_1}{\kappa_2}\sigma_3\right) &\\
        &+ \int_0^\infty\frac{1}{\varepsilon_1p_1(x) + \varepsilon_2p_2(x)}\bigg(
        \frac{2\sigma_1\sigma_3}{p_1(x)p_2(x)}
        - \frac{\varepsilon_2p_2(x)}{\varepsilon_1p_1(x)^3}\sigma_1^2
        - \frac{\varepsilon_1p_1(x)}{\varepsilon_2p_2(x)^3}\sigma_3^2\bigg)\d x &
}
\fleq{
    \tau_2
        &= \frac{\Phi_D}{\kappa_1\varepsilon_1+\kappa_2\varepsilon_2}
        \left(\frac{\varepsilon_2\kappa_2}{\kappa_1}\sigma_2 - \frac{\varepsilon_1\kappa_1}{\kappa_2}\sigma_4\right) &\\
        &+ \int_0^\infty\frac{1}{\varepsilon_1p_1(x) + \varepsilon_2p_2(x)}\bigg(
        \frac{2\sigma_2\sigma_4}{p_1(x)p_2(x)}
        - \frac{\varepsilon_2p_2(x)}{\varepsilon_1p_1(x)^3}\sigma_2^2
        - \frac{\varepsilon_1p_1(x)}{\varepsilon_2p_2(x)^3}\sigma_4^2\bigg)\d x &
}
\fleq{
    \omega_\tau(L)
        &= \frac{1}{L} \frac{1}{\kappa_1\varepsilon_1+\kappa_2\varepsilon_2} \left(\frac{(\sigma_1+\sigma_2)(\sigma_3 +\sigma_4)}{2\kappa_1\kappa_2}
        - \frac{1}{4\kappa_1^2}\frac{\kappa_2\varepsilon_2}{\kappa_1\varepsilon_1}(\sigma_1+\sigma_2)^2
        - \frac{1}{4\kappa_2^2}\frac{\kappa_1\varepsilon_1}{\kappa_2\varepsilon_2}(\sigma_3+\sigma_4)^2\right) &\\
        &+ \frac{1}{2L}\sum_{n=1}^\infty \frac{1}{\varepsilon_1p_1+\varepsilon_2p_2}\left(
        \frac{2(\sigma_1+(-1)^n\sigma_2)(\sigma_3+(-1)^n\sigma_4)}{p_1p_2}
        - \frac{1}{p_1}\frac{\varepsilon_2p_2}{\varepsilon_1p_1^2}(\sigma_1+(-1)^n\sigma_2)^2
        - \frac{1}{p_2}\frac{\varepsilon_1p_1}{\varepsilon_2p_2^2}(\sigma_3+(-1)^n\sigma_4)^2\right) &\\
        &- \frac{1}{2}\int_0^\infty\frac{1}{\varepsilon_1p_1(x) + \varepsilon_2p_2(x)}\bigg(
        \frac{2(\sigma_1\sigma_3 + \sigma_2\sigma_4)}{p_1(x)p_2(x)}
        - \frac{\varepsilon_2p_2(x)}{\varepsilon_1p_1(x)^3}\left(\sigma_1^2+\sigma_2^2\right)
        - \frac{\varepsilon_1p_1(x)}{\varepsilon_2p_2(x)^3}\left(\sigma_3^2 + \sigma_4^2\right)\bigg)\d x &
        \label{ote}
}
\subsection{Consistency with literature}
We test our expression for consistency with with the result of Ref. \cite{supp,lit2}. For this we have to set
\fleq{
    & L = 2L' &\\
    & \sigma_1'=\sigma_1=\sigma_2 &\\
    & \sigma_2'=\sigma_3=\sigma_4 &
}
since in Ref. \cite{supp} the walls are located at $z=\pm L$ and the system therefore has a length of $2L$.

\paragraph{Surface interaction energies}
\fleq{
    \bar{\omega}_{\gamma, 1}(L')
    &= \frac{\sigma_1'}{2\kappa_1\varepsilon_1}\left(\coth(2\kappa_1L') - 1
        + \frac{1}{\sinh(2\kappa_1L')}\right) &\\
    &= \frac{\sigma_1'}{2\kappa_1\varepsilon_1}\left(
        \frac{\cosh(2\kappa_1L') + 1}{\sinh(2\kappa_1L')} - 1\right) &\\
    &= \frac{\sigma_1'}{2\kappa_1\varepsilon_1}\bigg(
        \frac{\cosh^2(\kappa_1L') + \obr{\sinh^2(\kappa_1L') + 1}{=\cosh^2(\kappa_1L')}}
        {2\sinh(\kappa_1L')\cosh(\kappa_1L')} - 1\bigg) &\\
    &= \frac{\sigma_1'}{2\kappa_1\varepsilon_1}\left(\coth(\kappa_1L') - 1\right) \quad\checkmark &
}

\paragraph{Surface tensions}
\fleq{
    & \bar{\gamma_1}2L_xL_y = (\gamma_1 + \gamma_2)L_xL_y &\\
    & \f \bar{\gamma_1} = \frac{\gamma_1+\gamma_2}{2}
    = \frac{\sigma_1'^2}{4\kappa_1\varepsilon_1} + \frac{\sigma_2'^2}{4\kappa_1\varepsilon_1}
    = \frac{\sigma_1'^2}{2\kappa_1\varepsilon_1} \quad\checkmark &
}
\fleq{
    & \bar{\gamma_2}2L_xL_y = (\gamma_3 + \gamma_4)L_xL_y &\\
    & \f \bar{\gamma_2} = \frac{\gamma_3+\gamma_4}{2}
    = \frac{\sigma_2'^2}{4\kappa_2\varepsilon_2} + \frac{\sigma_2'\Phi_D}{2} + 
        \frac{\sigma_2'^2}{4\kappa_2\varepsilon_2} + \frac{\sigma_2'\Phi_D}{2}
    = \frac{\sigma_1'^2}{2\kappa_1\varepsilon_1} + \sigma_2'\Phi_D \quad\checkmark &
}

\paragraph{Interfacial tension}
\fleq{
    & \bar{\gamma_{1,2}}L_yL' = \gamma_{1,2}L_yL = \gamma_{1,2}2L_yL' &\\
    &\f \bar{\gamma_{1,2}} = 2\gamma_{1,2} = 
        -\frac{\varepsilon_1\varepsilon_2\kappa_1\kappa_2\Phi_D^2}{\kappa_1\varepsilon_1+\kappa_2\varepsilon_2}
        \quad\checkmark &
}

\paragraph{Line interaction energy}
\fleq{
    &\sigma_1 + (-1)^n \sigma_2 = \sigma_1' + (-1)^n \sigma_1' = \piecewise{2\sigma_1' & n \text{ even} \\ 0 & n \text{ odd}} &\\
    &\sigma_3 + (-1)^n \sigma_4 = \sigma_2' + (-1)^n \sigma_2' = \piecewise{2\sigma_2' & n \text{ even} \\ 0 & n \text{ odd}} &
}
so we can define $n' := \frac{n}{2}$ and write
\fleq{
    &p_i = \sqrt{\left(\frac{n\pi}{L}\right)^2 + \kappa_i^2}
    = \sqrt{\left(\frac{n\pi}{2L'}\right)^2 + \kappa_i^2}
    = \sqrt{\left(\frac{n'\pi}{L'}\right)^2 + \kappa_i^2} &
}
and
\fleq{
    \omega_\tau(L)
        &= \frac{1}{2L'} \frac{1}{\kappa_1\varepsilon_1+\kappa_2\varepsilon_2} \left(\frac{2\sigma_1'\sigma_2'}{\kappa_1\kappa_2}
        - \frac{1}{\kappa_1^2}\frac{\kappa_2\varepsilon_2}{\kappa_1\varepsilon_1}\sigma_1'^2
        - \frac{1}{\kappa_2^2}\frac{\kappa_1\varepsilon_1}{\kappa_2\varepsilon_2}\sigma_2'^2\right) &\\
        &+ \frac{1}{L'}\sum_{n'=1}^\infty \frac{1}{\varepsilon_1p_1+\varepsilon_2p_2}\left(
        \frac{2\sigma_1'\sigma_2'}{p_1p_2}
        - \frac{1}{p_1}\frac{\varepsilon_2p_2}{\varepsilon_1p_1^2}\sigma_1'^2
        - \frac{1}{p_2}\frac{\varepsilon_1p_1}{\varepsilon_2p_2^2}\sigma_2'^2\right) &\\
        &- \int_0^\infty\frac{1}{\varepsilon_1p_1(x) + \varepsilon_2p_2(x)}\bigg(
        \frac{2\sigma_1'\sigma_2'}{p_1(x)p_2(x)}
        - \frac{\varepsilon_2p_2(x)}{\varepsilon_1p_1(x)^3}\sigma_1'^2
        - \frac{\varepsilon_1p_1(x)}{\varepsilon_2p_2(x)^3}\sigma_1'^2\bigg)\d x &
}
We compare this with the expression given in \cite{supp}
\fleq{
    \omega_\tau
    &=\frac{\sigma_1'^2}{\kappa_1^2\varepsilon_1}\frac{1}{2\kappa_1L'}
    \frac{\frac{2\sigma_2'\kappa_1}{\sigma_1'\kappa_2}
    - \frac{\kappa_2\varepsilon_2}{\kappa_1\varepsilon_1}
    - \frac{\sigma_2'^2\kappa_1^3\varepsilon_1}{\sigma_1'^2\kappa_2^3\varepsilon_2}}
    {1+\frac{\kappa_2\varepsilon_2}{\kappa_1\varepsilon_1}} &\\
    &+ \frac{\sigma_1'^2}{\kappa_1^2\varepsilon_1}\frac{1}{\kappa_1L'}\sum_{n=1}^\infty
    \frac{\frac{\sigma_2'\varepsilon_1}{\sigma_1'\varepsilon_2}\frac{1}{\frac{n^2\pi^2}{\kappa_1^2L'^2}+\frac{\kappa_2^2}{\kappa_1^2}}
    - \frac{1}{\frac{n^2\pi^2}{\kappa_1^2L'^2}+ 1}}{1+\frac{\sqrt{\frac{n^2\pi_2}{\kappa_1^2L'^2}+1}}
    {\frac{\varepsilon_2}{\varepsilon_1}\sqrt{\frac{n^2\pi^2}{\kappa_1^2L'^2}+\frac{\kappa_2^2}{\kappa_1^2}}}}
    \frac{1}{\sqrt{\frac{n^2\pi^2}{\kappa_1^2L'^2}+1}} 
    + \frac{\frac{\sigma_2'}{\sigma_1'}\frac{1}{\frac{n^2\pi^2}{\kappa_1^2L'^2}+1}
    - \frac{\sigma_2'^2\varepsilon_1}{\sigma_1'^2\varepsilon_2}\frac{1}{\frac{n^2\pi^2}{\kappa_1^2L'^2}+ \frac{\kappa_2^2}{\kappa_1^2}}}
    {1+\frac{\frac{\varepsilon_2}{\varepsilon_1}\sqrt{\frac{n^2\pi^2}{\kappa_1^2L'^2}+\frac{\kappa_2^2}{\kappa_1^2}}}
    {\sqrt{\frac{n^2\pi_2}{\kappa_1^2L'^2}+1}}}
    \frac{1}{\sqrt{\frac{n^2\pi^2}{\kappa_1^2L'^2}+\frac{\kappa_2^2}{\kappa_1^2}}} &\\
    &+ \frac{\sigma_1'^2}{\kappa_1^2\varepsilon_1}\int_0^\infty\d x
    \frac{\frac{\sigma_2'\varepsilon_1}{\sigma_1'\varepsilon_2}\frac{1}{x^2\pi^2+\frac{\kappa_2^2}{\kappa_1^2}}
    - \frac{1}{x^2\pi^2+ 1}}{1+\frac{\sqrt{\frac{n^2\pi_2}{\kappa_1^2L'^2}+1}}
    {\frac{\varepsilon_2}{\varepsilon_1}\sqrt{x^2\pi^2+\frac{\kappa_2^2}{\kappa_1^2}}}}
    \frac{1}{\sqrt{x^2\pi^2+1}} 
    + \frac{\frac{\sigma_2'}{\sigma_1'}\frac{1}{x^2\pi^2+1}
    - \frac{\sigma_2'^2\varepsilon_1}{\sigma_1'^2\varepsilon_2}\frac{1}{x^2\pi^2+ \frac{\kappa_2^2}{\kappa_1^2}}}
    {1+\frac{\frac{\varepsilon_2}{\varepsilon_1}\sqrt{x^2\pi^2+\frac{\kappa_2^2}{\kappa_1^2}}}
    {\sqrt{\frac{n^2\pi_2}{\kappa_1^2L'^2}+1}}}
    \frac{1}{\sqrt{x^2\pi^2+\frac{\kappa_2^2}{\kappa_1^2}}} &\\
    &=\frac{1}{2\kappa_1L'}\frac{\sigma_1'^2}{\kappa_1}\frac{1}{\kappa_1\varepsilon_1+\kappa_2\varepsilon_2}
    \left(\frac{2\sigma_2'\kappa_1}{\sigma_1'\kappa_2}
    - \frac{\kappa_2\varepsilon_2}{\kappa_1\varepsilon_1}
    - \frac{\sigma_2'^2\kappa_1^3\varepsilon_1}{\sigma_1'^2\kappa_2^3\varepsilon_2}\right) &\\
    &+ \frac{\sigma_1'^2}{\varepsilon_1}\frac{1}{L'}\sum_{n=1}^\infty
    \frac{\frac{\sigma_2'\varepsilon_1}{\sigma_1'\varepsilon_2}\frac{1}{p_2^2}
    - \frac{1}{p_1^2}}{1+\frac{\varepsilon_1p_1}{\varepsilon_2p_2}}
    \frac{1}{p_1} 
    + \frac{\frac{\sigma_2'}{\sigma_1'}\frac{1}{p_1^2}
    - \frac{\sigma_2'^2\varepsilon_1}{\sigma_1'^2\varepsilon_2}\frac{1}{p_2^2}}
    {1+\frac{\varepsilon_2p_2}{\varepsilon_1p_1}}
    \frac{1}{p_2} &\\
    &+ \frac{\sigma_1'^2}{\varepsilon_1}\int_0^\infty\d x
    \frac{\frac{\sigma_2'\varepsilon_1}{\sigma_1'\varepsilon_2}\frac{1}{p_2(x)^2}
    - \frac{1}{p_1(x)^2}}{1+\frac{\varepsilon_1p_1(x)}{\varepsilon_2p_2(x)}}
    \frac{1}{p_1(x)} 
    + \frac{\frac{\sigma_2'}{\sigma_1'}\frac{1}{p_1(x)^2}
    - \frac{\sigma_2'^2\varepsilon_1}{\sigma_1'^2\varepsilon_2}\frac{1}{p_2(x)^2}}
    {1+\frac{\varepsilon_2p_2(x)}{\varepsilon_1p_1(x)}}
    \frac{1}{p_2(x)} &\\
    &=\frac{1}{2L'}\frac{1}{\kappa_1\varepsilon_1+\kappa_2\varepsilon_2}
    \left(\frac{2\sigma_1'\sigma_2'}{\kappa_1\kappa_2}
    - \frac{\sigma_1'^2\kappa_2\varepsilon_2}{\kappa_1^3\varepsilon_1}
    - \frac{\sigma_2'^2\kappa_1\varepsilon_1}{\kappa_2^3\varepsilon_2}\right) &\\
    &+ \frac{1}{L'}\sum_{n=1}^\infty
    \frac{1}{\varepsilon_1p_1+\varepsilon_2p_2}\left(
    \frac{\sigma_1'\sigma_2'}{p_1p_2} - \frac{\varepsilon_2\sigma_1'^2p_2}{\varepsilon_1p_1^3}
    + \frac{\sigma_1'\sigma_2'}{p_1p_2} - \frac{\varepsilon_1\sigma_2'^2p_1}{\varepsilon_2p_2^3}
    \right)&\\
    &+ \int_0^\infty\d x
    \frac{1}{\varepsilon_1p_1(x)+\varepsilon_2p_2(x)}\left(
    \frac{\sigma_1'\sigma_2'}{p_1(x)p_2(x)} - \frac{\varepsilon_2\sigma_1'^2p_2(x)}{\varepsilon_1p_1(x)^3}
    + \frac{\sigma_1'\sigma_2'}{p_1(x)p_2(x)} - \frac{\varepsilon_1\sigma_2'^2p_1(x)}{\varepsilon_2p_2(x)^3}
    \right) \quad\checkmark&
}
\paragraph{Line tension}
\fleq{
    & \bar{\tau}2l = (\tau_1 + \tau_2)l &
}
\fleq{
    \f \bar{\tau} &= \frac{\tau_1+\tau_2}{2} &\\
    &= \frac{\Phi_D}{2(\kappa_1\varepsilon_1+\kappa_2\varepsilon_2)}
    \left(\frac{\varepsilon_2\kappa_2}{\kappa_1}(\sigma_1+\sigma_2) - \frac{\varepsilon_1\kappa_1}{\kappa_2}(\sigma_3+\sigma_4)\right) &\\
    &+ \frac{1}{2}\int_0^\infty\frac{1}{\varepsilon_1p_1(x) + \varepsilon_2p_2(x)}\bigg(
    \frac{2(\sigma_1\sigma_3+\sigma_2\sigma_4)}{p_1(x)p_2(x)}
    - \frac{\varepsilon_2p_2(x)}{\varepsilon_1p_1(x)^3}(\sigma_1^2+\sigma_2^2)
    - \frac{\varepsilon_1p_1(x)}{\varepsilon_2p_2(x)^3}(\sigma_3^2+\sigma_4^2)\bigg)\d x &
}
Since the integral term in this expression es equal to the integral term in our expression for the line interaction energy
and the integral term appearing in \cite{supp} for line interaction and the term for line tension in \cite{lit2} are the same
and we have already checked the line interaction we only need to check the first term.
\fleq{
    &\frac{\Phi_D}{2(\kappa_1\varepsilon_1+\kappa_2\varepsilon_2)}
    \left(\frac{\varepsilon_2\kappa_2}{\kappa_1}(\sigma_1+\sigma_2) - \frac{\varepsilon_1\kappa_1}{\kappa_2}(\sigma_3+\sigma_4)\right) &\\
    &=\frac{\Phi_D}{\kappa_1\varepsilon_1+\kappa_2\varepsilon_2}
    \left(\frac{\varepsilon_2\kappa_2}{\kappa_1}\sigma_1' - \frac{\varepsilon_1\kappa_1}{\kappa_2}\sigma_2'\right)
    =\frac{\Phi_D}{1+\frac{\kappa_2\varepsilon_2}{\kappa_1\varepsilon_1}}
    \left(\frac{\kappa_2\varepsilon_2}{\kappa_1\varepsilon_1}\frac{\sigma_1'}{\kappa_1} - \frac{\sigma_2'}{\kappa_2}\right)
    \quad\checkmark &
}

\paragraph{Conclusion}
Therefore we can obtain the result for the case of
identical particles given in Ref. \cite{supp,lit2} from our general expressions.

\section{Superposition Solution}
First, we write the potentials calculated in chapter \ref{ch:pot}
(Eqs. \eqref{phis1} and \eqref{phis2}) by using some abbreviations:
\fleq{
    \Phi_1^s(x,z) &= a_1 e^{-\kappa_1 z} + b_1 e^{\kappa_1(z-L)} + c_1 e^{-\kappa_1 x} &\\
    &+ \frac{1}{\pi}\int_{-\infty}^\infty d_1(q) e^{-p_1 x} \cos(qz) \d q &\\
    &+ \frac{1}{\pi}\int_{-\infty}^\infty f_1(q) e^{-p_1 x} \cos(-(z-L)q) \d q &
}
\fleq{
    \partial_x\Phi_1^s(x,z) &= -\kappa_1c_1 e^{-\kappa_1 x} &\\
    &+ \frac{1}{\pi}\int_{-\infty}^\infty -p_1d_1(q) e^{-p_1 x} \cos(qz) \d q &\\
    &+ \frac{1}{\pi}\int_{-\infty}^\infty -p_1f_1(q) e^{-p_1 x} \cos(-(z-L)q) \d q &
}
\fleq{
    \Phi_2^s(x,z) &= a_2 e^{-\kappa_2 z} + b_2 e^{\kappa_2(z-L)} + 2\Phi_D + c_2 e^{\kappa_2 x} &\\
    &+ \frac{1}{\pi}\int_{-\infty}^\infty d_2(q) e^{p_2 x} \cos(qz) \d q &\\
    &+ \frac{1}{\pi}\int_{-\infty}^\infty f_2(q) e^{p_2 x} \cos(-(z-L)q) \d q &
}
Then
\fleq{
    &\sigma_1\Phi_1(x,0) + \sigma_2\Phi_1(x,L) &\\
    &= \sigma_1\bigg(a_1 + b_1e^{-\kappa_1L} + c_1 e^{-\kappa_1x}
    + \frac{1}{\pi}\int_{-\infty}^\infty d_1(q) e^{-p_1 x} \d q
    + \frac{1}{\pi}\int_{-\infty}^\infty f_1(q) e^{-p_1 x} \cos(qL) \d q \bigg) &\\
    &+ \sigma_2\bigg(a_1e^{-\kappa_1L} + b_1 + c_1 e^{-\kappa_1x}
    + \frac{1}{\pi}\int_{-\infty}^\infty d_1(q) e^{-p_1 x} \cos(qL) \d q
    + \frac{1}{\pi}\int_{-\infty}^\infty f_1(q) e^{-p_1 x} \d q \bigg) &
}
\fleq{
    &\sigma_3\Phi_2(x,0) + \sigma_4\Phi_2(x,L) &\\
    &= \sigma_3\bigg(a_2 + b_2e^{-\kappa_2L} + 2\Phi_D + c_2 e^{\kappa_2x}
    + \frac{1}{\pi}\int_{-\infty}^\infty d_2(q) e^{p_2 x} \d q
    + \frac{1}{\pi}\int_{-\infty}^\infty f_2(q) e^{p_2 x} \cos(qL) \d q \bigg) &\\
    &+ \sigma_4\bigg(a_2e^{-\kappa_2L} + b_2 + 2\Phi_D + c_2 e^{\kappa_2x}
    + \frac{1}{\pi}\int_{-\infty}^\infty d_2(q) e^{p_2 x} \cos(qL) \d q
    + \frac{1}{\pi}\int_{-\infty}^\infty f_2(q) e^{p_2 x} \d q \bigg) &
}
\fleq{
    & (\partial_x\Phi_1)(0,z) = -\kappa_1c_1
    + \frac{1}{\pi}\int_{-\infty}^\infty -p_1d_1(q) \cos(qz) \d q
    + \frac{1}{\pi}\int_{-\infty}^\infty -p_1f_1(q) \cos(-(z-L)q) \d q &
}
We then proceed to insert these expressions into Eq. \eqref{Hform} in order to calculate the energy.
\fleq{
    \mathcal{H}
    &= \frac{L_y}{2}\sigma_1\bigg(a_1L_z + b_1e^{-\kappa_1L}L_x - \frac{c_1}{\kappa_1}(e^{-\kappa_1L_x} - 1) &\\
        &\quad+ \frac{1}{\pi} \int_{-\infty}^\infty -\frac{d_1(q)}{p_1}(e^{-p_1L_x} - 1) \d q &\\
        &\quad+ \frac{1}{\pi} \int_{-\infty}^\infty -\frac{f_1(q)}{p_1}(e^{-p_1L_x} - 1) \cos(qL) \d q \bigg) &\\
    &+ \frac{L_y}{2}\sigma_2\bigg(a_1e^{-\kappa_1L}L_z + b_1L_x - \frac{c_1}{\kappa_1}(e^{-\kappa_1L_x} - 1) &\\
        &\quad+ \frac{1}{\pi} \int_{-\infty}^\infty -\frac{d_1(q)}{p_1}(e^{-p_1L_x} - 1) \cos(qL) \d q &\\
        &\quad+ \frac{1}{\pi} \int_{-\infty}^\infty -\frac{f_1(q)}{p_1}(e^{-p_1L_x} - 1) \d q \bigg) &\\
    &+ \frac{L_y}{2}\sigma_3\bigg(a_2L_z + b_2e^{-\kappa_2L}L_x + 2\Phi_DL_x + \frac{c_2}{\kappa_2}(1 - e^{-\kappa_2L_x}) &\\
        &\quad+ \frac{1}{\pi} \int_{-\infty}^\infty \frac{d_2(q)}{p_2}(1 - e^{-p_2L_x}) \d q &\\
        &\quad+ \frac{1}{\pi} \int_{-\infty}^\infty \frac{f_2(q)}{p_2}(1 - e^{-p_2L_x}) \cos(qL) \d q \bigg) &\\
    &+ \frac{L_y}{2}\sigma_4\bigg(a_2e^{-\kappa_2L}L_z + b_2L_x + 2\Phi_DL_x + \frac{c_2}{\kappa_2}(1 - e^{-\kappa_2L_x}) &\\
        &\quad+ \frac{1}{\pi} \int_{-\infty}^\infty \frac{d_2(q)}{p_2}(1 - e^{-p_2L_x}) \cos(qL) \d q &\\
        &\quad+ \frac{1}{\pi} \int_{-\infty}^\infty \frac{f_2(q)}{p_2}(1 - e^{-p_2L_x}) \d q \bigg) &\\
    &+ \frac{L_xL_y\Phi_D}{2} (\sigma_3 + \sigma_4) &\\
    &+ \frac{\varepsilon_1L_y\Phi_D}{2}\bigg(-\kappa_1c_1L &\\
        &\quad+ \frac{1}{\pi} \int_{-\infty}^\infty -p_1d_1(q)\ubr{\left[\frac{\sin(qz)}{q}\right]_0^L}{=\frac{\sin(qL)}{q}}\d q &\\
        &\quad+ \frac{1}{\pi} \int_{-\infty}^\infty -p_1f_1(q)\ubr{\left[\frac{\sin(qz+qL)}{-q}\right]_0^L}{=\frac{\sin(qL)}{q}}\d q
        \bigg)
}
We are interested in the case of large $L_x$ so we approx. $e^{-aL_x}\approx 0$
for all factors $a\in\R$ not depending on $L_x$.
\fleq{
    \mathcal{H}
    &= \frac{L_y}{2}\sigma_1\bigg(a_1L_z + b_1e^{-\kappa_1L}L_x + \frac{c_1}{\kappa_1}
        + \frac{1}{\pi} \int_{-\infty}^\infty\frac{d_1(q)}{p_1}\d q
        + \frac{1}{\pi} \int_{-\infty}^\infty \frac{f_1(q)}{p_1} \cos(qL) \d q \bigg) &\\
    &+ \frac{L_y}{2}\sigma_2\bigg(a_1e^{-\kappa_1L}L_z + b_1L_x + \frac{c_1}{\kappa_1}
        + \frac{1}{\pi} \int_{-\infty}^\infty \frac{d_1(q)}{p_1} \cos(qL) \d q
        + \frac{1}{\pi} \int_{-\infty}^\infty \frac{f_1(q)}{p_1} \d q \bigg) &\\
    &+ \frac{L_y}{2}\sigma_3\bigg(a_2L_z + b_2e^{-\kappa_2L}L_x + 2\Phi_DL_x + \frac{c_2}{\kappa_2}
        + \frac{1}{\pi} \int_{-\infty}^\infty \frac{d_2(q)}{p_2} \d q
        + \frac{1}{\pi} \int_{-\infty}^\infty \frac{f_2(q)}{p_2} \cos(qL) \d q \bigg) &\\
    &+ \frac{L_y}{2}\sigma_4\bigg(a_2e^{-\kappa_2L}L_z + b_2L_x + 2\Phi_DL_x + \frac{c_2}{\kappa_2}
        + \frac{1}{\pi} \int_{-\infty}^\infty \frac{d_2(q)}{p_2} \cos(qL) \d q
        + \frac{1}{\pi} \int_{-\infty}^\infty \frac{f_2(q)}{p_2} \d q \bigg) &\\
    &+ \frac{L_xL_y\Phi_D}{2} (\sigma_3 + \sigma_4) &\\
    &+ \frac{\varepsilon_1L_y\Phi_D}{2}\bigg(-\kappa_1c_1L
        + \frac{1}{\pi} \int_{-\infty}^\infty -p_1d_1(q)\frac{\sin(qL)}{q}\d q
        + \frac{1}{\pi} \int_{-\infty}^\infty -p_1f_1(q)\frac{\sin(qL)}{q}\d q
        \bigg) &\\
    &= \left[(a_1 + b_1e^{-\kappa_1L})\frac{\sigma_1}{4}\right]2L_xL_y &\\
    &+ \left[(a_1e^{-\kappa_1L} + b_1)\frac{\sigma_2}{4}\right]2L_xL_y &\\
    &+ \left[(a_2 + b_2e^{-\kappa_2L})\frac{\sigma_3}{4}\right]2L_xL_y &\\
    &+ \left[(a_2e^{-\kappa_2L} + b_2)\frac{\sigma_4}{4}\right]2L_xL_y &\\
    &+ \left[\frac{3}{4}\Phi_D(\sigma_3+\sigma_4)\right]2L_xL_y &\\
    &+ \left[\frac{\sigma_1c_1}{4\kappa_1}
        + \frac{1}{4\pi}\int_{-\infty}^\infty \sigma_1\frac{d_1}{p_1} \d q
        + \frac{1}{4\pi}\int_{-\infty}^\infty \sigma_1\frac{f_1}{p_1}\cos(qL) \d q\right] 2L_y &\\
    &+ \left[\frac{\sigma_2c_1}{4\kappa_1}
        + \frac{1}{4\pi}\int_{-\infty}^\infty \sigma_2\frac{d_1}{p_1}\cos(qL) \d q
        + \frac{1}{4\pi}\int_{-\infty}^\infty \sigma_2\frac{f_1}{p_1} \d q\right] 2L_y &\\
    &+ \left[\frac{\sigma_3c_2}{4\kappa_2}
        + \frac{1}{4\pi}\int_{-\infty}^\infty \sigma_3\frac{d_2}{p_2} \d q
        + \frac{1}{4\pi}\int_{-\infty}^\infty \sigma_3\frac{f_2}{p_2}\cos(qL) \d q\right] 2L_y &\\
    &+ \left[\frac{\sigma_4c_2}{4\kappa_2}
        + \frac{1}{4\pi}\int_{-\infty}^\infty \sigma_4\frac{d_2}{p_2}\cos(qL) \d q
        + \frac{1}{4\pi}\int_{-\infty}^\infty \sigma_4\frac{f_2}{p_2} \d q\right] 2L_y &\\
    &+ \left[\left(\frac{1}{\pi}\int_{-\infty}^\infty -\frac{p_1(d_1+f_1)}{q}\sin(qL)\d q\right)
        \frac{\varepsilon_1\Phi_D}{4}\right] 2L_y &\\
    &+ \left[-\frac{\kappa_1\varepsilon_1\Phi_Dc_1}{2}\right]L_yL &
}
\fleq{
    & a_1 = \frac{\sigma_1}{\varepsilon_1\kappa_1} &\\
    & b_1 = \frac{\sigma_2}{\varepsilon_1\kappa_1} &\\
    & c_1 = \frac{2\varepsilon_2\kappa_2\Phi_D}{\varepsilon_1\kappa_1+\varepsilon_2\kappa_2} &\\
    & d_1 = \frac{\varepsilon_2p_2}{\varepsilon_1p_1+\varepsilon_2p_2}
        \left(-\frac{\sigma_1}{\varepsilon_1}\frac{1}{p_1^2}+ \frac{\sigma_3}{\varepsilon_2}\frac{1}{p_2^2}\right) &\\
    & f_1 = \frac{\varepsilon_2p_2}{\varepsilon_1p_1+\varepsilon_2p_2}
        \left(-\frac{\sigma_2}{\varepsilon_1}\frac{1}{p_1^2}+ \frac{\sigma_4}{\varepsilon_2}\frac{1}{p_2^2}\right) &
}

\fleq{
    & a_2 = \frac{\sigma_3}{\varepsilon_2\kappa_2} &\\
    & b_2 = \frac{\sigma_4}{\varepsilon_2\kappa_2} &\\
    & c_2 = -\frac{2\varepsilon_1\kappa_1\Phi_D}{\varepsilon_1\kappa_1+\varepsilon_2\kappa_2} &\\
    & d_2 = -\frac{\varepsilon_1p_1}{\varepsilon_1p_1+\varepsilon_2p_2}
        \left(-\frac{\sigma_1}{\varepsilon_1}\frac{1}{p_1^2}+ \frac{\sigma_3}{\varepsilon_2}\frac{1}{p_2^2}\right) &\\
    & f_2 = -\frac{\varepsilon_1p_1}{\varepsilon_1p_1+\varepsilon_2p_2}
        \left(-\frac{\sigma_2}{\varepsilon_1}\frac{1}{p_1^2}+ \frac{\sigma_4}{\varepsilon_2}\frac{1}{p_2^2}\right) &
}
As explained earlier, we now identify the different terms by their proportionality,
$\sigma$ and $L$-dependence:
\fleq{
    & \tilde{\gamma_1} = \frac{\sigma_1^2}{2\varepsilon_1\kappa_1} &\\
    & \tilde{\gamma_2} = \frac{\sigma_2^2}{2\varepsilon_1\kappa_1} &\\
    & \tilde{\gamma_3} = \frac{\sigma_3^2}{2\varepsilon_2\kappa_2} + \frac{3}{2}\Phi_D\sigma_3 &\\
    & \tilde{\gamma_4} = \frac{\sigma_4^2}{2\varepsilon_2\kappa_2} + \frac{3}{2}\Phi_D\sigma_4 &
}
\fleq{
    \tilde{\omega}_{\gamma,1} &= \frac{\sigma_1\sigma_2}{4\varepsilon_1\kappa_1}e^{-\kappa_1L}
        + \frac{\sigma_1\sigma_2}{4\varepsilon_1\kappa_1}e^{-\kappa_1L} &\\
    &= \frac{\sigma_1\sigma_2}{2\varepsilon_1\kappa_1}e^{-\kappa_1L} &\\
    \tilde{\omega}_{\gamma,2} &= \frac{\sigma_3\sigma_4}{4\varepsilon_2\kappa_2}e^{-\kappa_2L}
        + \frac{\sigma_3\sigma_4}{4\varepsilon_2\kappa_2}e^{-\kappa_2L} &\\
    &= \frac{\sigma_3\sigma_4}{2\varepsilon_2\kappa_2}e^{-\kappa_2L} &
}
\fleq{
    & \gamma_{1,2} = -\frac{\kappa_1\varepsilon_1\kappa_2\varepsilon_2\Phi_D^2}
        {\kappa_1\varepsilon_1+\kappa_2\varepsilon_2} &
}
\fleq{
    \tilde{\tau_1}
    &= \frac{\sigma_1}{2\kappa_1}
        + \frac{\sigma_3c_2}{2\kappa_2}
        + \frac{1}{2\pi}\int_{-\infty}^\infty \sigma_1\frac{d_1}{p_1} \d q
        + \frac{1}{2\pi}\int_{-\infty}^\infty \sigma_3\frac{d_2}{p_2} \d q &\\
    &\color{gray}= \frac{\sigma_1c_1}{2\kappa_1}
        \frac{2\varepsilon_2\kappa_2\Phi_D}{\varepsilon_1\kappa_1+\varepsilon_2\kappa_2}
        -\frac{\sigma_3c_2}{2\kappa_2}
        \frac{2\varepsilon_1\kappa_1\Phi_D}{\varepsilon_1\kappa_1+\varepsilon_2\kappa_2} &\\
        &\color{gray}+ \frac{1}{2\pi}\int_{-\infty}^\infty \sigma_1\frac{1}{p_1}
        \frac{\varepsilon_2p_2}{\varepsilon_1p_1+\varepsilon_2p_2}
        \left(-\frac{\sigma_1}{\varepsilon_1}\frac{1}{p_1^2} + \frac{\sigma_3}{\varepsilon_2}\frac{1}{p_2^2}\right) \d q &\\
        &\color{gray}+ \frac{1}{2\pi}\int_{-\infty}^\infty -\sigma_3\frac{1}{p_2} 
        \frac{\varepsilon_1p_1}{\varepsilon_1p_1+\varepsilon_2p_2}
        \left(-\frac{\sigma_1}{\varepsilon_1}\frac{1}{p_1^2} + \frac{\sigma_3}{\varepsilon_2}\frac{1}{p_2^2}\right) \d q &\\
    &= \frac{\Phi_D}{\varepsilon_1\kappa_1+\varepsilon_2\kappa_1}
        \left(\sigma_1\varepsilon_2\frac{\kappa_2}{\kappa_1} - \sigma_3\varepsilon_1\frac{\kappa_1}{\kappa_2}\right) &\\
        &+ \frac{1}{2\pi}\int_{-\infty}^\infty \frac{1}{\varepsilon_1p_1+\varepsilon_2p_2}
        \left(\frac{2\sigma_1\sigma_3}{p_1p_2} - \sigma_1^2\frac{\varepsilon_2}{\varepsilon_1}\frac{p_2}{p_1^3}
        - \sigma_3^2\frac{\varepsilon_1}{\varepsilon_2}\frac{p_1}{p_2^3}\right)\d q &\\
    \tilde{\tau_2}
    &= \frac{\sigma_2c_1}{2\kappa_1}
        + \frac{\sigma_4c_2}{2\kappa_2}
        + \frac{1}{2\pi}\int_{-\infty}^\infty \sigma_2\frac{f_1}{p_1} \d q
        + \frac{1}{2\pi}\int_{-\infty}^\infty \sigma_4\frac{f_2}{p_2} \d q &\\
    &\color{gray}= \frac{\sigma_2}{2\kappa_1}
        \frac{2\varepsilon_2\kappa_2\Phi_D}{\varepsilon_1\kappa_1+\varepsilon_2\kappa_2}
        - \frac{\sigma_4c_2}{2\kappa_2}
        \frac{2\varepsilon_1\kappa_1\Phi_D}{\varepsilon_1\kappa_1+\varepsilon_2\kappa_2} &\\
        &\color{gray}+ \frac{1}{2\pi}\int_{-\infty}^\infty \sigma_2\frac{1}{p_1} 
        \frac{\varepsilon_2p_2}{\varepsilon_1p_1+\varepsilon_2p_2}
        \left(-\frac{\sigma_2}{\varepsilon_1}\frac{1}{p_1^2}+ \frac{\sigma_4}{\varepsilon_2}\frac{1}{p_2^2}\right)\d q &\\
        &\color{gray}- \frac{1}{2\pi}\int_{-\infty}^\infty \sigma_4\frac{1}{p_2} 
        \frac{\varepsilon_1p_1}{\varepsilon_1p_1+\varepsilon_2p_2}
        \left(-\frac{\sigma_2}{\varepsilon_1}\frac{1}{p_1^2}+ \frac{\sigma_4}{\varepsilon_2}\frac{1}{p_2^2}\right) \d q &\\
    &= \frac{\Phi_D}{\varepsilon_1\kappa_1+\varepsilon_2\kappa_1}
        \left(\sigma_2\varepsilon_2\frac{\kappa_2}{\kappa_1} - \sigma_4\varepsilon_1\frac{\kappa_1}{\kappa_2}\right) &\\
        &+ \frac{1}{2\pi}\int_{-\infty}^\infty \frac{1}{\varepsilon_1p_1+\varepsilon_2p_2}
        \left(\frac{2\sigma_2\sigma_4}{p_1p_2} - \sigma_2^2\frac{\varepsilon_2}{\varepsilon_1}\frac{p_2}{p_1^3}
        - \sigma_4^2\frac{\varepsilon_1}{\varepsilon_2}\frac{p_1}{p_2^3}\right)\d q &\\
}
\fleq{
    \tilde\omega_\tau
    &= \frac{1}{4\pi}\int_{-\infty}^\infty \sigma_1\frac{f_1}{p_1}\cos(qL) \d q
        + \frac{1}{4\pi}\int_{-\infty}^\infty \sigma_2\frac{d_1}{p_1}\cos(qL) \d q
        + \frac{1}{4\pi}\int_{-\infty}^\infty \sigma_3\frac{f_2}{p_2}\cos(qL) \d q &\\
        &+ \frac{1}{4\pi}\int_{-\infty}^\infty \sigma_4\frac{d_2}{p_2}\cos(qL) \d q
        + \left(\frac{1}{\pi}\int_{-\infty}^\infty -\frac{p_1(d_1+f_1)}{q}\sin(qL)\d q\right)
        \frac{\varepsilon_1\Phi_D}{4} &\\
    &\color{gray}= \frac{1}{4\pi}\int_{-\infty}^\infty \sigma_1\frac{1}{p_1}
        \frac{\varepsilon_2p_2}{\varepsilon_1p_1+\varepsilon_2p_2}
        \left(-\frac{\sigma_2}{\varepsilon_1}\frac{1}{p_1^2}+ \frac{\sigma_4}{\varepsilon_2}\frac{1}{p_2^2}\right)\cos(qL) \d q &\\
        &\color{gray}+ \frac{1}{4\pi}\int_{-\infty}^\infty \sigma_2\frac{1}{p_1}
        \frac{\varepsilon_2p_2}{\varepsilon_1p_1+\varepsilon_2p_2}
        \left(-\frac{\sigma_1}{\varepsilon_1}\frac{1}{p_1^2}+ \frac{\sigma_3}{\varepsilon_2}\frac{1}{p_2^2}\right)\cos(qL) \d q &\\
        &\color{gray}+ \frac{1}{4\pi}\int_{-\infty}^\infty -\sigma_3\frac{1}{p_2}
        \frac{\varepsilon_1p_1}{\varepsilon_1p_1+\varepsilon_2p_2}
        \left(-\frac{\sigma_2}{\varepsilon_1}\frac{1}{p_1^2}+ \frac{\sigma_4}{\varepsilon_2}\frac{1}{p_2^2}\right)\cos(qL) \d q &\\
        &\color{gray}+ \frac{1}{4\pi}\int_{-\infty}^\infty -\sigma_4\frac{1}{p_2}
        \frac{\varepsilon_1p_1}{\varepsilon_1p_1+\varepsilon_2p_2}
        \left(-\frac{\sigma_1}{\varepsilon_1}\frac{1}{p_1^2}+ \frac{\sigma_3}{\varepsilon_2}\frac{1}{p_2^2}\right)\cos(qL) \d q &\\
        &\color{gray}+ \bigg(\frac{1}{\pi}\int_{-\infty}^\infty -p_1\bigg(
        \frac{\varepsilon_2p_2}{\varepsilon_1p_1+\varepsilon_2p_2}
        \left(-\frac{\sigma_1}{\varepsilon_1}\frac{1}{p_1^2}+ \frac{\sigma_3}{\varepsilon_2}\frac{1}{p_2^2}\right) &\\
        &\color{gray}+ \frac{\varepsilon_2p_2}{\varepsilon_1p_1+\varepsilon_2p_2}
        \left(-\frac{\sigma_2}{\varepsilon_1}\frac{1}{p_1^2}+ \frac{\sigma_4}{\varepsilon_2}\frac{1}{p_2^2}\right)\bigg)
        \frac{\sin(qL)}{q}\d q\bigg) \frac{\varepsilon_1\Phi_D}{4} &\\
    &= \frac{1}{2\pi}\int_{-\infty}^\infty\frac{1}{\varepsilon_1p_1+\varepsilon_2p_2}\left(
        \frac{\sigma_2\sigma_3 + \sigma_1\sigma_4}{p_1p_2} - \sigma_1\sigma_2\frac{\varepsilon_2}{\varepsilon_1}\frac{p_2}{p_1^3}
        - \sigma_3\sigma_4\frac{\varepsilon_1}{\varepsilon_2}\frac{p_1}{p_2^3}\right)\cos(qL) \d q &\\
        &- \frac{\Phi_D}{4\pi}\int_{-\infty}^\infty \frac{\varepsilon_1\varepsilon_2p_1p_2}{\varepsilon_1p_1+\varepsilon_2p_3}
        \left(-\frac{\sigma_1+\sigma_2}{\varepsilon_1}\frac{1}{p_1^2}+\frac{\sigma_3+\sigma_4}{\varepsilon_2}\frac{1}{p_2^2}
        \right)\frac{\sin(qL)}{q} \d q
}
We then proceed to calculate the limits of the $L$-dependent terms so we can obtain our final quantities
\fleq{
    & \lim_{L\to\infty}\tilde\omega_{\gamma,i} = 0 &
}
with this limit and the expression calculated above we can use Eq. \eqref{limsubtract} to obtain
\fleq{
    & \gamma_1 = \tilde{\gamma_1} = \frac{\sigma_1^2}{2\varepsilon_1\kappa_1} \quad
    \gamma_2 = \tilde{\gamma_2} = \frac{\sigma_2^2}{2\varepsilon_1\kappa_1} \quad
    \gamma_3 = \tilde{\gamma_3} = \frac{\sigma_3^2}{2\varepsilon_2\kappa_2} + \frac{3}{2}\Phi_D\sigma_3 \quad
    \gamma_4 = \tilde{\gamma_4} = \frac{\sigma_4^2}{2\varepsilon_2\kappa_2} + \frac{3}{2}\Phi_D\sigma_4 &
}
\fleq{
    & \omega_{\gamma,1} = \frac{\sigma_1\sigma_2}{2\varepsilon_1\kappa_1}e^{-\kappa_1L} \quad
    \omega_{\gamma,2} = \frac{\sigma_3\sigma_4}{2\varepsilon_2\kappa_2}e^{-\kappa_2L} &
}
\fleq{
    \lim_{L\to\infty}\tilde\omega_\tau
    &= -\frac{\varepsilon_1\varepsilon_2\kappa_1\kappa_2\Phi_D}{4(\varepsilon_1\kappa_1+\varepsilon_2\kappa_3)}
        \left(-\frac{\sigma_1+\sigma_2}{\varepsilon_1}\frac{1}{\kappa_1^2}+\frac{\sigma_3+\sigma_4}{\varepsilon_2}\frac{1}{\kappa_2^2}
        \right)&
}
with this limit and the expression calculated above we can use Eq. \eqref{limsubtract} to obtain
\fleq{
    \tau_1
    &= \frac{\Phi_D}{\varepsilon_1\kappa_1+\varepsilon_2\kappa_1}
        \left(\sigma_1\varepsilon_2\frac{\kappa_2}{\kappa_1} - \sigma_3\varepsilon_1\frac{\kappa_1}{\kappa_2}\right) &\\
        &+ \frac{1}{2\pi}\int_{-\infty}^\infty \frac{1}{\varepsilon_1p_1+\varepsilon_2p_2}
        \left(\frac{2\sigma_1\sigma_3}{p_1p_2} - \sigma_1^2\frac{\varepsilon_2}{\varepsilon_1}\frac{p_2}{p_1^3}
        - \sigma_3^2\frac{\varepsilon_1}{\varepsilon_2}\frac{p_1}{p_2^3}\right)\d q &\\
        &- \frac{\varepsilon_1\varepsilon_2\kappa_1\kappa_2\Phi_D}{2(\varepsilon_1\kappa_1+\varepsilon_2\kappa_3)}
        \left(-\frac{\sigma_1}{\varepsilon_1}\frac{1}{\kappa_1^2}+\frac{\sigma_3}{\varepsilon_2}\frac{1}{\kappa_2^2}
        \right)&\\
    &= \frac{3}{2}\frac{\Phi_D}{\varepsilon_1\kappa_1+\varepsilon_2\kappa_1}
        \left(\sigma_1\varepsilon_2\frac{\kappa_2}{\kappa_1} - \sigma_3\varepsilon_1\frac{\kappa_1}{\kappa_2}\right) &\\
        &+ \frac{1}{2\pi}\int_{-\infty}^\infty \frac{1}{\varepsilon_1p_1+\varepsilon_2p_2}
        \left(\frac{2\sigma_1\sigma_3}{p_1p_2} - \sigma_1^2\frac{\varepsilon_2}{\varepsilon_1}\frac{p_2}{p_1^3}
        - \sigma_3^2\frac{\varepsilon_1}{\varepsilon_2}\frac{p_1}{p_2^3}\right)\d q &\\
    \tau_2
    &= \frac{\Phi_D}{\varepsilon_1\kappa_1+\varepsilon_2\kappa_1}
        \left(\sigma_2\varepsilon_2\frac{\kappa_2}{\kappa_1} - \sigma_4\varepsilon_1\frac{\kappa_1}{\kappa_2}\right) &\\
        &+ \frac{1}{2\pi}\int_{-\infty}^\infty \frac{1}{\varepsilon_1p_1+\varepsilon_2p_2}
        \left(\frac{2\sigma_2\sigma_4}{p_1p_2} - \sigma_2^2\frac{\varepsilon_2}{\varepsilon_1}\frac{p_2}{p_1^3}
        - \sigma_4^2\frac{\varepsilon_1}{\varepsilon_2}\frac{p_1}{p_2^3}\right)\d q &\\
        &- \frac{\varepsilon_1\varepsilon_2\kappa_1\kappa_2\Phi_D}{2(\varepsilon_1\kappa_1+\varepsilon_2\kappa_3)}
        \left(-\frac{\sigma_2}{\varepsilon_1}\frac{1}{\kappa_1^2}+\frac{\sigma_4}{\varepsilon_2}\frac{1}{\kappa_2^2}
        \right)&\\
    &= \frac{3}{2}\frac{\Phi_D}{\varepsilon_1\kappa_1+\varepsilon_2\kappa_1}
        \left(\sigma_2\varepsilon_2\frac{\kappa_2}{\kappa_1} - \sigma_4\varepsilon_1\frac{\kappa_1}{\kappa_2}\right) &\\
        &+ \frac{1}{2\pi}\int_{-\infty}^\infty \frac{1}{\varepsilon_1p_1+\varepsilon_2p_2}
        \left(\frac{2\sigma_2\sigma_4}{p_1p_2} - \sigma_2^2\frac{\varepsilon_2}{\varepsilon_1}\frac{p_2}{p_1^3}
        - \sigma_4^2\frac{\varepsilon_1}{\varepsilon_2}\frac{p_1}{p_2^3}\right)\d q &
}
\fleq{
    \omega_\tau
    &= \frac{\varepsilon_1\varepsilon_2\kappa_1\kappa_2\Phi_D}{4(\varepsilon_1\kappa_1+\varepsilon_2\kappa_3)}
        \left(-\frac{\sigma_1+\sigma_2}{\varepsilon_1}\frac{1}{\kappa_1^2}+\frac{\sigma_3+\sigma_4}{\varepsilon_2}\frac{1}{\kappa_2^2}
        \right)&\\
        &+ \frac{1}{2\pi}\int_{-\infty}^\infty\frac{1}{\varepsilon_1p_1+\varepsilon_2p_2}\left(
        \frac{\sigma_2\sigma_3 + \sigma_1\sigma_4}{p_1p_2} - \sigma_1\sigma_2\frac{\varepsilon_2}{\varepsilon_1}\frac{p_2}{p_1^3}
        - \sigma_3\sigma_4\frac{\varepsilon_1}{\varepsilon_2}\frac{p_1}{p_2^3}\right)\cos(qL) \d q &\\
        &- \frac{\Phi_D}{4\pi}\int_{-\infty}^\infty \frac{\varepsilon_1\varepsilon_2p_1p_2}{\varepsilon_1p_1+\varepsilon_2p_3}
        \left(-\frac{\sigma_1+\sigma_2}{\varepsilon_1}\frac{1}{p_1^2}+\frac{\sigma_3+\sigma_4}{\varepsilon_2}\frac{1}{p_2^2}
        \right)\frac{\sin(qL)}{q} \d q &
        \label{ots}
}
\subsection{Consistency with literature}
We test our expression for consistency with with the result of Ref. \cite{supp}. For this we have to set
\fleq{
    & L = 2L' &\\
    & \sigma_1'=\sigma_1=\sigma_2 &\\
    & \sigma_2'=\sigma_3=\sigma_4 &
}
since in Ref. \cite{supp} the walls are located at $z=\pm L$ and the system therefore has a length of $2L$.
\paragraph{Surface interaction energy}
\fleq{
    & \omega_{\gamma,1} = \frac{\sigma_1\sigma_2}{2\varepsilon_1\kappa_1}e^{-\kappa_1L}
    = \frac{\sigma_1'^2}{2\varepsilon_1\kappa_1}e^{-2\kappa_1L'} &
}
we compare with the expression from \cite{lit}
\fleq{
    \omega_{\gamma,1} &= \frac{\sigma_1'^2}{2\kappa_1\varepsilon_1}
        \left(2e^{-\kappa_1L'}\cosh(\kappa_1L') - 1\right) &\\
        &= \frac{\sigma_1'^2}{2\kappa_1\varepsilon_1}\left(1 + e^{-2\kappa_1L} - 1\right) &\\
        &= \frac{\sigma_1'^2}{2\kappa_1\varepsilon_1}e^{-2\kappa_1L} \quad\checkmark &
}
\paragraph{Line interaction energy}
\fleq{
    \omega_\tau
    &= \frac{\varepsilon_1\varepsilon_2\kappa_1\kappa_2\Phi_D}{2(\varepsilon_1\kappa_1+\varepsilon_2\kappa_3)}
        \left(-\frac{\sigma_1'}{\varepsilon_1}\frac{1}{\kappa_1^2}+\frac{\sigma_2'}{\varepsilon_2}\frac{1}{\kappa_2^2}
        \right)&\\
        &+ \frac{1}{2\pi}\int_{-\infty}^\infty\frac{1}{\varepsilon_1p_1+\varepsilon_2p_2}\left(
        \frac{2\sigma_1'\sigma_2'}{p_1p_2} - \sigma_1'^2\frac{\varepsilon_2}{\varepsilon_1}\frac{p_2}{p_1^3}
        - \sigma_2'^2\frac{\varepsilon_1}{\varepsilon_2}\frac{p_1}{p_2^3}\right)\cos(2qL') \d q &\\
        &- \frac{\Phi_D}{2\pi}\int_{-\infty}^\infty \frac{\varepsilon_1\varepsilon_2p_1p_2}{\varepsilon_1p_1+\varepsilon_2p_3}
        \left(-\frac{\sigma_1'}{\varepsilon_1}\frac{1}{p_1^2}+\frac{\sigma_2'}{\varepsilon_2}\frac{1}{p_2^2}
        \right)\frac{\sin(2qL')}{q} \d q &
}
we compare with the expression from \cite{supp}
\fleq{
    \omega_\tau
    &= \frac{\sigma_1'^2}{\kappa_1^2\varepsilon_1}\bigg[
        \frac{\kappa_2}{\kappa_1\left(1+\frac{\varepsilon_2\kappa_2}{\varepsilon_1\kappa_1}\right)}\left(
        \frac{\sigma_2'\kappa_1^2}{\sigma_1'\kappa_2^2} - \frac{\varepsilon_2}{\varepsilon_1}\right)
        \frac{\Phi_D\kappa_1\varepsilon_1}{2\sigma_1'} &\\
    &- \frac{1}{\pi}\frac{\Phi_D\kappa_1\varepsilon_1}{\sigma_1'}\int_0^\infty\d q
        \frac{\sqrt{q^2+1}\sqrt{q^2+\frac{\kappa_2^2}{\kappa_1^2}}}
        {\sqrt{q^2+1}+\frac{\varepsilon_2}{\varepsilon_1}\sqrt{q^2+\frac{\kappa_2^2}{\kappa_1^2}}}
        \left(\frac{\frac{\sigma_2'}{\sigma_1'}}{q^2+\frac{\kappa_2^2}{\kappa_1^2}}
        - \frac{\frac{\varepsilon_2}{\varepsilon_1}}{q^2+1}\right)
        \frac{\sin(2q\kappa_1L')}{q} &\\
    &+ \frac{1}{\pi}\int_0^\infty\d q
        \frac{\sqrt{q^2+\frac{\kappa_2^2}{\kappa_1^2}}}
        {\sqrt{q^2+1}+\frac{\varepsilon_2}{\varepsilon_1}\sqrt{q^2+\frac{\kappa_2^2}{\kappa_1^2}}}
        \left(\frac{\frac{\sigma_2'}{\sigma_1'}}{q^2+\frac{\kappa_2^2}{\kappa_1^2}}
        - \frac{\frac{\varepsilon_2}{\varepsilon_1}}{q^2+1}\right)
        \frac{\cos(2q\kappa_1L')}{\sqrt{q^2 + 1}} &\\
    &- \frac{1}{\pi}\int_0^\infty\d q
        \frac{\sqrt{q^2+1}}
        {\sqrt{q^2+1}+\frac{\varepsilon_2}{\varepsilon_1}\sqrt{q^2+\frac{\kappa_2^2}{\kappa_1^2}}}
        \left(\frac{\frac{\sigma_2'}{\sigma_1'}}{q^2+\frac{\kappa_2^2}{\kappa_1^2}}
        - \frac{\frac{\varepsilon_2}{\varepsilon_1}}{q^2+1}\right)\frac{\sigma_2'\varepsilon_1}{\sigma_1'\varepsilon_2}
        \frac{\cos(2q\kappa_1L')}{\sqrt{q^2 + \frac{\kappa_2^2}{\kappa_1^2}}}
        \bigg] &\\
    &=  \frac{\sigma_1'^2}{\kappa_1^2\varepsilon_1}\bigg[
        \frac{\Phi_D}{2(\varepsilon_1\kappa_1+\varepsilon_2\kappa_2)} 
        \left(\frac{\sigma_2'\kappa_1}{\sigma_1'\kappa_2} - \frac{\varepsilon_2\kappa_2}{\varepsilon_1\kappa_1}\right)
        \frac{\kappa_1^2\varepsilon_1^2}{\sigma_1'} &\\
    &- \frac{1}{\kappa_1}\frac{1}{\pi}\frac{\Phi_D\kappa_1\varepsilon_1}{\sigma_1'}\int_0^\infty\d q
        \frac{\sqrt{\frac{q^2}{\kappa_1^2}+1}\sqrt{\frac{q^2}{\kappa_1^2}+\frac{\kappa_2^2}{\kappa_1^2}}}
        {\sqrt{\frac{q^2}{\kappa_1^2}+1}+\frac{\varepsilon_2}{\varepsilon_1}\sqrt{\frac{q^2}{\kappa_1^2}+\frac{\kappa_2^2}{\kappa_1^2}}}
        \left(\frac{\frac{\sigma_2'}{\sigma_1'}}{\frac{q^2}{\kappa_1^2}+\frac{\kappa_2^2}{\kappa_1^2}}
        - \frac{\frac{\varepsilon_2}{\varepsilon_1}}{\frac{q^2}{\kappa_1^2}+1}\right)
        \frac{\sin(2qL')}{\frac{q}{\kappa_1}} &\\
    &+ \frac{1}{\kappa_1}\frac{1}{\pi}\int_0^\infty\d q
        \frac{\sqrt{\frac{q^2}{\kappa_1^2}+\frac{\kappa_2^2}{\kappa_1^2}}}
        {\sqrt{\frac{q^2}{\kappa_1^2}+1}+\frac{\varepsilon_2}{\varepsilon_1}\sqrt{\frac{q^2}{\kappa_1^2}+\frac{\kappa_2^2}{\kappa_1^2}}}
        \left(\frac{\frac{\sigma_2'}{\sigma_1'}}{\frac{q^2}{\kappa_1^2}+\frac{\kappa_2^2}{\kappa_1^2}}
        - \frac{\frac{\varepsilon_2}{\varepsilon_1}}{\frac{q^2}{\kappa_1^2}+1}\right)
        \frac{\cos(2qL')}{\sqrt{\frac{q^2}{\kappa_1^2} + 1}} &\\
    &- \frac{1}{\kappa_1}\frac{1}{\pi}\int_0^\infty\d q
        \frac{\sqrt{\frac{q^2}{\kappa_1^2}+1}}
        {\sqrt{\frac{q^2}{\kappa_1^2}+1}+\frac{\varepsilon_2}{\varepsilon_1}\sqrt{\frac{q^2}{\kappa_1^2}+\frac{\kappa_2^2}{\kappa_1^2}}}
        \left(\frac{\frac{\sigma_2'}{\sigma_1'}}{\frac{q^2}{\kappa_1^2}+\frac{\kappa_2^2}{\kappa_1^2}}
        - \frac{\frac{\varepsilon_2}{\varepsilon_1}}{\frac{q^2}{\kappa_1^2}+1}\right)\frac{\sigma_2'\varepsilon_1}{\sigma_1'\varepsilon_2}
        \frac{\cos(2qL')}{\sqrt{\frac{q^2}{\kappa_1^2} + \frac{\kappa_2^2}{\kappa_1^2}}}
        \bigg] &\\
    &=  \frac{\Phi_D}{2(\varepsilon_1\kappa_1+\varepsilon_2\kappa_2)} 
        \left(\frac{\sigma_2'\kappa_1\varepsilon_1}{\kappa_2} - \frac{\sigma_1'\varepsilon_2\kappa_2}{\kappa_1}\right) &\\
    &- \frac{1}{\pi}\frac{\sigma_1'^2}{\varepsilon_1}\frac{\Phi_D\varepsilon_1}{\sigma_1'}\int_0^\infty\d q
        \frac{p_1p_2}{p_1+\frac{\varepsilon_2}{\varepsilon_1}p_2}
        \left(\frac{\frac{\sigma_2'}{\sigma_1'}}{p_2^2} - \frac{\frac{\varepsilon_2}{\varepsilon_1}}{p_1^2}\right)
        \frac{\sin(2qL')}{q} &\\
    &+ \frac{\sigma_1'^2}{\varepsilon_1}\frac{1}{\pi}\int_0^\infty\d q
        \frac{p_2}{p_1+\frac{\varepsilon_2}{\varepsilon_1}p_2}
        \left(\frac{\frac{\sigma_2'}{\sigma_1'}}{p_2^2} - \frac{\frac{\varepsilon_2}{\varepsilon_1}}{p_1^2}\right)
        \frac{\cos(2qL')}{p_1} &\\
    &- \frac{\sigma_1'^2}{\varepsilon_1}\frac{1}{\pi}\int_0^\infty\d q
        \frac{p_1}{p_1+\frac{\varepsilon_2}{\varepsilon_1}p_2}
        \left(\frac{\frac{\sigma_2'}{\sigma_1'}}{p_2^2} - \frac{\frac{\varepsilon_2}{\varepsilon_1}}{p_1^2}\right)
        \frac{\sigma_2'\varepsilon_1}{\sigma_1'\varepsilon_2} \frac{\cos(2qL')}{p_2} &\\
    &=  \frac{\Phi_D\varepsilon_1\varepsilon_2\kappa_1\kappa_2}{2(\varepsilon_1\kappa_1+\varepsilon_2\kappa_2)} 
        \left(\frac{\sigma_2'}{\varepsilon_2}\frac{1}{\kappa_2^2} - \frac{\sigma_1'}{\varepsilon_1}\frac{1}{\kappa_1^2}\right) &\\
    &- \frac{\Phi_D}{\pi}\int_0^\infty\d q
        \frac{p_1p_2}{p_1+\frac{\varepsilon_2}{\varepsilon_1}p_2}
        \left(\frac{\sigma_2'}{p_2^2} - \frac{\sigma_1'\frac{\varepsilon_2}{\varepsilon_1}}{p_1^2}\right)
        \frac{\sin(2qL')}{q} &\\
    &+ \frac{1}{\pi}\int_0^\infty\d q
        \frac{p_2}{\varepsilon_1p_1+\varepsilon_2p_2}
        \left(\frac{\sigma_1'\sigma_2'}{p_2^2} - \frac{\frac{\varepsilon_2}{\varepsilon_1}\sigma_1'^2}{p_1^2}\right)
        \frac{\cos(2qL')}{p_1} &\\
    &- \frac{1}{\pi}\int_0^\infty\d q
        \frac{p_1}{\varepsilon_1p_1+\varepsilon_2p_2}
        \left(\frac{\sigma_1'\sigma_2'}{p_2^2} - \frac{\frac{\varepsilon_2}{\varepsilon_1}\sigma_1'^2}{p_1^2}\right)
        \frac{\sigma_2'\varepsilon_1}{\sigma_1'\varepsilon_2} \frac{\cos(2qL')}{p_2} &\\
    &=  \frac{\Phi_D\varepsilon_1\varepsilon_2\kappa_1\kappa_2}{2(\varepsilon_1\kappa_1+\varepsilon_2\kappa_2)} 
        \left(\frac{\sigma_2'}{\varepsilon_2}\frac{1}{\kappa_2^2} - \frac{\sigma_1'}{\varepsilon_1}\frac{1}{\kappa_1^2}\right) &\\
    &- \frac{\Phi_D}{\pi}\int_0^\infty\d q
        \frac{p_1p_2}{\varepsilon_1p_1+\varepsilon_2p_2}
        \left(\frac{\sigma_2'}{\varepsilon_2}\frac{1}{p_2^2} - \frac{\sigma_1'}{\varepsilon_1}\frac{1}{p_1^2}\right)
        \frac{\sin(2qL')}{q} &\\
    &+ \frac{1}{\pi}\int_0^\infty\d q
        \frac{1}{\varepsilon_1p_1+\varepsilon_2p_2}
        \left(\frac{\sigma_1'\sigma_2'}{p_1p_2} - \frac{\varepsilon_2}{\varepsilon_1}\frac{p_2}{p_1^3}\right)
        \cos(2qL') &\\
    &- \frac{1}{\pi}\int_0^\infty\d q
        \frac{1}{\varepsilon_1p_1+\varepsilon_2p_2}
        \left(\sigma_2'^2\frac{\varepsilon_1}{\varepsilon_2}\frac{p_1}{p_2^3} - \frac{\sigma_1'\sigma_2'}{p_1p_2}\right)
        \cos(2qL') &\\
    &= \frac{\varepsilon_1\varepsilon_2\kappa_1\kappa_2\Phi_D}{2(\varepsilon_1\kappa_1+\varepsilon_2\kappa_3)}
        \left(-\frac{\sigma_1'}{\varepsilon_1}\frac{1}{\kappa_1^2}+\frac{\sigma_2'}{\varepsilon_2}\frac{1}{\kappa_2^2}
        \right)&\\
        &+ \frac{1}{2\pi}\int_{-\infty}^\infty\frac{1}{\varepsilon_1p_1+\varepsilon_2p_2}\left(
        \frac{2\sigma_1'\sigma_2'}{p_1p_2} - \sigma_1'^2\frac{\varepsilon_2}{\varepsilon_1}\frac{p_2}{p_1^3}
        - \sigma_2'^2\frac{\varepsilon_1}{\varepsilon_2}\frac{p_1}{p_2^3}\right)\cos(2qL') \d q &\\
        &- \frac{\Phi_D}{2\pi}\int_{-\infty}^\infty \frac{\varepsilon_1\varepsilon_2p_1p_2}{\varepsilon_1p_1+\varepsilon_2p_3}
        \left(-\frac{\sigma_1'}{\varepsilon_1}\frac{1}{p_1^2}+\frac{\sigma_2'}{\varepsilon_2}\frac{1}{p_2^2}
        \right)\frac{\sin(2qL')}{q} \d q \quad\checkmark&
}

\paragraph{Conclusion}
Therefore we can obtain the result for the case of
identical particles given in Ref. \cite{supp,lit2} from our general expressions.

\newpage
\begingroup
\setlength{\abovedisplayskip}{0pt}
\setlength{\belowdisplayskip}{0pt}
\setlength{\abovedisplayshortskip}{0pt}
\setlength{\belowdisplayshortskip}{0pt}
\section{Solution Summary}
\subsection{Exact Solution}
\subsubsection{Potential}
\fleq{
    \Phi_1^e(x,z) &= \frac{\sigma_1\cosh(\kappa_1(L-z)) + \sigma_2\cosh(\kappa_1z)}{\varepsilon_1\kappa_1\sinh(\kappa_1L)} &\\
    &+ \frac{\kappa_2\varepsilon_2\Phi_D}{\kappa_1\varepsilon_1+\kappa_2\varepsilon_2}e^{-\kappa_1x} &\\
    &+ \frac{1}{L} \frac{1}{\kappa_1\varepsilon_1+\kappa_2\varepsilon_2}\left(\frac{\sigma_3 + \sigma_4}{\kappa_2} - \frac{1}{\kappa_1} \frac{\kappa_2\varepsilon_2}{\kappa_1\varepsilon_1}(\sigma_1+\sigma_2)\right)e^{-\kappa_1x} &\\
    &+ \sum_{n=1}^\infty \frac{2}{L} \frac{\varepsilon_2p_2}{\varepsilon_1p_1 + \varepsilon_2p_2}\left(\frac{\sigma_3 + (-1)^n \sigma_4}{\varepsilon_2}\frac{1}{p_2^2} - \frac{\sigma_1 + (-1)^n \sigma_2}{\varepsilon_1}\frac{1}{p_1^2}\right)e^{-p_1x}\cos\left(\frac{n\pi z}{L}\right)
}
\fleq{
    \Phi_2^e(x,z) &= \frac{\sigma_3\cosh(\kappa_2(L-z)) + \sigma_4\cosh(\kappa_2z)}{\varepsilon_2\kappa_2\sinh(\kappa_2L)} &\\
    &+ \Phi_D  - \frac{\kappa_1\varepsilon_1\Phi_D}{\kappa_1\varepsilon_1+\kappa_2\varepsilon_2}e^{\kappa_2x} &\\
    &+ \frac{1}{L} \frac{1}{\kappa_1\varepsilon_1+\kappa_2\varepsilon_2}\left(\frac{\sigma_1 + \sigma_2}{\kappa_1} - \frac{1}{\kappa_2} \frac{\kappa_1\varepsilon_1}{\kappa_2\varepsilon_2}(\sigma_3+\sigma_4)\right)e^{\kappa_2x} &\\
    &+ \sum_{n=1}^\infty \frac{2}{L} \frac{\varepsilon_1p_1}{\varepsilon_1p_1 + \varepsilon_2p_2}\left(\frac{\sigma_1 + (-1)^n \sigma_2}{\varepsilon_1}\frac{1}{p_1^2} - \frac{\sigma_3 + (-1)^n \sigma_4}{\varepsilon_2}\frac{1}{p_2^2}\right)e^{p_2x}\cos\left(\frac{n\pi z}{L}\right)
}
\subsubsection{Energy}
\fleq{
    &\gamma_1 = \frac{\sigma_1^2}{2\kappa_1\varepsilon_1} \qquad
     \gamma_2 = \frac{\sigma_2^2}{2\kappa_1\varepsilon_1} &\\
    &\gamma_3 = \frac{\sigma_3^2}{2\kappa_2\varepsilon_2}+\sigma_3\Phi_D \qquad
     \gamma_4 = \frac{\sigma_4^4}{2\kappa_2\varepsilon_2}+\sigma_4\Phi_D &
}
\fleq{
    &\omega_{\gamma, 1}(L) = \frac{1}{4\kappa_1\varepsilon_1}\left((\sigma_1^2 + \sigma_2^2)(\coth(\kappa_1L) - 1)
        + \frac{2\sigma_1\sigma_2}{\sinh(\kappa_1L)}\right) &\\
    &\omega_{\gamma, 2}(L) = \frac{1}{4\kappa_2\varepsilon_2}\left((\sigma_3^2 + \sigma_4^2)(\coth(\kappa_2L) - 1)
        + \frac{2\sigma_3\sigma_4}{\sinh(\kappa_2L)}\right) &
}
\fleq{
    & \gamma_{1,2}
        = -\frac{\varepsilon_1\varepsilon_2\kappa_1\kappa_2\Phi_D^2}{2(\kappa_1\varepsilon_1+\kappa_2\varepsilon_2)}&
}
\fleq{
    \tau_1
        &= \frac{\Phi_D}{\kappa_1\varepsilon_1+\kappa_2\varepsilon_2}
        \left(\frac{\varepsilon_2\kappa_2}{\kappa_1}\sigma_1 - \frac{\varepsilon_1\kappa_1}{\kappa_2}\sigma_3\right) &\\
        &+ \int_0^\infty\frac{1}{\varepsilon_1p_1(x) + \varepsilon_2p_2(x)}\bigg(
        \frac{2\sigma_1\sigma_3}{p_1(x)p_2(x)}
        - \frac{\varepsilon_2p_2(x)}{\varepsilon_1p_1(x)^3}\sigma_1^2
        - \frac{\varepsilon_1p_1(x)}{\varepsilon_2p_2(x)^3}\sigma_3^2\bigg)\d x &
}
\fleq{
    \tau_2
        &= \frac{\Phi_D}{\kappa_1\varepsilon_1+\kappa_2\varepsilon_2}
        \left(\frac{\varepsilon_2\kappa_2}{\kappa_1}\sigma_2 - \frac{\varepsilon_1\kappa_1}{\kappa_2}\sigma_4\right) &\\
        &+ \int_0^\infty\frac{1}{\varepsilon_1p_1(x) + \varepsilon_2p_2(x)}\bigg(
        \frac{2\sigma_2\sigma_4}{p_1(x)p_2(x)}
        - \frac{\varepsilon_2p_2(x)}{\varepsilon_1p_1(x)^3}\sigma_2^2
        - \frac{\varepsilon_1p_1(x)}{\varepsilon_2p_2(x)^3}\sigma_4^2\bigg)\d x &
}
\fleq{
    \omega_\tau(L)
        &= \frac{1}{L} \frac{1}{\kappa_1\varepsilon_1+\kappa_2\varepsilon_2} \left(\frac{(\sigma_1+\sigma_2)(\sigma_3 +\sigma_4)}{2\kappa_1\kappa_2}
        - \frac{1}{4\kappa_1^2}\frac{\kappa_2\varepsilon_2}{\kappa_1\varepsilon_1}(\sigma_1+\sigma_2)^2
        - \frac{1}{4\kappa_2^2}\frac{\kappa_1\varepsilon_1}{\kappa_2\varepsilon_2}(\sigma_3+\sigma_4)^2\right) &\\
        &+ \frac{1}{2L}\sum_{n=1}^\infty \frac{1}{\varepsilon_1p_1+\varepsilon_2p_2}\left(
        \frac{2(\sigma_1+(-1)^n\sigma_2)(\sigma_3+(-1)^n\sigma_4)}{p_1p_2}
        - \frac{1}{p_1}\frac{\varepsilon_2p_2}{\varepsilon_1p_1^2}(\sigma_1+(-1)^n\sigma_2)^2
        - \frac{1}{p_2}\frac{\varepsilon_1p_1}{\varepsilon_2p_2^2}(\sigma_3+(-1)^n\sigma_4)^2\right) &\\
        &- \frac{1}{2}\int_0^\infty\frac{1}{\varepsilon_1p_1(x) + \varepsilon_2p_2(x)}\bigg(
        \frac{2(\sigma_1\sigma_3 + \sigma_2\sigma_4)}{p_1(x)p_2(x)}
        - \frac{\varepsilon_2p_2(x)}{\varepsilon_1p_1(x)^3}\left(\sigma_1^2+\sigma_2^2\right)
        - \frac{\varepsilon_1p_1(x)}{\varepsilon_2p_2(x)^3}\left(\sigma_3^2 + \sigma_4^2\right)\bigg)\d x &
}
\subsubsection{Notation}
\fleq{&p_i = \sqrt{\left(\frac{n\pi}{L}\right)^2 + \kappa_i^2}&}
\fleq{&p_i(x) = \sqrt{x^2\pi^2 + \kappa_i^2}&} 

\newpage
\subsection{Superposition Solution}
\subsubsection{Potential}
\fleq{
    \Phi_1^s(x,z) &= \frac{\sigma_1}{\varepsilon_1\kappa_1}e^{-\kappa_1z} + \frac{\sigma_2}{\varepsilon_1\kappa_1}e^{\kappa_1(z-L)}&\\
    &+ \frac{2\kappa_2\varepsilon_2\Phi_D}{\kappa_1\varepsilon_1+\kappa_2\varepsilon_2}e^{-\kappa_1x} &\\
    &+ \frac{1}{\pi}\int_{-\infty}^{\infty} \frac{\varepsilon_2p_2}{\varepsilon_1p_1+\varepsilon_2p_2}\left(-\frac{\sigma_1}{\varepsilon_1}\frac{1}{p_1^2} + \frac{\sigma_3}{\varepsilon_2}\frac{1}{p_2^2}\right)e^{-p_1x}\cos(qz)\d q &\\
    &+ \frac{1}{\pi}\int_{-\infty}^{\infty} \frac{\varepsilon_2p_2}{\varepsilon_1p_1+\varepsilon_2p_2}\left(-\frac{\sigma_2}{\varepsilon_1}\frac{1}{p_1^2} + \frac{\sigma_4}{\varepsilon_2}\frac{1}{p_2^2}\right)e^{-p_1x}\cos(-q(z-L))\d q &
}
\fleq{
    \Phi_2^s(x,z) &= \frac{\sigma_3}{\varepsilon_2\kappa_2}e^{-\kappa_2z} + \frac{\sigma_4}{\varepsilon_2\kappa_2}e^{\kappa_2(z-L)}&\\
    &+ 2\Phi_D  - 2\frac{\kappa_1\varepsilon_1\Phi_D}{\kappa_1\varepsilon_1+\kappa_2\varepsilon_2}e^{\kappa_2x} &\\
    &+ \frac{1}{\pi}\int_{-\infty}^{\infty} -\frac{\varepsilon_1p_1}{\varepsilon_1p_1+\varepsilon_2p_2}\left(-\frac{\sigma_1}{\varepsilon_1}\frac{1}{p_1^2} + \frac{\sigma_3}{\varepsilon_2}\frac{1}{p_2^2}\right)e^{p_2x}\cos(qz)\d q &\\
    &+ \frac{1}{\pi}\int_{-\infty}^{\infty} -\frac{\varepsilon_1p_1}{\varepsilon_1p_1+\varepsilon_2p_2}\left(-\frac{\sigma_2}{\varepsilon_1}\frac{1}{p_1^2} + \frac{\sigma_4}{\varepsilon_2}\frac{1}{p_2^2}\right)e^{p_2x}\cos(-q(z-L))\d q &
}
\subsubsection{Energy}
\fleq{
    &\gamma_1 = \frac{\sigma_1^2}{2\varepsilon_1\kappa_1} \qquad
     \gamma_2 = \frac{\sigma_2^2}{2\varepsilon_1\kappa_1} &\\
    &\gamma_3 = \frac{\sigma_3^2}{2\varepsilon_2\kappa_2} + \frac{3}{2}\Phi_D\sigma_3 \qquad
     \gamma_4 = \frac{\sigma_4^2}{2\varepsilon_2\kappa_2} + \frac{3}{2}\Phi_D\sigma_4 &
}
\fleq{
    &\omega_{\gamma, 1}(L) = \frac{\sigma_1\sigma_2}{2\varepsilon_1\kappa_1}e^{-\kappa_1L}&\\
    &\omega_{\gamma, 2}(L) = \frac{\sigma_3\sigma_4}{2\varepsilon_2\kappa_2}e^{-\kappa_2L}&
}
\fleq{
    &\gamma_{1,2} = - \frac{\varepsilon_1\varepsilon_2\kappa_1\kappa_2\Phi_D^2}{\varepsilon_1\kappa_1+\varepsilon_2\kappa_2} &
}
\fleq{
    \tau_1
    &= \frac{3}{2}\frac{\Phi_D}{\varepsilon_1\kappa_1+\varepsilon_2\kappa_1}
        \left(\sigma_1\varepsilon_2\frac{\kappa_2}{\kappa_1} - \sigma_3\varepsilon_1\frac{\kappa_1}{\kappa_2}\right) &\\
        &+ \frac{1}{2\pi}\int_{-\infty}^\infty \frac{1}{\varepsilon_1p_1+\varepsilon_2p_2}
        \left(\frac{2\sigma_1\sigma_3}{p_1p_2} - \sigma_1^2\frac{\varepsilon_2}{\varepsilon_1}\frac{p_2}{p_1^3}
        - \sigma_3^2\frac{\varepsilon_1}{\varepsilon_2}\frac{p_1}{p_2^3}\right)\d q &\\
    \tau_2
    &= \frac{3}{2}\frac{\Phi_D}{\varepsilon_1\kappa_1+\varepsilon_2\kappa_1}
        \left(\sigma_2\varepsilon_2\frac{\kappa_2}{\kappa_1} - \sigma_4\varepsilon_1\frac{\kappa_1}{\kappa_2}\right) &\\
        &+ \frac{1}{2\pi}\int_{-\infty}^\infty \frac{1}{\varepsilon_1p_1+\varepsilon_2p_2}
        \left(\frac{2\sigma_2\sigma_4}{p_1p_2} - \sigma_2^2\frac{\varepsilon_2}{\varepsilon_1}\frac{p_2}{p_1^3}
        - \sigma_4^2\frac{\varepsilon_1}{\varepsilon_2}\frac{p_1}{p_2^3}\right)\d q &
}
\fleq{
    \omega_\tau
    &= \frac{\varepsilon_1\varepsilon_2\kappa_1\kappa_2\Phi_D}{4(\varepsilon_1\kappa_1+\varepsilon_2\kappa_3)}
        \left(-\frac{\sigma_1+\sigma_2}{\varepsilon_1}\frac{1}{\kappa_1^2}+\frac{\sigma_3+\sigma_4}{\varepsilon_2}\frac{1}{\kappa_2^2}
        \right)&\\
        &+ \frac{1}{2\pi}\int_{-\infty}^\infty\frac{1}{\varepsilon_1p_1+\varepsilon_2p_2}\left(
        \frac{\sigma_2\sigma_3 + \sigma_1\sigma_4}{p_1p_2} - \sigma_1\sigma_2\frac{\varepsilon_2}{\varepsilon_1}\frac{p_2}{p_1^3}
        - \sigma_3\sigma_4\frac{\varepsilon_1}{\varepsilon_2}\frac{p_1}{p_2^3}\right)\cos(qL) \d q &\\
        &- \frac{\Phi_D}{4\pi}\int_{-\infty}^\infty \frac{\varepsilon_1\varepsilon_2p_1p_2}{\varepsilon_1p_1+\varepsilon_2p_3}
        \left(-\frac{\sigma_1+\sigma_2}{\varepsilon_1}\frac{1}{p_1^2}+\frac{\sigma_3+\sigma_4}{\varepsilon_2}\frac{1}{p_2^2}
        \right)\frac{\sin(qL)}{q} \d q &
}
\subsubsection{Notation}
\fleq{&p_i = \sqrt{q^2 + \kappa_i^2}&}

\endgroup
\newpage

\section{Discussion}
\subsection{Surface interaction $\omega_{\gamma, 1}(L)$}
\subsubsection{Exact Solution}
\fleq{
    &\omega_{\gamma, 1}(L) = \frac{1}{4\kappa_1\varepsilon_1}\left((\sigma_1^2 + \sigma_2^2)(\coth(\kappa_1L) - 1) + \frac{2\sigma_1\sigma_2}{\sinh(\kappa_1L)}\right)
}
\paragraph{Limits}
\fleq{
    \lim_{L\to\infty}\omega_{\gamma, 1}(L) &= \frac{1}{4\kappa_1\varepsilon_1}\left((\sigma_1^2 + \sigma_2^2)(\lim_{L\to\infty}\coth(\kappa_1L) - 1) + \lim_{L\to\infty}\frac{2\sigma_1\sigma_2}{\sinh(\kappa_1L)}\right) &\\
    &=\frac{1}{4\kappa_1\varepsilon_1}\left((\sigma_1^2 + \sigma_2^2)(1 - 1) + 0 \right) & \\
    &= 0 &
}
\fleq{
    \lim_{L\to 0}\omega_{\gamma, 1}(L) &= \lim_{L\to 0}\frac{1}{4\kappa_1\varepsilon_1}\frac{(\sigma_1^2 + \sigma_2^2)(\cosh(k_1L) - \sinh(\kappa_1L)) +2\sigma_1\sigma_2 }{\sinh(\kappa_1L)}{{\longrightarrow (\sigma_1 + \sigma_2)^2}\atop{\longrightarrow 0\qquad\qquad}}&\\
    &= \piecewise{ \lim\limits_{L\to 0}\frac{2\sigma_1^2}{4\kappa_1\varepsilon_1}\frac{\cosh(\kappa_1L) - \sinh(\kappa_1L) - 1}{\sinh(\kappa_1L)} & \sigma_1 = - \sigma_2\\ +\infty & \sigma_1\neq -\sigma_2} &\\
    &\overset{\text{L'Hopital}}{=} \piecewise{ \lim\limits_{L\to 0}\frac{2\sigma_1^2}{4\kappa_1\varepsilon_1}\frac{\sin(\kappa_1L) - \cosh(\kappa_1L)}{\cosh(\kappa_1L)} & \sigma_1 = - \sigma_1\\ +\infty & \sigma_1\neq -\sigma_2} &\\
    &= \piecewise{ -\frac{2\sigma_1^2}{4\kappa_1\varepsilon_1} & \sigma_1 = - \sigma_2\\ +\infty & \sigma_1\neq -\sigma_2}&
}
\paragraph{Asymptotics}
If both $\sigma_1$ and $\sigma_2$ are zero, then $\omega_{\gamma, 1}(L)$ is constantly zero.
This trivial case shall be excluded from the following analysis.
\fleq{
    \omega_{\gamma, 1}(L) &= \frac{1}{4\kappa_1\varepsilon_1}\frac{(\sigma_1^2 + \sigma_2^2)(\cosh(k_1L) - \sinh(\kappa_1L)) +2\sigma_1\sigma_2 }{\sinh(\kappa_1L)}&\\
    &= \frac{1}{4\kappa_1\varepsilon_1}\frac{(\sigma_1^2 + \sigma_2^2)(e^{\kappa_1 L} + e^{-\kappa_1L} - e^{\kappa_1L} + e^{-\kappa_1L}) +4\sigma_1\sigma_2 }{e^{\kappa_1L} - e^{-\kappa_1L}}&\\
    &= \frac{2}{4\kappa_1\varepsilon_1}\frac{(\sigma_1^2 + \sigma_2^2)e^{-\kappa_1L} +2\sigma_1\sigma_2}{e^{\kappa_1L} - e^{-\kappa_1L}}&\\
    &= \frac{2}{4\kappa_1\varepsilon_1}\Bigg((\sigma_1^2 + \sigma_2^2)\ubr{\frac{1}{e^{2\kappa_1L}-1}}{\in\landauO(e^{-2\kappa_1L})} +2\sigma_1\sigma_2\ubr{\frac{1}{e^{\kappa_1L} + e^{-\kappa_1L}}}{\in\landauO(e^{-\kappa_1L})}\Bigg)&\\
    &\in\piecewise{\landauO(e^{-\kappa_1L}) & \sigma_1\sigma_2\neq 0 \\ \landauO(e^{-2\kappa_1L}) & \sigma_1\sigma_2 = 0}
}
So, in general, for large $L$ the interaction energy decays as $e^{-\kappa_1L}$ if both $\sigma_1$ and $\sigma_2$ are non-zero.
In case one of them is zero, the energy decays exponentially twice as fast due to the finite size of our system.
\paragraph{Zeros}
\fleq{
    &\omega_{\gamma, 1}(L) = \frac{1}{4\kappa_1\varepsilon_1}\frac{(\sigma_1^2 + \sigma_2^2)(\cosh(k_1L) - \sinh(\kappa_1L)) +2\sigma_1\sigma_2 }{\sinh(\kappa_1L)} = 0 &\\
    &\gdw (\sigma_1^2 + \sigma_2^2)(\cosh(k_1L) - \sinh(\kappa_1L)) +2\sigma_1\sigma_2 = 0 \qquad\text{ since } \omega_{\gamma,1}(0)\neq 0 \text{ we only need to look at } L>0 &\\
    &\gdw \cosh(k_1L) - \sinh(\kappa_1L) = -\frac{2\sigma_1\sigma_2}{\sigma_1^2 + \sigma_2^2}&\\
    &\gdw e^{-\kappa_1L} = -\frac{2\sigma_1\sigma_2}{\sigma_1^2 + \sigma_2^2}&
}
So there is a zero if and only if
\eq{0 < -\frac{2\sigma_1\sigma_2}{\sigma_1^2 + \sigma_2^2} < 1 \label{wgzc1}}
$\sigma_1\sigma_2 < 0$ is a necessary condition for this. Under this condition we have
\fleq{
    \eqref{wgzc1} &\gdw -2\sigma_1\sigma_2  < \sigma_1^2 + \sigma_2^2  &\\
    &\gdw \sigma_1^2 + \sigma_2^2 + 2\sigma_1\sigma_2 > 0 &\\
    &\gdw (\sigma_1 + \sigma_2)^2 > 0 &\\
    &\gdw \sigma_1 \neq -\sigma_2
}
Thus if $\sigma_1\sigma_2 \ge 0$ or $\sigma_1 = -\sigma_2$ there are no zeroes and if $\sigma_1\sigma_2 < 0$ and $\sigma_1\neq -\sigma_2$ there is exactly one zero at $L = -\frac{1}{\kappa_1}\ln\left(-\frac{2\sigma_1\sigma_2}{\sigma_1^2 + \sigma_2^2}\right)$.
\paragraph{Local extrema}
\fleq{
    \frac{\d}{\d L} \omega_{\gamma, 1}(L) &= \frac{1}{4\varepsilon_1\kappa_1}\left((\sigma_1^2 + \sigma_2^2)\frac{-\kappa_1}{\sinh^2(\kappa_1L)} + 2 \sigma_1\sigma_2\kappa_1\cosh(\kappa_1L)\left(-\frac{1}{\sinh^2(\kappa_1L)}\right)\right)&\\
    &= -\frac{1}{4\varepsilon_1}\left((\sigma_1^2 + \sigma_2^2)\frac{1}{\sinh^2(\kappa_1L)} + 2\sigma_1\sigma_2\frac{\cosh(\kappa_1L)}{\sinh^2(\kappa_1L)}\right) &\\
    &= -\frac{1}{4\varepsilon_1\sinh^2(\kappa_1L)}\left(\sigma_1^2 + \sigma_2^2 + 2\sigma_1\sigma_2\cosh(\kappa_1L)\right) &
}
\fleq{
    \frac{\d}{\d L} \omega_{\gamma, 1}(L) = 0 &\gdw \sigma_1^2 + \sigma_2^2 + 2\sigma_1\sigma_2\cosh(\kappa_1L) = 0&\\
    &\gdw \cosh(\kappa_1L) = -\frac{\sigma_1^2 + \sigma_2^2}{2\sigma_1\sigma_2}&
}
this is solvable if and only if
\eq{-\frac{\sigma_1^2 + \sigma_2^2}{2\sigma_1\sigma_2} \ge 1 \label{wgec1}}
$\sigma_1\sigma_2 < 0$ is a necessary condition for this. Under this condition we have
\fleq{
    \eqref{wgec1} &\gdw -\sigma_1^2 -\sigma_2^2 \le 2\sigma_1\sigma_2 &\\
    &\gdw -\sigma_1^2 -\sigma_2^2 -2\sigma_1\sigma_2 \le 0 &\\
    &\gdw \sigma_1^2 + \sigma_2^2 + 2\sigma_1\sigma_2 \ge 0 &\\
    &\gdw (\sigma_1 + \sigma_2)^2 \ge 0 &
}
i.e., $\sigma_1\sigma_2 < 0$ is both necessary and sufficient.
If $\sigma_1\neq - \sigma_2$ we know that $\omega_{\gamma, 1}(L)$ has exactly one zero
and goes to zero in the limit to infinity, it must have at least one local minimum.
Since we only have one local extremum this extremum has to be a minimum.
If $\sigma_1 = -\sigma_2$ we know that $\omega_{\gamma, 1} < 0$ and $\omega_{\gamma, 1}(L)$
goes to zero in the limit $L\to\infty$ either $L=0$ is a minimum or there is a local minimum with $L>0$.
Since putting $\sigma_1 = -\sigma_2$ into our equation yields only $L=0$ we know there is no local
minimum with $L>0$ so there is a minimum at $L=0$.\\

So we conclude if $\sigma_1\sigma_2 < 0$ we have one local minimum at 
\eq{ L = \frac{1}{\kappa_1}\arcosh\left(-\frac{1}{2}\left(\frac{\sigma_1}{\sigma_2} + \frac{\sigma_1}{\sigma_2}\right)\right)}
and no other local extrema and if $\sigma_1\sigma_2 \ge 0$ we have no local extrema.

\subsubsection{Superposition solution}
\fleq{
    &\omega_{\gamma, 1}(L) = \frac{\sigma_1\sigma_2}{2\kappa_1\varepsilon_1}e^{-\kappa_1L}
}
\paragraph{Limits}
We see that in contrast to the exact solution, the superposition approximation result
does always converge in the limit $L\to 0$.
\paragraph{Asymptotics}
We see that in the case $\sigma_1\sigma_2=0$ the superposition approximation result is zero,
while the exact solution is of order $\landauO(e^{-2\kappa_1L})$. In the case $\sigma_1\sigma_2\neq 0$
the superposition solution predicts the exponential order correctly ($\landauO(e^{-\kappa_1L})$
in accordance with the exact solution) but the prefactor is too small by a factor of two
when compared to the exact solution, i.e.
\eq{\lim_{L\to\infty} \frac{\omega_{\gamma,1}^e(L)}{\omega_{\gamma,1}^s(L)} = 2}
So the superposition solution is unable to give
an asymptotically correct solution for $L\to\infty$.
Furthermore, we can also conclude that the factor of 2
that was also found in Ref. \cite{lit} for identical particles
is not a consequence of the symmetry of the system.
\paragraph{Zeros}
Except for the constant zero cases ($\sigma_1\sigma_2=0$) there are no further zeros.
\paragraph{Local extrema}
There are no local extremas.

\subsubsection{Plots}
Below, we show one plot for the surface interaction energy $\omega_{\gamma,1}(L)$
as a function of the separation distance $L$ for each of the distinctive cases outlined above.
For all the following plots of this chapter we use
$\varepsilon_1=\phieEi$, $\varepsilon_2=\phieEii$, $\kappa_1=\phieKi$, $\kappa_2=\phieKii$, and $\Phi_D=\phiePhiD$.
\\

\textbf{Case 1: $\sigma_1\sigma_2 > 0$}
\nopagebreak

    \begin{figure}[H]
        \centering \small\input{en_pppp_1.tex}\input{en_pppp_3.tex}\normalsize\caption{$\omega_{\gamma,1}(L)$ for $\sigma_1=\phieSi$, $\sigma_2=\phieSii$} \label{plot:og1.1}
    \end{figure}

\textbf{Case 2: $\sigma_1\sigma_2 = 0$}
\nopagebreak

    \begin{figure}[H]
        \centering \small\input{en_p000_1.tex}\input{en_p000_3.tex}\normalsize\caption{$\omega_{\gamma,1}(L)$ for $\sigma_1=\phieSi$, $\sigma_2=0$} \label{plot:og1_2}
    \end{figure}

Since $\omega_{\gamma,1}^s(L)$ is constantly zero, the value for the curve
on the right plot of Fig. \ref{plot:og1_2} is undefined.\\
\textbf{Case 3: $\sigma_1\sigma_2 < 0, \sigma_1 \neq -\sigma_2$}
\nopagebreak

    \begin{figure}[H]
        \centering \small\input{en_pmpp_1.tex}\input{en_pmpp_3.tex}\normalsize\caption{$\omega_{\gamma,1}(L)$ for $\sigma_1=\phieSi$, $\sigma_2=-\phieSii$} \label{plot:og1_3}
    \end{figure}

\textbf{Case 4: $\sigma_1 = -\sigma_2$}
\nopagebreak

    \begin{figure}[H]
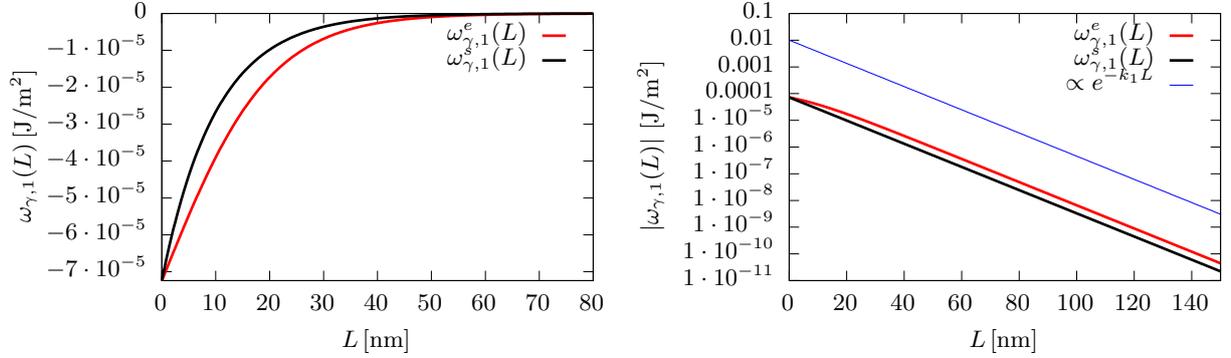

        \centering \small\input{en_anti1_1.tex}\input{en_anti1_3.tex}\normalsize\caption{$\omega_{\gamma,1}(L)$  for $\sigma_1=0.02$, $\sigma_2=-0.02$} \label{plot:og1_4}
    \end{figure}

\subsection{Line tension $\tau_1$}
When comparing the expressions for $\tau_1$ we see that while the integral term is the same
for both the superposition approximation and the exact solution, they differ in the prefactor
of the first term, where the superposition solution has an additional factor of $\frac{3}{2}$
compared to the exact solution. This is due to the fact that for $\sigma_2=\sigma_4=0$ and $L\to\infty$
the exact solution for the potential $\Phi^e$ converges to $\Phi^{\sigma_1,\sigma_3}$
(the Fourier series becomes a Fourier integral in the limit $L\to\infty$) while the superposition
approximation does not because all terms involving $\Phi_D$ are doubled when
we add $\Phi^{\sigma_2,\sigma_4}(x,-(z-L))$ and these terms do not depend on the $\sigma$'s.
So the superposition approximation does not correctly predict $\tau_1$, despite $\tau_1$ being
a property of a semi-infinite system, while the exact solution does.

\subsection{Line interaction $\omega_\tau(L)$}
Since an analytical analysis of the expression is hard, we will look at several different cases
numerically.
Since the parameter space of $\omega_\tau^e(L)$ ($\omega_\tau^s(L)$) is 8 (9) dimensional,
our following analysis can of course be nothing more than a sytematic exploration of a
part of the possible behaviours,
but it already shows a lot of interesting behaviour.
\\

First, we plot $\omega_\tau^e(L)$ as a function of $L$ for different values of the parameters
$\sigma_4$, $\kappa_1$, $\kappa_2$, $\varepsilon_1$, $\varepsilon_2$, and $\Phi_D$.
Then we compare the exact solution $\omega_\tau^e(L)$ and the superposition solution
$\omega_\tau^s(L)$ for different combinations of the charge densities $\sigma_1$,
$\sigma_2$, $\sigma_3$, and $\sigma_4$. This also includes the special case
of oppositely charged plates. And finally, we discuss the case where
one or more of the four charge densities equal to zero.
\\

As shown by the plots the line interaction energy can be positive as well as negative,
depending upon the parameters of the system. For a given system, it can even change
sign with changing separation. Also, it is evident from the plots that for small
separations the superposition approximation predicts a qualitatively wrong behaviour
at short separations.

\subsubsection{Variation of a charge ($\sigma_4$)}

    \begin{figure}[H]
        \centering \small\input{en_var_sigma_ppp_1.tex}\input{en_var_sigma_ppp_2.tex}\normalsize\caption{$\omega_\tau(L)$ (left) and location and magnitude of its Maximum (right) for $\sigma_1=\phieSi$, $\sigma_2=\phieSii$, $\sigma_3=\phieSiii$} \label{plot:ot_var_sigma_ppp_1}
    \end{figure}

So we see from Fig. \ref{plot:ot_var_sigma_ppp_1} that we already have qualitatively different behaviour from
from that seen in \cite{lit}: For slightly positive values and for all negative values we see non-monotonic behaviour and a maximum.

\subsubsection{Variation of the first inverse Debye length $\kappa_1$}

    \begin{figure}[H]
        \centering \small\input{en_var_kappa_ppp_2.tex}\input{en_var_kappa_ppp_3.tex}\normalsize\caption{$\omega_\tau(L)$ (left) and location and magnitude of its Maximum (right) for $\sigma_1=\phieSi$, $\sigma_2=-\phieSii$, $\sigma_3=-\phieSiii$, $\sigma_4=\phieSiv$} \label{plot:ot_var_kappa_ppp_2}
    \end{figure}

\subsubsection{Variation of inverse Debye length $\kappa_2$}

    \begin{figure}[H]
        \centering \small\input{en_var_kappa2_pmpm_1.tex}\input{en_var_kappa2_pmpm_4.tex}\normalsize\caption{$\omega_\tau(L)$ (left) and location and magnitude of its minimum (right) for
$\sigma_1=\phieSi$, $\sigma_2=-\phieSii$, $\sigma_3=\SI{0.004}{e/nm^2}$, $\sigma_4=-\SI{0.002}{e/nm^2}$,
$\varepsilon_1=\SI{2}{\varepsilon_0}$, $\varepsilon_2=\SI{20}{\varepsilon_0}$, $\kappa_1=\SI{0.3}{nm^{-1}}$} \label{plot:ot_var_kappa2_ppp_3}
    \end{figure}

    \begin{figure}[H]
        \centering \small\input{en_var_kappa2_pmpm_2.tex}\input{en_var_kappa2_pmpm_3.tex}\normalsize\caption{First (left) and Second (right) Maximum of $\omega_\tau(L)$ for
$\sigma_1=\phieSi$, $\sigma_2=-\phieSii$, $\sigma_3=\SI{0.004}{e/nm^2}$, $\sigma_4=-\SI{0.002}{e/nm^2}$,
$\varepsilon_1=\SI{2}{\varepsilon_0}$, $\varepsilon_2=\SI{20}{\varepsilon_0}$, $\kappa_1=\SI{0.3}{nm^{-1}}$} \label{plot:ot_var_kappa2_ppp_4}
    \end{figure}

Here we see another qualitative difference from \cite{lit}:
We see in Fig. \ref{plot:ot_var_kappa2_ppp_3}, \ref{plot:ot_var_kappa2_ppp_4} that it is possible to
have a minimum and a maximum, but also a minimum and two maxima depending on the parameters.

\subsubsection{Variation of permittivity $\varepsilon_1$}

    \begin{figure}[H]
        \centering \small\input{en_var_epsilon_ppp_2.tex}\input{en_var_epsilon_ppp_4.tex}\normalsize\caption{$\omega_\tau(L)$ and location and magnitude of its first maximum for $\sigma_1=\phieSi$, $\sigma_2=-\phieSii$, $\sigma_3=-\phieSiii$, $\sigma_4=\phieSiv$} \label{plot:ot_var_epsilon_ppp_2}
    \end{figure}

    \begin{figure}[H]
        \centering \small\input{en_var_epsilon_ppp_7.tex}\input{en_var_epsilon_ppp_6.tex}\normalsize\caption{$\omega_\tau(L)$ and location and magnitude of its minimum for $\sigma_1=\phieSi$, $\sigma_2=-\phieSii$, $\sigma_3=-\phieSiii$, $\sigma_4=\phieSiv$} \label{plot:ot_var_epsilon_ppp_6}
    \end{figure}

    \begin{figure}[H]
        \centering \small\input{en_var_epsilon_ppp_3.tex}\input{en_var_epsilon_ppp_5.tex}\normalsize\caption{$\omega_\tau(L)$ and location and magnitude of its second maximum for $\sigma_1=\phieSi$, $\sigma_2=-\phieSii$, $\sigma_3=-\phieSiii$, $\sigma_4=\phieSiv$} \label{plot:ot_var_epsilon_ppp_3}
    \end{figure}

Here we have depending on the parameter either a maximum and a minimum, two maxima and a minimum or just a maximum.

\subsubsection{Variation of permittivity $\varepsilon_2$}

    \begin{figure}[H]
        \centering \small\input{en_var_epsilon2_ppp_2.tex}\input{en_var_epsilon2_ppp_6.tex}\normalsize\caption{$\omega_\tau(L)$ and location and magnitude of its first minimum for $\sigma_1=\phieSi$, $\sigma_2=-\phieSii$, $\sigma_3=-\phieSiii$, $\sigma_4=\phieSiv$} \label{plot:ot_var_epsilon2_ppp_2}
    \end{figure}

    \begin{figure}[H]
        \centering \small\input{en_var_epsilon2_ppp_4.tex}\input{en_var_epsilon2_ppp_5.tex}\normalsize\caption{Location and magnitude of the first (left) and second (right) maximum of $\omega_\tau(L)$ for $\sigma_1=\phieSi$, $\sigma_2=-\phieSii$, $\sigma_3=-\phieSiii$, $\sigma_4=\phieSiv$} \label{plot:ot_var_epsilon2_ppp_3}
    \end{figure}

Here we have depending on the parameter either a maximum or a minimum and two maxima.

\subsubsection{Variation of the Donnan potential $\Phi_D$}
\label{en_disc_var_phid}
It is clear from the analytic expressions that the exact solution for the line interaction energy
does not depend upon the Donnan potential $\Phi_D$.
However, quite surprisingly, the superposition expression
for the line interaction energy does depend upon $\Phi_D$.
Below we show the variation of
the line interaction energy $\omega_\tau^s(L)$ with the variation of $\Phi_D$
and discuss the asymptotic behaviour of the ratio $\omega_\tau^e(L)/\omega_\tau^s(L)$.
\\

From this discussion we exclude the cases that do not have a three phase contact line
and therefore have zero line interaction energies, as will be detailed in \ref{en_disc_zero}.
Furthermore we distinguish two cases here.
The finite size effect case and the non finite size effect case.
The finite size effect occurs if one of the walls in not charged,
and is analyzed in \ref{en_disc_zero}.

\paragraph{Finite size effect case}
If $\Phi_D$ is zero then the superposition solution is constantly zero
(see \ref{en_disc_zero}).

If $\Phi_D$ is not zero, then the ratio converges to
zero because of the different asymptotics outlined in \ref{en_disc_zero},
as can be seen in Figs. \ref{plot:ot_p0p0} and \ref{plot:ot_0p0p}.

\paragraph{Non finite size effect case}
If $\Phi_D$ is zero then in all cases we observe,
the ratio of exact solution to superposition solution converges against two.\\

    \begin{figure}[H]
        \centering \small\input{en_var_psid_ppp_1.tex}\input{en_var_psid_ppp_3.tex}\normalsize\caption{$\omega_\tau(L)$ for $\sigma_1=\phieSi$, $\sigma_2=\phieSii$, $\sigma_3=\phieSiii$, $\sigma_4=\phieSiv$} \label{plot:ot_var_psid_ppp_1}
    \end{figure}

    \begin{figure}[H]
        \centering \small\input{en_var_psid_pp0_1.tex}\input{en_var_psid_pp0_3.tex}\normalsize\caption{$\omega_\tau(L)$ for $\sigma_1=\phieSi$, $\sigma_2=\phieSii$, $\sigma_3=0$, $\sigma_4=\phieSiv$} \label{plot:ot_var_psid_pp0_1}
    \end{figure}

For most other non finite size effect cases the ratio tends to increasingly deviate from the factor of two with raising $\Phi_D$,
but the exact behaviour tends to differ. The cases outlined in \ref{en_disc_all_signs} show behaviour similar
to that displayed in Fig. \ref{plot:ot_var_psid_ppp_1}, while those in \ref{en_disc_zero} display behaviour
similar to that displayed in Fig. \ref{plot:ot_var_psid_pp0_1}.

\subsubsection{All signs}
\label{en_disc_all_signs}
In the following plots we consider all 16 possible sign combinations for the charge densities
$\sigma_1$, $\sigma_2$, $\sigma_3$ and $\sigma_4$.
If we regard the exact solution $\omega_\tau^e(L)$ as a function of the $\sigma$'s, we have
\eq{\fa{s\in\R}\omega_\tau^e(s\sigma_1,s\sigma_2,s\sigma_3,s\sigma_4;L) = s^2\omega_\tau^e(\sigma_1,\sigma_2,\sigma_3,\sigma_4;L)}
or more specifically for case $s=-1$
\eq{\omega_\tau^e(-\sigma_1,-\sigma_2,-\sigma_3,-\sigma_4;L) = \omega_\tau^e(\sigma_1,\sigma_2,\sigma_3,\sigma_4;L)}
i.e., the exact solution is invariant regarding inversion of all signs.
For the superposition solution this in not true and can be seen in the following plot
where we have a sign combination on the left and the combination with all signs inverted on the right hand side.

    \begin{figure}[H]
        \centering \small\input{en_pppp_2.tex}\input{en_mmmm_2.tex}\normalsize\caption{$\omega_\tau(L)$ for  $\sigma_1=\phieSi$, $\sigma_2=\phieSii$, $\sigma_3=\phieSiii$, $\sigma_4=\phieSiv$
(left) and inverted signs (right)} \label{plot:ot_pppp}
    \end{figure}

    \begin{figure}[H]
        \centering \small\input{en_pppm_2.tex}\input{en_mmmp_2.tex}\normalsize\caption{$\omega_\tau(L)$ for  $\sigma_1=\phieSi$, $\sigma_2=\phieSii$, $\sigma_3=\phieSiii$, $\sigma_4=-\phieSiv$
(left) and inverted signs (right)} \label{plot:ot_pppm}
    \end{figure}

    \begin{figure}[H]
        \centering \small\input{en_ppmp_2.tex}\input{en_mmpm_2.tex}\normalsize\caption{$\omega_\tau(L)$ for  $\sigma_1=\phieSi$, $\sigma_2=\phieSii$, $\sigma_3=-\phieSiii$, $\sigma_4=\phieSiv$
(left) and inverted signs (right)} \label{plot:ot_ppmp}
    \end{figure}

    \begin{figure}[H]
        \centering \small\input{en_ppmm_2.tex}\input{en_mmpp_2.tex}\normalsize\caption{$\omega_\tau(L)$ for  $\sigma_1=\phieSi$, $\sigma_2=\phieSii$, $\sigma_3=-\phieSiii$, $\sigma_4=-\phieSiv$
(left) and inverted signs (right)} \label{plot:ot_ppmm}
    \end{figure}

    \begin{figure}[H]
        \centering \small\input{en_pmpp_2.tex}\input{en_mpmm_2.tex}\normalsize\caption{$\omega_\tau(L)$ for  $\sigma_1=\phieSi$, $\sigma_2=-\phieSii$, $\sigma_3=\phieSiii$, $\sigma_4=\phieSiv$
(left) and inverted signs (right)} \label{plot:ot_pmpp}
    \end{figure}

    \begin{figure}[H]
        \centering \small\input{en_pmpm_2.tex}\input{en_mpmp_2.tex}\normalsize\caption{$\omega_\tau(L)$ for  $\sigma_1=\phieSi$, $\sigma_2=-\phieSii$, $\sigma_3=\phieSiii$, $\sigma_4=-\phieSiv$
(left) and inverted signs (right)} \label{plot:ot_pmpm}
    \end{figure}

    \begin{figure}[H]
        \centering \small\input{en_pmmp_2.tex}\input{en_mppm_2.tex}\normalsize\caption{$\omega_\tau(L)$ for  $\sigma_1=\phieSi$, $\sigma_2=-\phieSii$, $\sigma_3=-\phieSiii$, $\sigma_4=\phieSiv$
(left) and inverted signs (right)} \label{plot:ot_pmmp}
    \end{figure}

    \begin{figure}[H]
        \centering \small\input{en_pmmm_2.tex}\input{en_mppp_2.tex}\normalsize\caption{$\omega_\tau(L)$ for  $\sigma_1=\phieSi$, $\sigma_2=-\phieSii$, $\sigma_3=-\phieSiii$, $\sigma_4=-\phieSiv$
(left) and inverted signs (right)} \label{plot:ot_pmmm}
    \end{figure}

In all the 16 cases discussed above, both $\omega_\tau^e(L)$ and $\omega_\tau^s(L)$ show similar asymptotic
behaviour to that displayed in Fig. \ref{plot:ot_as1}, i.e., they vary $\sim \exp(-\kappa_2L)$ for $L\to\infty$.

    \begin{figure}[H]
        \centering \small\input{en_pppp_4.tex}\input{en_pppp_6.tex}\normalsize\caption{$\omega_\tau(L)$ for  $\sigma_1=\phieSi$, $\sigma_2=\phieSii$, $\sigma_3=\phieSiii$, $\sigma_4=\phieSiv$} \label{plot:ot_as1}
    \end{figure}

\subsubsection{Lines}

    \begin{figure}[H]
        \centering \small\input{en_line1_2.tex}\normalsize\caption{$\omega_\tau(L)$ for $\sigma_1=\phieSi$, $\sigma_2=\phieSii$, $\sigma_3=\phieSi$, $\sigma_4=\phieSii$
} \label{plot:ot_line1}
    \end{figure}

In the case depicted in Fig. \ref{plot:ot_line1} we still have a three-phase contact line,
even though both walls are homogeneously charged because we still have non-identical fluids
on both sides of the interface.
If we additionally set $\kappa_1=\kappa_2$ and $\varepsilon_1=\varepsilon_2$ we can see directly from
the formula given in Eqs. \eqref{ote} and \eqref{ots}, that the line interaction energy is constantly zero for both the exact solution and
superposition approximation,
which is to be expected since we in that case have two identical fluids and two homogeneously charged
walls, i.e. there is no line and therefore no line interaction.
The same holds more generally in any case with $\sigma_1=\sigma_3$, $\sigma_2=\sigma_4$, 
$\kappa_1=\kappa_2$ and $\varepsilon_1=\varepsilon_2$, since all these cases have no line.

    \begin{figure}[H]
        \centering \small\input{en_line3_2.tex}\input{en_line4_2.tex}\normalsize\caption{$\omega_\tau(L)$ for $\sigma_1=\phieSi$, $\sigma_2=\phieSii$, $\sigma_3=\phieSii$, $\sigma_4=\phieSi$
(left) and additionally $\varepsilon_1=\varepsilon_2=\SI{2}{\varepsilon_0}$, $\kappa_1 = \kappa_2 = \SI{0.1}{nm^{-1}}$ (right)} \label{plot:ot_line3}
    \end{figure}

In the case of the right hand side plot we still have a line, since each wall has two different charges
on it, although it is not a three phase contact line anymore since the liquids on each ``side'' are the same.
\\

All four of these cases have (except obviously the case of zero energy) the usual asymptotics for $L\to\infty$.
\\

The exact solution $\omega_\tau^e(L)$ usually diverges because of its first term that is proportional to $\frac{1}{L}$,
but there are two principal ways to overcome this. Either all the terms in it vanish separately,
i.e. $\sigma_1=-\sigma_2$ and $\sigma_3=-\sigma_4$ (for a discussion of this case see \ref{antisym})
or the terms cancel each other. The latter can be achieved by $\kappa_1=\kappa_2$, $\varepsilon_1=\varepsilon_2$
and $2(\sigma_1+\sigma_2)(\sigma_3+\sigma_4)-(\sigma_1+\sigma_2)^2-(\sigma_3+\sigma_4)^2=-(\sigma_1+\sigma_2-\sigma_3-\sigma_4)^2=0$,
i.e. $\sigma_1+\sigma_2-\sigma_3-\sigma_4=0$, as is the case in Fig. \ref{plot:ot_line3} on the right side
and we see indeed that $\omega_\tau^e(L)$ converges this case.
\\
It should also be noted that the $\frac{1}{L}$ term in the exact solution will always cause convergence
to negative infinity since the coefficient is always $\le 0$:
\eq{
    \frac{1}{L} \ubr{\frac{1}{\kappa_1\varepsilon_1+\kappa_2\varepsilon_2}}{\displaystyle >0}
    \ubr{\left(\frac{(\sigma_1+\sigma_2)(\sigma_3 +\sigma_4)}{2\kappa_1\kappa_2}
    - \frac{1}{4\kappa_1^2}\frac{\kappa_2\varepsilon_2}{\kappa_1\varepsilon_1}(\sigma_1+\sigma_2)^2
    - \frac{1}{4\kappa_2^2}\frac{\kappa_1\varepsilon_1}{\kappa_2\varepsilon_2}(\sigma_3+\sigma_4)^2\right)}
    {\displaystyle
        = -\left(\frac{\sigma_1+\sigma_2}{2\kappa_1\sqrt{\frac{\kappa_1\varepsilon_1}{\kappa_2\varepsilon_2}}}
                -\frac{\sigma_3+\sigma_4}{2\kappa_2\sqrt{\frac{\kappa_2\varepsilon_2}{\kappa_1\varepsilon_1}}}
              \right)^2
      \le 0}
}
\subsubsection{Antisymmetric Charges}
\label{antisym}
As discussed previously, $\omega_\tau^e(L)$ converges in the case $\sigma_1=-\sigma_2$ and $\sigma_3=-\sigma_4$.
This is obviously an effect that can not occur in the case of identical particles discussed in Ref. \cite{lit}.
We also notice that in this case $\omega_\tau^s(L)$ does not depend on $\Phi_D$ and $\omega_\tau^e(L)/\omega_\tau^s(L)$
always converges to 2 in the limit $L\to\infty$.

    \begin{figure}[H]
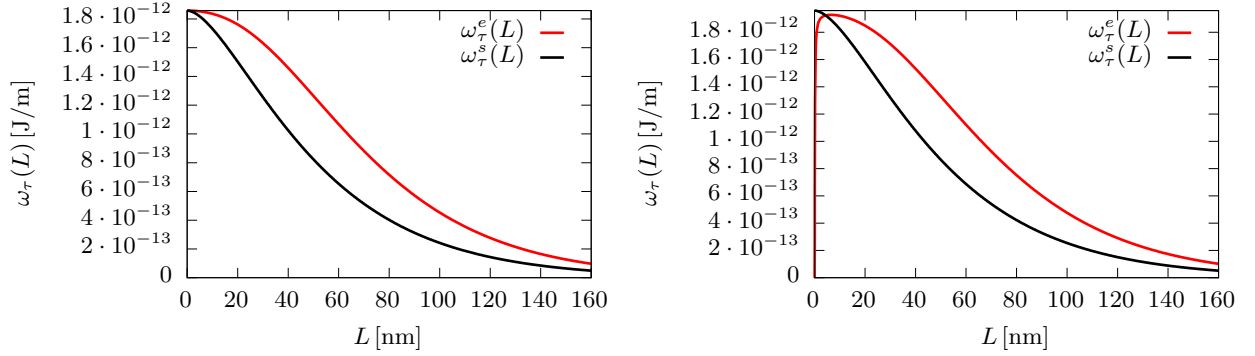

        \centering \small\input{en_anti12_2.tex}\input{en_anti13_2.tex}\normalsize\caption{$\omega_\tau(L)$ for $\sigma_1=\SI{0.02}{e/nm^2}$, $\sigma_2=\SI{-0.02}{e/nm^2}$, $\sigma_3=\SI{0.002}{e/nm^2}$, $\sigma_4=\SI{-0.002}{e/nm^2}$
(left) and $\sigma_1=\SI{0.02}{e/nm^2}$, $\sigma_2=\SI{-0.02}{e/nm^2}$, $\sigma_3=\SI{0.0021}{e/nm^2}$, $\sigma_4=\SI{-0.002}{e/nm^2}$ (right)} \label{plot:ot_anti12}
    \end{figure}

In Fig. \ref{plot:ot_anti12} we see such a converging case on the left side. On the right side, we see the behaviour
for a small deviation in $\sigma_3$ from the previous case.
\\

All these cases have the usual exponential decay in the asymptotic limit $L\to\infty$.

\subsubsection{Zero Charges}
\label{en_disc_zero}

    \begin{figure}[H]
        \centering
        \scalebox{0.67}{\input{en_p0p0_2.tex}}
        \scalebox{0.67}{\input{en_p0p0_4.tex}}
        \scalebox{0.67}{\input{en_p0p0_6.tex}}
        \caption{$\omega_\tau(L)$ for $\sigma_1=\phieSi$, $\sigma_2=0$, $\sigma_3=\phieSiii$, $\sigma_4=0$} \label{plot:ot_p0p0}
    \end{figure}

    \begin{figure}[H]
        \centering
        \scalebox{0.67}{\input{en_0p0p_2.tex}}
        \scalebox{0.67}{\input{en_0p0p_4.tex}}
        \scalebox{0.67}{\input{en_0p0p_6.tex}}
        \caption{$\omega_\tau(L)$ for $\sigma_1=0$, $\sigma_2=\phieSii$, $\sigma_3=0$, $\sigma_4=\phieSiv$} \label{plot:ot_0p0p}
    \end{figure}

If only one wall is charged, then we have an exponential decrease in the exact solution
with exponent $-2\kappa_2L$, as seen in Figs. \ref{plot:ot_p0p0} and \ref{plot:ot_0p0p}.
This is again a finite size effect,
because the relevant distance of interaction is now twice as long.
The superposition solution can of course not show a finite size effect and still shows
a exponential decrease with exponent $-\kappa_2L$ if $\Phi_D\neq 0$.
If $\Phi_D$ is zero then the superposition solution is a constant zero, as can be seen
from Eq. \eqref{ots}.

    \begin{figure}[H]
        \centering
        \scalebox{0.67}{\input{en_p000_2.tex}}
        \scalebox{0.67}{\input{en_p000_4.tex}}
        \scalebox{0.67}{\input{en_p000_6.tex}}
        \caption{$\omega_\tau(L)$ for $\sigma_1=\phieSi$, $\sigma_2=0$, $\sigma_3=0$, $\sigma_4=0$} \label{plot:ot_p000}
    \end{figure}

The same behaviour can be seen even if only one of the four wall parts is changed, and
even if the charged wall is not on the side of the medium with the smaller $\kappa$,
as can be seen in Fig. \ref{plot:ot_p000}.

    \begin{figure}[H]
        \centering
        \scalebox{0.67}{\input{en_p00p_2.tex}}
        \scalebox{0.67}{\input{en_p00p_4.tex}}
        \scalebox{0.67}{\input{en_p00p_6.tex}}
        \caption{$\omega_\tau(L)$ for $\sigma_1=\phieSi$, $\sigma_2=0$, $\sigma_3=0$, $\sigma_4=\phieSiv$} \label{plot:ot_p00p}
    \end{figure}

The same finite size effect does not occur if the charged walls are diagonally aligned, as can be seen
in Fig. \ref{plot:ot_p00p}. In this case a zero $\Phi_D$ leads to
a convergence of $\frac{\omega_\tau^e(L)}{\omega_\tau^s(L)}$ to two,
similar to the behaviour presented in \ref{en_disc_var_phid}.

    \begin{figure}[H]
        \centering
        \scalebox{0.67}{\input{en_pm00_2.tex}}
        \scalebox{0.67}{\input{en_pm00_4.tex}}
        \scalebox{0.67}{\input{en_pm00_6.tex}}
        \caption{$\omega_\tau(L)$ for  $\sigma_1=\phieSi$, $\sigma_2=-\phieSii$, $\sigma_3=0$, $\sigma_4=0$} \label{plot:ot_pm00}
    \end{figure}

Since $\omega_\tau^s(L)$ changes sign around $\simeq\SI{30}{nm}$ we see a discontinuity
in the plot of $\omega_\tau^e(L)/\omega_\tau^s(L)$ at that position.

    \begin{figure}[H]
        \centering
        \scalebox{0.67}{\input{en_00pp_2.tex}}
        \scalebox{0.67}{\input{en_00pp_4.tex}}
        \scalebox{0.67}{\input{en_00pp_6.tex}}
        \caption{$\omega_\tau(L)$ for  $\sigma_1=0$, $\sigma_2=0$, $\sigma_3=\phieSiii$, $\sigma_4=\phieSiv$} \label{plot:ot_00pp}
    \end{figure}

The cases where the only the walls touching one of the media are charged
show no significantly different behaviour than the previous case,
as can be seen in Fig. \ref{plot:ot_pm00} and \ref{plot:ot_00pp}.

    \begin{figure}[H]
        \centering
        \scalebox{0.67}{\input{en_pp0p_2.tex}}
        \scalebox{0.67}{\input{en_pp0p_4.tex}}
        \scalebox{0.67}{\input{en_pp0p_6.tex}}
        \caption{$\omega_\tau(L)$ for  $\sigma_1=\phieSi$, $\sigma_2=\phieSii$, $\sigma_3=0$, $\sigma_4=\phieSiv$} \label{plot:ot_pp0p}
    \end{figure}

Even the case with just one zero charge in Fig. \ref{plot:ot_pp0p} shows behaviour similar to the previous cases
in $\frac{\omega_\tau^e(L)}{\omega_\tau^s(L)}$.

\chapter{Conclusion}
\label{ch:conclusion}
In summary, using a simplified model system with parallel plates in contact
with two immiscible fluids, we have derived analytical expressions
for surface and line interaction energies exactly as well as under the superposition
approximation. Our results can be used to calculate the interaction between
closely separated, not too strongly charged particles trapped at an electrolyte interface.
We consider a general situation where the particles and therefore, the plates
of our model system are not identical. As a result, they can carry different
charge densities even when in contact with the same liquid.
Our results clearly show a rich behaviour concerning both the surface
and the line interaction energies that can not be seen for identical particles.

\paragraph{Surface interaction}
The exact solution for surface interaction $\omega_{\gamma,i}^e(L)$ ($i\in\m{1,2}$)
can be non-monotonic and can have a minimum.
The superposition solution always converges for $L\to0$, but the exact
solution only converges if and only if the charges are of opposite sign and equal
magnitude. Otherwise the exact solution diverges against (positive) infinity.
Furthermore, the exact solution can have qualitatively different
asymptotic behaviour from the superposition solution $\omega_{\gamma,i}^s(L)$.
In case only one of the walls is charged, the superposition solution
predicts a zero surface interaction energy,
while the exact solution predicts a nonzero energy
that decays as $\landauO(e^{-2\kappa_i L})$ for $L\to\infty$.
In the cases where both walls are charged both the exact solution and the
superposition approximation decays as $\landauO(e^{-\kappa_i L})$,
but we still see a factor of two asymptotically in the ratio
$\omega_{\gamma,i}^e(L)/\omega_{\gamma,i}^s(L)$.
The factor of two is thus not a result of the symmetry of charges in Ref. \cite{lit}.

\paragraph{Line interaction}
Both the exact and the superposition solution for
the line interaction can show non-monotonic behaviour
and can show minima and maxima depending upon the parameters of the system.
The exact solution does not depend on the Donnan potential $\Phi_D$,
while the superposition solution generally does.
The superposition solution always converges for $L\to 0$, while
the exact solution usually diverges against negative infinity.
If ($\sigma_1=-\sigma_2$ and $\sigma_3=-\sigma_4$) or
($\kappa_1=\kappa_2$ and $\varepsilon_1=\varepsilon_2$ and $\sigma_1+\sigma_2=\sigma_3+\sigma_4$)
the exact solution does converge for $L\to 0$.
In these cases the superposition solution does not depend on $\Phi_D$.
If $\kappa_1=\kappa_2$, $\varepsilon_1=\varepsilon_2$, $\sigma_1=\sigma_3$ and $\sigma_2=\sigma_4$
there is no three-phase contact line and both exact and superposition solution are
constantly zero, as expected.
In the cases where both walls have a charge
both the exact solution and the
superposition approximation asymptotically (for $L\to\infty$)
decay exponentially with the smaller of the two inverse Debye lengths,
which we will denote with $\kappa_i=\min\m{\kappa_1,\kappa_2}$ for
the remainder of this paragraph. So, if both walls are charged,
which is true for most cases, both exact and superposition solution
are of order $\landauO(e^{-\kappa_i L})$. If $\Phi_D$ is zero
the ratio $\omega_\tau^e(L)/\omega_\tau^s(L)$ converges to two in
these cases and if $\Phi_D$ is nonzero we usually see
increasingly different behaviour with increasing $\Phi_D$.
However, if only one of the walls is charged,
the exact solution has qualitatively different
asymptotic behaviour from the superposition solution:
In case only one of the walls is charged, the superposition solution
predicts a zero line interaction energy if $\Phi_D$ is zero and
a line interaction energy in $\landauO(e^{-\kappa_i L})$ otherwise,
while the exact solution always predicts a nonzero energy
that is in $\landauO(e^{-2\kappa_i L})$.
In these cases $\omega_\tau^e(L)/\omega_\tau^s(L)$ therefore converges
to zero if $\Phi_D$ is non-zero and is undefined otherwise.
\\

Possible topics of interest beyond the scope of this work:
\begin{itemize}
    \item
        It would be interesting to compare the this linear model of nonidentical particles
        with a non-linear numerical model.
    \item
        It would be interesting to compare our solutions with a superposition solution
        derived from using the spherical solution for one particle from Ref. \cite{sph}.
    \item 
        An analytic discussion of the line interaction would be desirable.
\end{itemize}

\newpage
\section*{Acknowledgment}
First and foremost, I want to thank my supervisor Arghya Majee for
his guidance, assistance, a lot of helpful suggestions and corrections
without whom this work would be a lot more taciturn.
\\

Furthermore, I want to thank Markus Bier for the admission
at the Max Planck Institute, helpful discussions and support in all
official matters.
\\

I also want to thank the staff of the Max Planck
Institute and department Dietrich for providing me with a
nice and friendly work environment.
\\

And last but not least I want to thank my family for their support.

\newpage
\section*{Offical statement}
Hiermit erkl\"are ich, Timo Schmetzer,
\begin{itemize}
    \item
        dass ich die vorliegende Arbeit selbstst\"andig verfasst habe
    \item
        dass ich keine anderen als die angegeben Quellen benutzt und alle
        w\"ortlich oder sinngem\"a\ss aus anderen Werken \"ubernommenen
        Aussagen als solche gekennzeichnet habe
    \item
        dass die eingereichte Arbeit weder vollst\"andig noch in wesentlichen
        Teilen Gegenstand eines anderen Prüfungsverfahrens gewesen ist,
    \item
        dass ich die Arbeit weder vollst\"andig noch in Teilen bereits
        ver\"offentlicht habe
    \item
        und dass der Inhalt des elektronischen Exemplars mit dem des
        Druckexemplars \"ubereinstimmt.
\end{itemize}
\vspace{2cm}
Timo Schmetzer

\end{document}